\documentclass[usenatbib,useAMS]{mn2e}
\usepackage{epsfig,psfig,graphicx,epstopdf,float,color}

\def\mnras{MNRAS}
\def\apj{ApJ}
\def\apjl{ApJL}

\def\aj{AJ}
\def\araa{Annu. Rev. Astron. Astrophys.}
\def\prd{PRD}
\def\aap{Astron. \& Astrophys.}

\def\physrep{Physics Reports}
\def\aaps{A\&AS}
\def\E{\textit{E}}
\def\B{\textit{B}}

\title[Radio Weak Lensing with VLA and MERLIN]{Radio Weak Gravitational Lensing with VLA and MERLIN}
\author[Patel et al.]{P. Patel$^1$\thanks{Prina.Patel@port.ac.uk},
  D. J. Bacon$^1$, R. J. Beswick$^{2,3}$, T.
  W. B. Muxlow$^{2,3}$ \& \newauthor B. Hoyle$^{1}$\\\\$^1$Institute of Cosmology \& Gravitation,
  University of Portsmouth, Dennis Sciama Building, Portsmouth, PO1
  3FX, UK.\\$^2${\it e}-MERLIN/VLBI National Radio Astronomy Facility, Jodrell Bank Observatory, The University of Manchester, \\Macclesfield, Cheshire, SK11 9DL, UK.\\$^3$Jodrell Bank Centre for Astrophysics, School of Physics and Astronomy, The University of Manchester, Oxford Road,\\ Manchester, M13 9PL, UK}
\begin{document}
\date{Accepted ----. Received ----; in original form ----.}
\pagerange{\pageref{firstpage}--\pageref{lastpage}} \pubyear{2009}
\maketitle
\label{firstpage}

\begin{abstract}
We carry out an exploratory weak gravitational lensing analysis on a combined VLA and MERLIN radio data set: a deep ($3.3\mu$Jy beam$^{-1}$ rms noise) 1.4\,GHz image of the Hubble Deep Field North. We measure the shear estimator distribution at this radio sensitivity for the first time, finding a similar distribution to that of optical shear estimators for HST ACS data in this field. We examine the residual systematics in shear estimation for the radio data, and give cosmological constraints from radio-optical shear cross-correlation functions. We emphasize the utility of cross-correlating shear estimators from radio and optical data in order to reduce the impact of systematics. Unexpectedly we find no evidence of correlation between optical and radio intrinsic ellipticities of matched objects; this result improves the properties of optical-radio lensing cross-correlations. We explore the ellipticity distribution of the radio counterparts to optical sources statistically, confirming the lack of correlation; as a result we suggest a connected statistical approach to radio shear measurements.
\end{abstract}

\section{Introduction}

\label{intro}

Gravitational lensing is a powerful probe of the large scale structure of the Universe, due to its ability to study the matter distribution without differentiating between luminous, baryonic matter and dark matter \citep[e.g.][]{2001PhR...340..291B,2003ARA&A..41..645R, 2008PhR...462...67M}. The effect has the added benefit of being sensitive to the geometry of the universe, making it useful for the study of the dark energy \citep{2002PhRvD..65f3001H}.

Almost all cosmic shear analyses so far have been carried out using optical data. However, future radio surveys could be competitive for lensing when compared to their optical counterparts, for several reasons. Firstly, radio sensitivities are rapidly reaching a level at which the radio emission from ordinary galaxies will be routinely resolved (e.g. with {\it e}-MERLIN\footnote{http://www.e-merlin.ac.uk/}, LOFAR\footnote{http://www.lofar.org/}, and eventually SKA\footnote{http://www.skatelescope.org/}; c.f. \citealp{2004MNRAS.352..131S}), leading to a comparable source density to that found in the optical. An advantage of radio over optical is that radio interferometers have well known and deterministic beam patterns which may allow for more accurate modeling of the effective PSF.

The only detection of cosmic shear at radio wavelengths so far was carried out by \citet{2004ApJ...617..794C}, using the FIRST survey \citep{1995ApJ...450..559B,1997ApJ...475..479W}. This has a detection threshold of 1\,mJy, with a consequent  $\simeq 20$ resolved sources per square degree useable for weak lensing; this is a much lower number density than found in deep optical shear surveys, with a concomitant lower signal-to-noise on the final cosmological constraints. However, the differential radio source counts at 1.4\,GHz show an increase at flux densities below 1\,mJy, e.g. \citet{2004MNRAS.352..131S}, and it is this increase in the number density at the micro-Jansky level that makes future radio weak lensing attractive.

Given this, it is of interest to make an initial foray into deep radio weak lensing with some of the most sensitive radio data currently available.  Using MERLIN and VLA deep observations of the Hubble Deep Field North (HDF-N) \citep{2005MNRAS.358.1159M}, we seek to understand the properties of the radio source ellipticity distribution down to $\simeq 50\,\mu$Jy, including systematic effects, and to make radio shear measurements in this field, providing cosmological constraints. Along the way, we will compare the radio shapes with corresponding shapes measured in the HDF-N optical image observed with the HST ACS \citep{2004ApJ...600L..93G}, in order to examine the power of cross-correlating shear estimators from the optical and the radio. This modest investigation will therefore pave the way for more ambitious radio weak shear surveys (and joint radio-optical surveys) in the near future.

Our paper is organised as follows: in \S\ref{basics} we outline the theoretical background and notation for weak gravitational lensing which we will use throughout, and then introduce the data used in \S\ref{data}. In \S\ref{measurements} we describe the methods used to measure shear estimators for each source detected in our image; then in \S\ref{results} we present the results of this process, giving shear estimator histograms for radio and optical data, and examining the impact of systematic effects. We present shear correlation functions for the radio and optical data, together with optical-radio shear cross-correlation functions, providing constraints on cosmological parameters. In \S\ref{cross} we compare the optical and radio shear estimators, finding little correlation between the intrinsic ellipticities of objects in the two bands; we show how this adds to the usefulness of cross-correlating optical and radio shear estimators. In \S\ref{blind} we present an alternative statistical method of cross-correlating the two datasets, before drawing conclusions in \S\ref{conclusions}. 


\section{Weak Gravitational Lensing}
\label{basics}

In this section we briefly describe the relevant theory behind weak gravitational lensing and the cosmic shear statistics used in this work (e.g. \citet{2001PhR...340..291B}, \citet{2003ARA&A..41..645R}, \citet{2008PhR...462...67M} for comprehensive reviews).

As light travels from a distant source to an observer it is deflected by the tidal gravitational field of the intervening matter. This deflection angle maps the position of the source $\vec{\beta}$ on the source plane to position $\vec{\theta}$ on the image plane. Since the deflection angle varies from place to place, extended sources like galaxies suffer a change in their observed shape; in the weak lensing limit this effect can be quantified by a $2\times 2$ distortion matrix
\begin{equation}
\mathcal{A}\equiv\frac{\partial\vec{\beta}}{\partial\vec{\theta}}=\delta_{ij}-\partial_i\partial_j\phi=\left( \begin{array}{cc} 1-\kappa-\gamma_{1} & -\gamma_{2} \\ -\gamma_{2} & 1-\kappa+\gamma_{2} \\ \end{array} \right)
\end{equation}
where $\kappa$ is the convergence, which describes the isotropic change in the object size, and $\gamma_{1}$ and $\gamma_{2}$ are shear components describing anisotropic stretching of the object along the $x$-axis and at 45$^{\circ}$ to the $x$-axis. $\partial_i$ is the partial derivative with respect to $\theta_i$, and $\phi$ is the lensing potential, related to the Newtonian gravitational potential $\Phi$ via the line of sight of sight integral
\begin{equation}
\label{eq:lenspot}
\phi(\vec{\chi})=\frac{2}{c^2}\int_{0}^{\chi}\frac{(\chi-\chi^{\prime})}{\chi\chi^{\prime}}\Phi(\vec{\chi^{\prime}})\,d\chi^{\prime},
\end{equation}
where $\chi$ is the comoving distance. The shear and convergence can then be written as  
\begin{equation}
\label{eq:shearcon}
\kappa=\frac{1}{2}(\partial_1^2+\partial_2^2)\phi,\,\,\,\gamma_{1}=\frac{1}{2}(\partial_1^2-\partial_2^2)\phi,\,\,\,\gamma_{2}=\partial_1\partial_2\phi.
\end{equation}
For the statistical analysis of cosmic shear we require suitable 2-point statistics. Here we only consider a 2-dimensional shear field (i.e. projected on the sky), although a full 3-dimensional field can be used, e.g. \citet{2003MNRAS.343.1327H}. Following \citet{1991ApJ...380....1M} we define tangential and $45^{\circ}$ rotated shear components, $\gamma_{t}$ and $\gamma_{r}$, in relation to the line connecting a pair of galaxies:
\begin{equation}
\gamma_{t}=\gamma_{1}\cos{2\theta}+\gamma_{2}\sin{2\theta} \\
\gamma_{r}=\gamma_{2}\cos{2\theta}-\gamma_{1}\sin{2\theta},
\end{equation}
where $\theta$ is the position angle between the $x$-axis and the line that connects the two galaxies. There are three pair-wise shear correlation functions 
\begin{eqnarray}
\label{eq:cfn}
\xi_{tt}(\theta)&\equiv&\langle\gamma_{t}(\vec{\theta^{\prime}})\gamma_{t}(\vec{\theta^{\prime}}+\vec{\theta})\rangle \nonumber \\  
\xi_{rr}(\theta)&\equiv&\langle\gamma_{r}(\vec{\theta^{\prime}})\gamma_{r}(\vec{\theta^{\prime}}+\vec{\theta})\rangle \nonumber \\
\xi_{tr}(\theta)&\equiv&\langle\gamma_{t}(\vec{\theta^{\prime}})\gamma_{r}(\vec{\theta^{\prime}}+\vec{\theta})\rangle
\end{eqnarray}
where averaging is done over all pairs of galaxies separated by an angle $\theta=|\vec\theta|$; in practice, we will average over pairs in a bin, i.e. pairs with separation $\theta\pm\Delta\theta$. Due to parity symmetry $\xi_{tr}=\xi_{rt}$ should equal zero, and this can be used to test for residual systematics. We further define the correlation functions $\xi_{\pm}(\theta)$ which are related to the convergence power spectrum $P_{\kappa}$ by 
\begin{equation}
\label{eq:rotcfn}
\xi_{\pm}(\theta)\equiv\xi_{tt}(\theta)\pm\xi_{rr}(\theta)=\int_{0}^{\infty}\,\frac{\ell}{2\pi}P_{\kappa}(\ell)J_{0,4}(\ell\theta)\,d\ell
\end{equation} 
where $J_{n}$ is a Bessel function of order $n$. The convergence power spectrum is a projection of the matter power spectrum $P_{\delta}$ along the line of sight e.g. \citet{2001PhR...340..291B},
\begin{eqnarray}
\label{eq:pkpdel}
P_{\kappa}(\ell)&=&\frac{9}{4}\frac{\Omega_{m}^{2}H_{0}^{4}}{c^{4}}\int_{0}^{\chi_{l}}\,\frac{1}{a^{2}(\chi)}P_{\delta}\left(\frac{\ell}{f_{K}(\chi)};\chi\right)\,d\chi \nonumber \\ &\times&\left[\int_{\chi}^{\chi_{l}}\,n(\chi^{\prime})\frac{f_{K}(\chi^{\prime}-\chi)}{f_{K}(\chi^{\prime})}\,d\chi^{\prime}\right]^{2},
\end{eqnarray}
where $\chi$ is the comoving distance along the line of sight and $\chi_{l}$ is the maximum comoving distance of the survey. $f_{K}(\chi)$ is the comoving angular diameter distance, $n(\chi)$ is the source distribution and $\ell$ is the wavenumber on the sky.

Weak gravitational lensing arises from scalar perturbations of the space-time metric, and therefore the shear field is expected to possess no handedness. The statistics of this scalar field can thus be described by a single correlation function, meaning that there must exist a degeneracy between the two correlation functions in equation (\ref{eq:rotcfn}). \citet{2002ApJ...568...20C} have shown how to transform these two correlation functions into gradient (\E-mode) and curl (\B-mode) components. The \E-mode contains the true lensing signal while the \B-mode is expected to be zero in the absence of noise. Clearly, a non-zero \B-mode indicates a non-gravitational contribution to the shear field and can therefore be used as a test for systematics. \citet{2002ApJ...568...20C} and \citet{2002ApJ...567...31P} derive the decomposition for the shear correlation functions,
\begin{equation}
\label{eq:cecb}
\xi_{E,B}(\theta)=\frac{\xi_{+}(\theta)\pm\xi^{\prime}(\theta)}{2},
\end{equation}
where the definition of $\xi^{\prime}(\theta)$ is also given in \citet{2002A&A...389..729S}
\begin{equation}
\label{eq:cprime}
\xi^{\prime}(\theta)=\xi_{-}(\theta)+4\int_{\theta}^{\infty}\frac{d\theta^{\prime}}{\theta^{\prime}}\,\xi_{-}(\theta^{\prime})-12\theta^{2}\int_{\theta}^{\infty}\frac{d\theta^{\prime}}{\theta^{\prime\,3}}\,\xi_{-}(\theta^{\prime}).
\end{equation}
In practice data will only contain information out to some radius $\theta_{1}$, and so the integrals will be truncated. 


\section{Data and Catalogue Creation}
\label{data}
In this section we describe the radio and optical observations we will analyse, and give details of how the source catalogues were prepared. 
\subsection{Radio data}
\label{imagemaking}

The HDF-N has been subject to some of the deepest radio observations
made to date using both the VLA and MERLIN
\citep{2005MNRAS.358.1159M}. In this study we have used data combined
from these two arrays to provide both high sensitivity and
sub-arcsecond imaging across a
large field-of-view. These combined data were originally presented by
\citet{2005MNRAS.358.1159M}. For completeness in the following section
we include a summary of the basic characteristics of these
observations, the data processing and creation of the science images
used for this study.

\subsubsection{Radio Observations}

The VLA in A-array observed the HDF-N for a total of 50\,h at 1.4-GHz
in November 1996. A detailed description of these observations and
their calibration can be found in \citet{2000ApJ...533..611R} and
\citet{2005MNRAS.358.1159M}. To provide wide-field imaging
capabilities and minimize the effects of chromatic aberration these
data were observed using the VLA in `pseudo-continuum mode',
correlating parallel hands of polarisaton into  7$\times$3.125-MHz
channels, centred on intermediate frequencies of 1365 and 1435
MHz. These data were subsequently time averaged to 13\,s and the full
field to the first side-lobe of the primary beam was imaged at low
angular resolution, employing significant spatial tapering, to identify
bright, confusing sources which might contaminate the inner portion
of the field. Each identified confusing source, above 0.5\,mJy, was
then carefully imaged in full spectral-line mode to account for
changes in the primary beam response over the passband and then
subtracted from the {\it uv} data. Over the course of the 50\,h
observations several isolated time segments were identified where it
was not possible to adequately subtract these confusing sources. This
incomplete subtraction of confusing sources in individual time
segments was diagnosed as probably associated with time variable
telescope point errors, or possibly low-level interference. All data
associated with these times was deleted, resulting in a total of 42\,h
of high-quality data which was used for all subsequent imaging and
combination with the MERLIN data.

Between 1996 February and 1997 April the MERLIN array, including the
76-m Lovell telescope, was used to observe the same location for a
total of 18.23\,d \cite[see][]{2005MNRAS.358.1159M}. All observations
were made at 1.40\,GHz, with the parallel hands of circular
polarisation correlated into 32 channels of width 0.5-MHz with 4\,s
time averaging. This configuration allowed imaging over a
10-by-10\,arcmin field, comparable to the half-power beamwidth of the
Lovell telescope (12.4\,arcmin).  Due to the lack of short
interferometer spacings in the MERLIN data and the primary beam of the
Lovell telescope, confusing sources beyond the 10-arcmin field caused
no significant problems. However to mitigate any potential problems
the four brightest sources within the 10-arcmin field were mapped and
subtracted from the visibility data prior to combining these MERLIN
data with data from the VLA. 

\subsubsection{Data Combination and Image Creation}

For both the MERLIN and the VLA data the observations and correlator
configurations were optimised for each of the separate arrays;
maximising the sensitivity and field-of-view available and minimising
the effects of time and chromatic aberration. Consequently each of
these two data-sets contains fundamentally different configurations
including different intermediate frequencies and channel
configurations. Due to these data structure differences, a combination
of computing and software limitations, and the scientific requirements
to image a very large field-of-view the combination of these two
data-sets in the visibility plane was not feasible. 

Following and extending the method outlined in \citet{2005MNRAS.358.1159M} these MERLIN and VLA data were imaged and combined in the sky plane completely covering a mosaic area of 8.5-by-8.5\,arcmin centred on the HDF-N \cite[see also][]{2007A&A...472..805R, 2008MNRAS.385.1143B}. Initially small individual image facets were created from both the VLA and MERLIN data at a series of positions completely covering the 8.5-by-8.5\,arcmin field separately and with no deconvolution applied. In each case these individual, `dirty' map and beam facets were created by naturally weighting the data and using the optimum griding spacing for each of the arrays (MERLIN cellsize 0\farcs05 and VLA 0\farcs4).  Following the re-griding of the coarse sampled VLA `dirty' maps and beams, the individual `dirty' map and beam facets of the MERLIN and VLA data were averaged to produce combined images with no deconvolution applied. The central quarter of these combined dirty map facets were then deconvolved with the combined-array dirty beams using a conventional {\sc clean} algorithm \citep{1974A&AS...15..417H} as implemented within the {\sc aips} routine {\sc aplcn}. The resultant facets were restored using a circularly symmetric Gaussian beam of width 0\farcs4. Each individual deconvolved facet was then mosaiced together, fully accounting for sky-curvature effects, to create a single clean image of the field. 

Extensive tests comparing this image based data combination with a visibility plane combination of smaller mosaics from the same data are detailed in \citet[][see their Figure\,1]{2005MNRAS.358.1159M}. These tests show that these two approaches produce essentially identical results.


\subsubsection{Radio Catalogue}

For both radio and optical source extraction, we use the freely available software SExtractor \citep{1996A&AS..117..393B}. Although SExtractor was developed for the analysis of optical data, several authors \citep{2003A&A...403..857B,2008MNRAS.383...75G,2008MNRAS.387.1037G,2007AJ....133.1331H} have shown that it is able to generate reliable noise maps and locate objects within radio images. It should be noted that SExtractor is highly sensitive to input parameters and there is no single output catalogue that will be suitable for all applications.

We make two radio catalogues. Firstly we create a more conservative `gold' set
based on the catalogue of \citet{2006MNRAS.371..963B}. That work used the 
$10^{\prime}\times10^{\prime}$ VLA pointing alone with an independent reduction, detecting objects at $5\sigma$ in the VLA image alone. Their
catalogue contains 537 sources, 83 of which are contained within the
co-added images we use here; the others are either outside the field
or resolved away by the much higher MERLIN resolution. We used
SExtractor in {\tt ASSOC} mode to produce an output catalogue of these
sources.

In order to produce a larger, fainter catalogue we use SExtractor in the standard mode to detect islands of flux above a given threshold. This requires a reliable noise map, which is made by estimating the local background noise at each mesh point of a grid across the image \citep{1996A&AS..117..393B}. The mesh size is an important input parameter; if chosen to be too small the background estimation is  affected by the presence of real objects; if chosen to be too large the small scale variations in the background cannot be reproduced. We adopt a size of 32 pixels, which corresponds to a 2\arcsec scale. We find 3.45$\mu$Jy beam$^{-1}$ rms noise in the resulting map, in close agreement with  \citet{2005MNRAS.358.1159M}. Using this noise map, sources with a total flux greater than 10$\mu$Jy $(\sim\!\!3\sigma)$ were extracted. This catalogue contains 691 objects and we refer to as the `silver' set of objects.

\subsection{Optical Data}

The optical data we use forms part of the Great Observatories Origins Deep Survey (GOODS) based on multiband HST imaging of the HDF and CDF. The HDF-N has been imaged in the ACS F435W, F606W, F814W and F850LP bands ($B, V, i \mbox{ and } z$ respectively) although for the purposes of this work we make use of only the $z$-band image and catalogue.

The $z-$band images were observed in 5 epochs separated by 40-50 days. In the odd numbered epochs each $10\arcmin\times16\arcmin$ field was tiled with a grid of $3\times5$ individual ACS pointings. In the even numbered epochs the field was rotated by $45^{\circ}$ and tiled with 16 separate pointings. The $z-$band image exposure time was typically 2100s divided into 4 exposures to ensure good cosmic ray rejection. In each exposure the telescope field of view was shifted by a small amount to allow optimal sampling of the PSF. The multiple epochs were then combined into a single mosaic. The observations and image reduction are described in detail in \citet{2004ApJ...600L..93G}. We use the publicly available SExtractor configuration files (specified for each band at http://archive.stsci.edu/pub/hlsp/goods/$\mbox{catalog}_{-}$r2/) to make our catalogues; these configuration files have been fine tuned to minimise the number of false detections. 


\section{Shear Measurement Methodology}
\label{measurements}
In this section we describe the methods used to make estimators of the shear for all of our radio and optical sources.
\subsection{Shapelets}
\label{shapelets}
\begin{figure*}
\centering
\psfig{figure=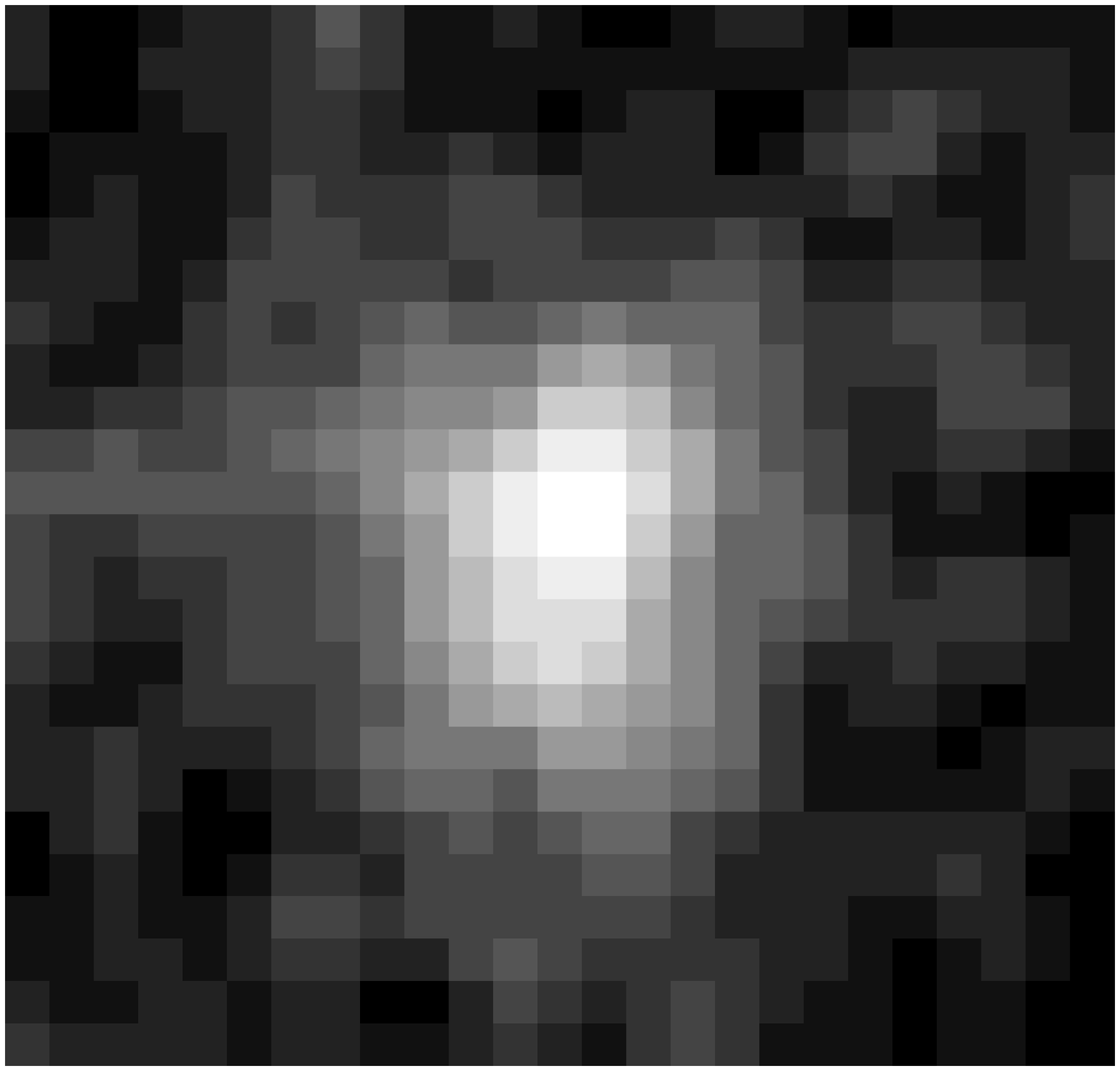,width=20mm}\psfig{figure=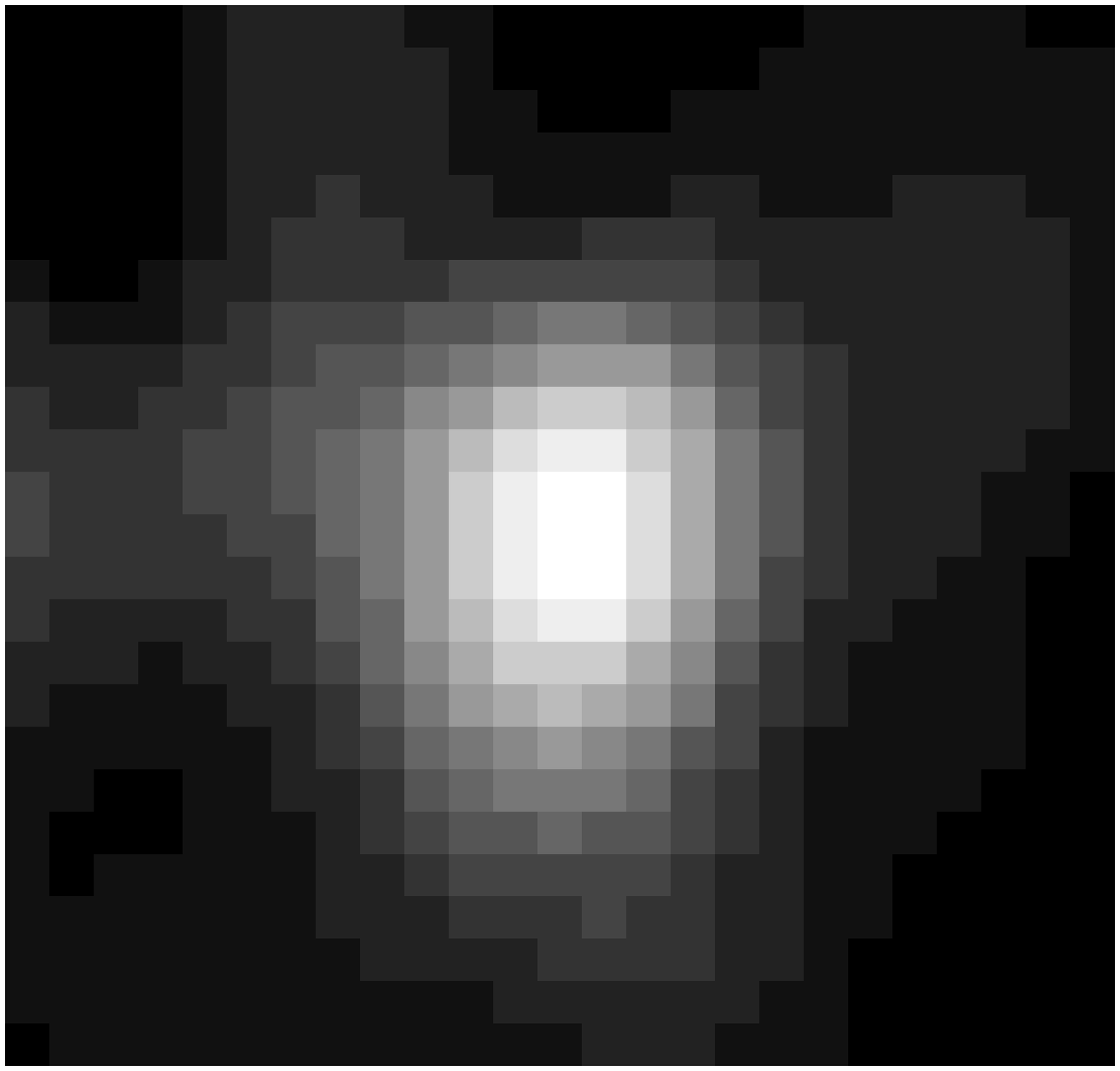,width=20mm}\hspace{2mm}
\psfig{figure=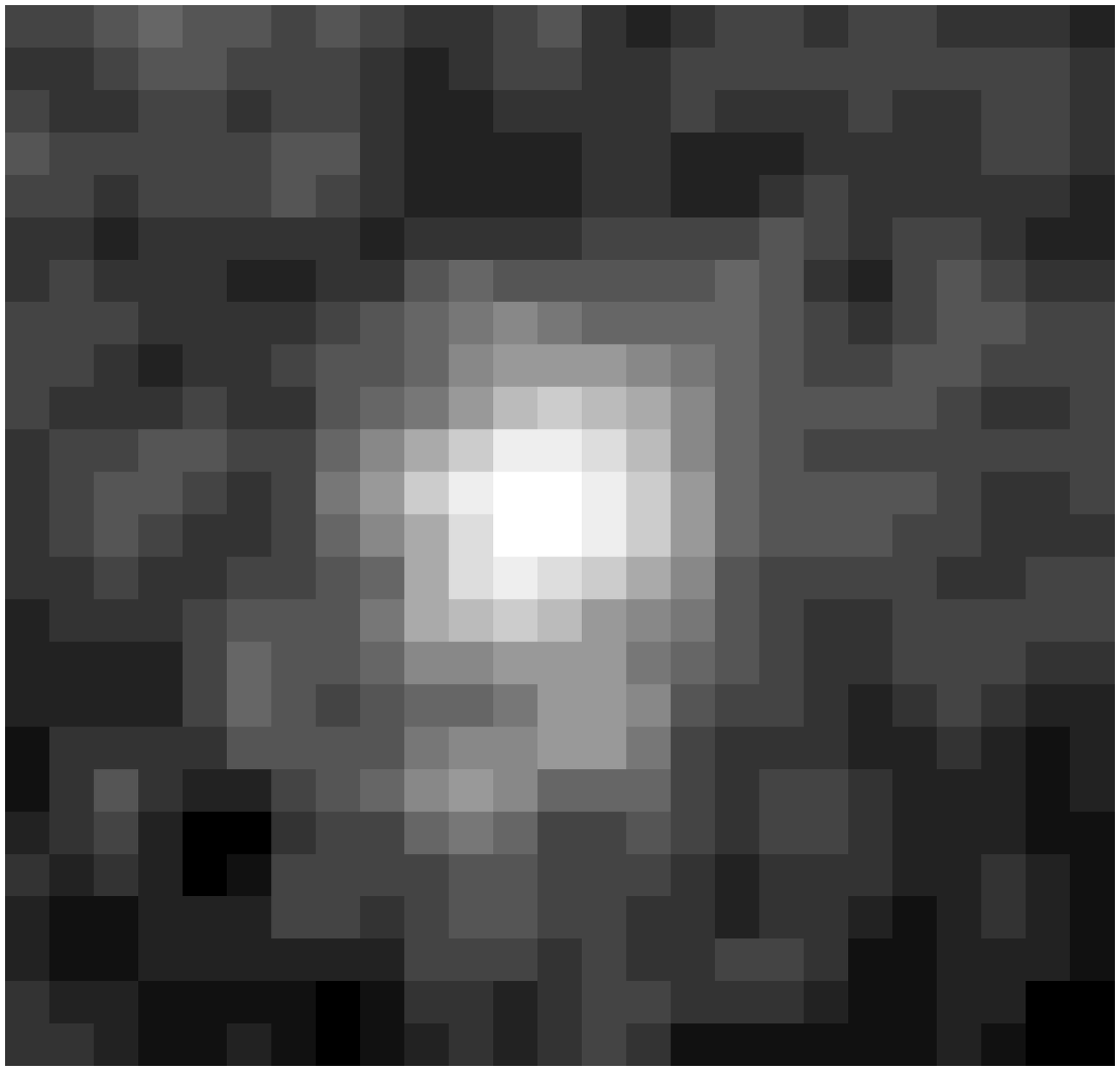,width=20mm}\psfig{figure=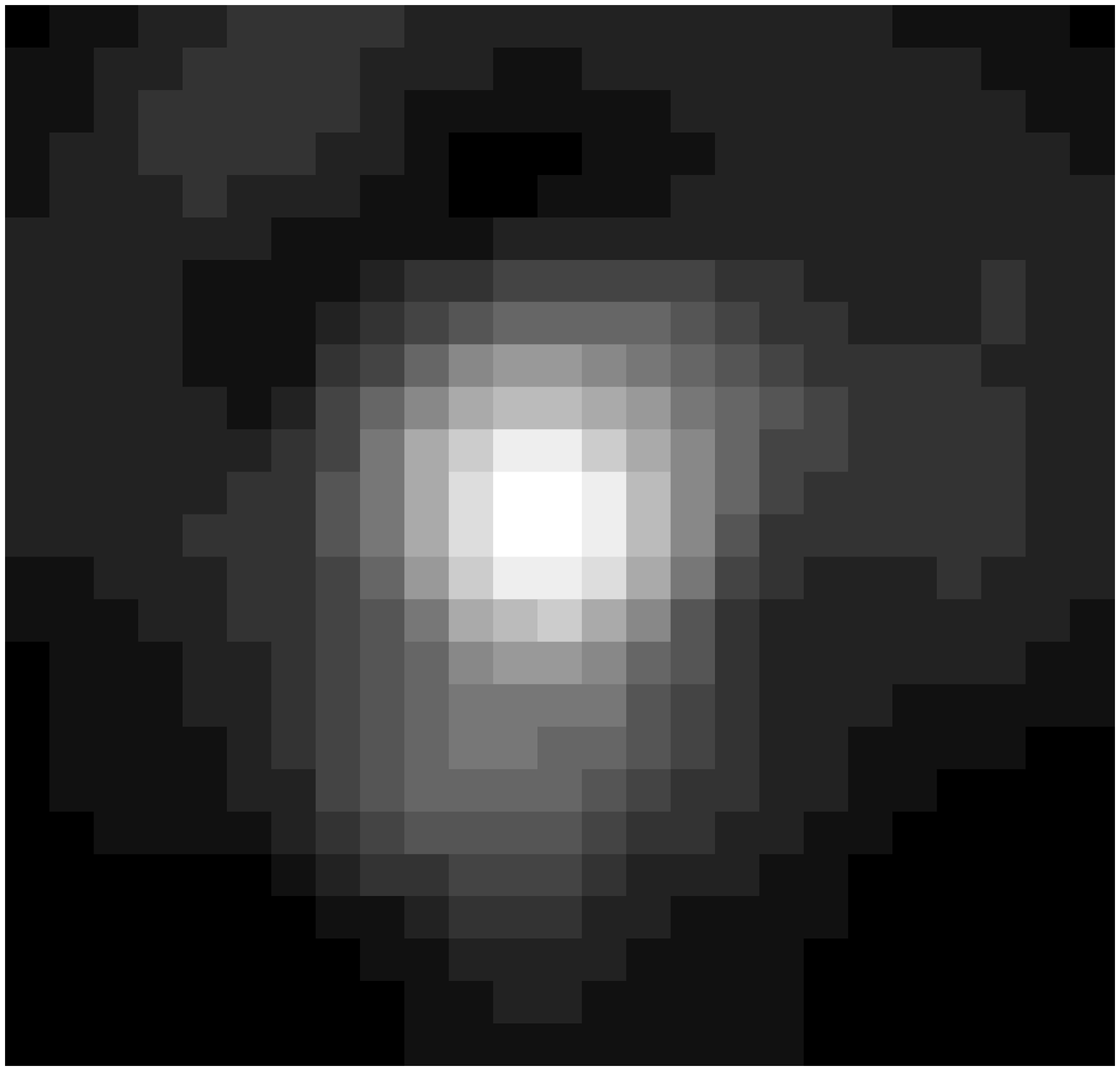,width=20mm}\hspace{2mm}
\psfig{figure=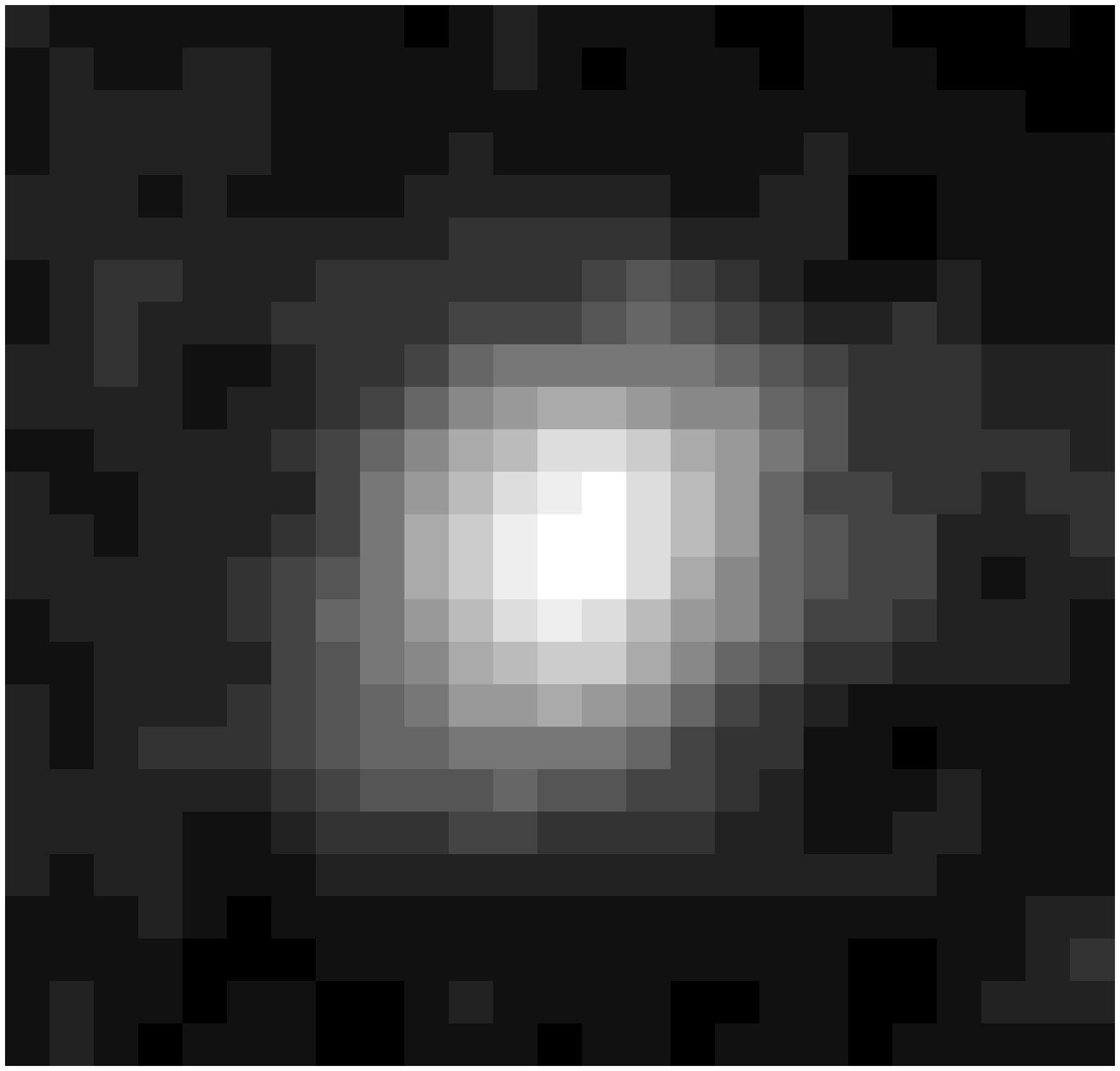,width=20mm}\psfig{figure=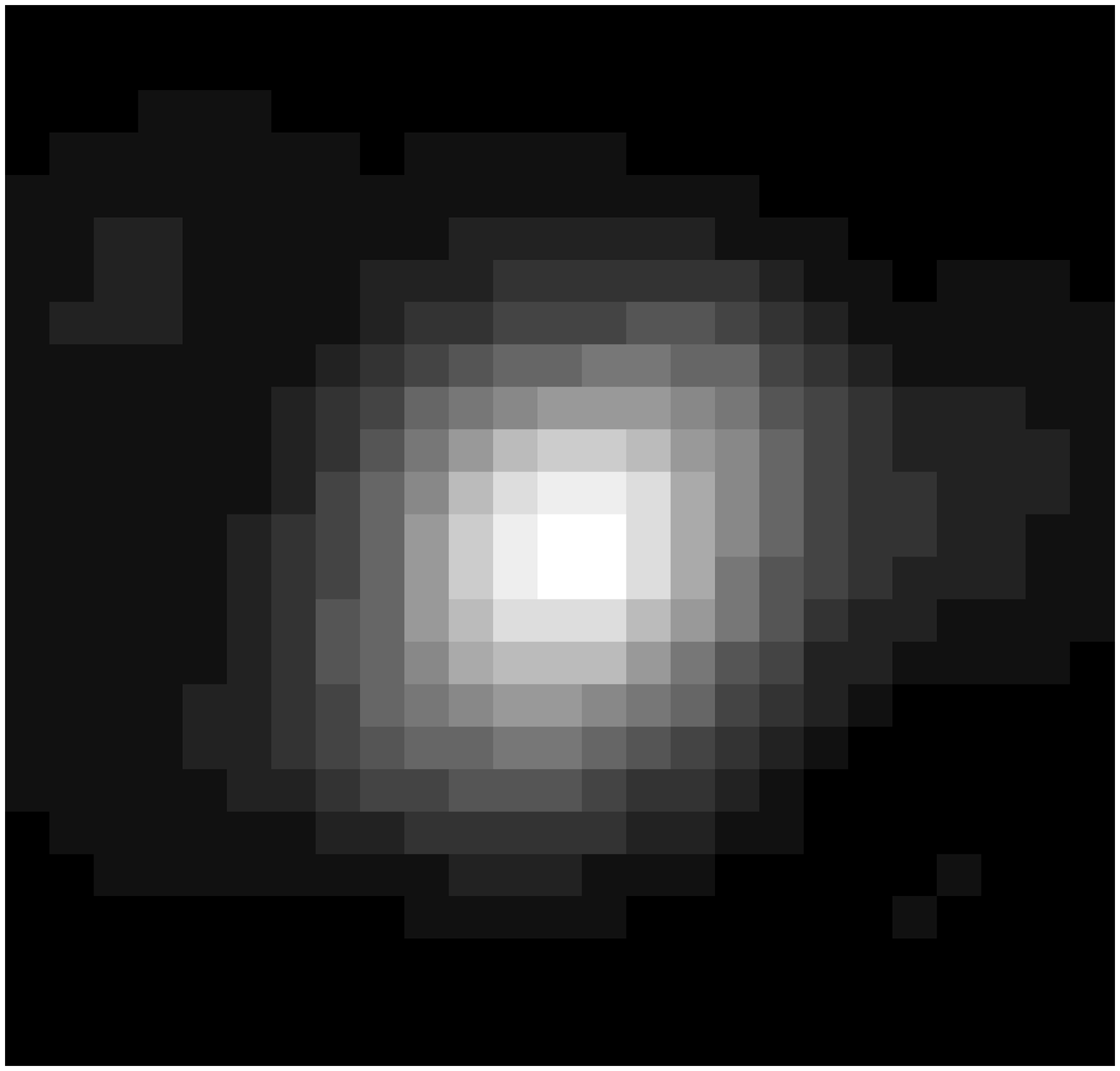,width=20mm}\hspace{2mm}
\psfig{figure=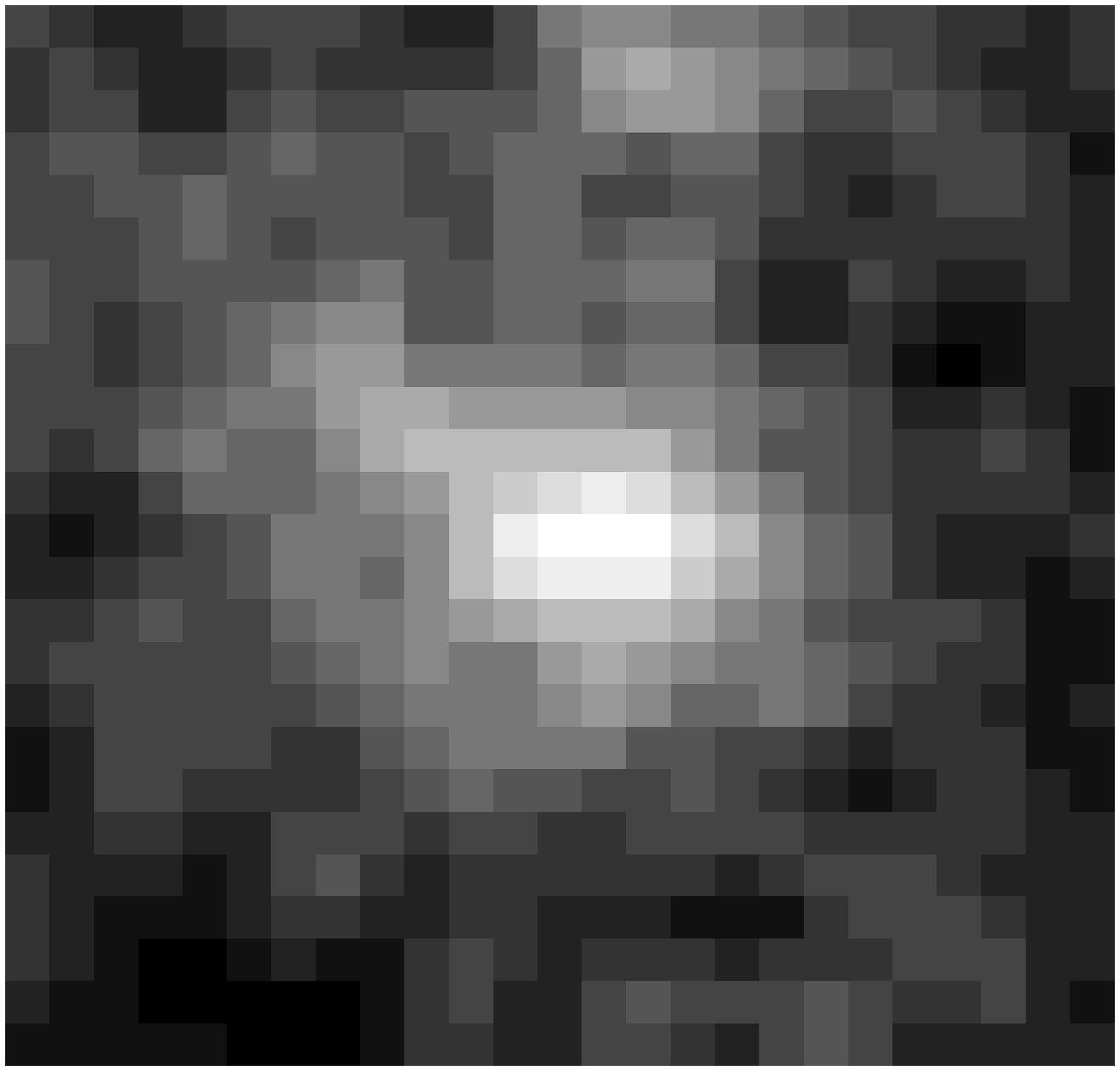,width=20mm}\psfig{figure=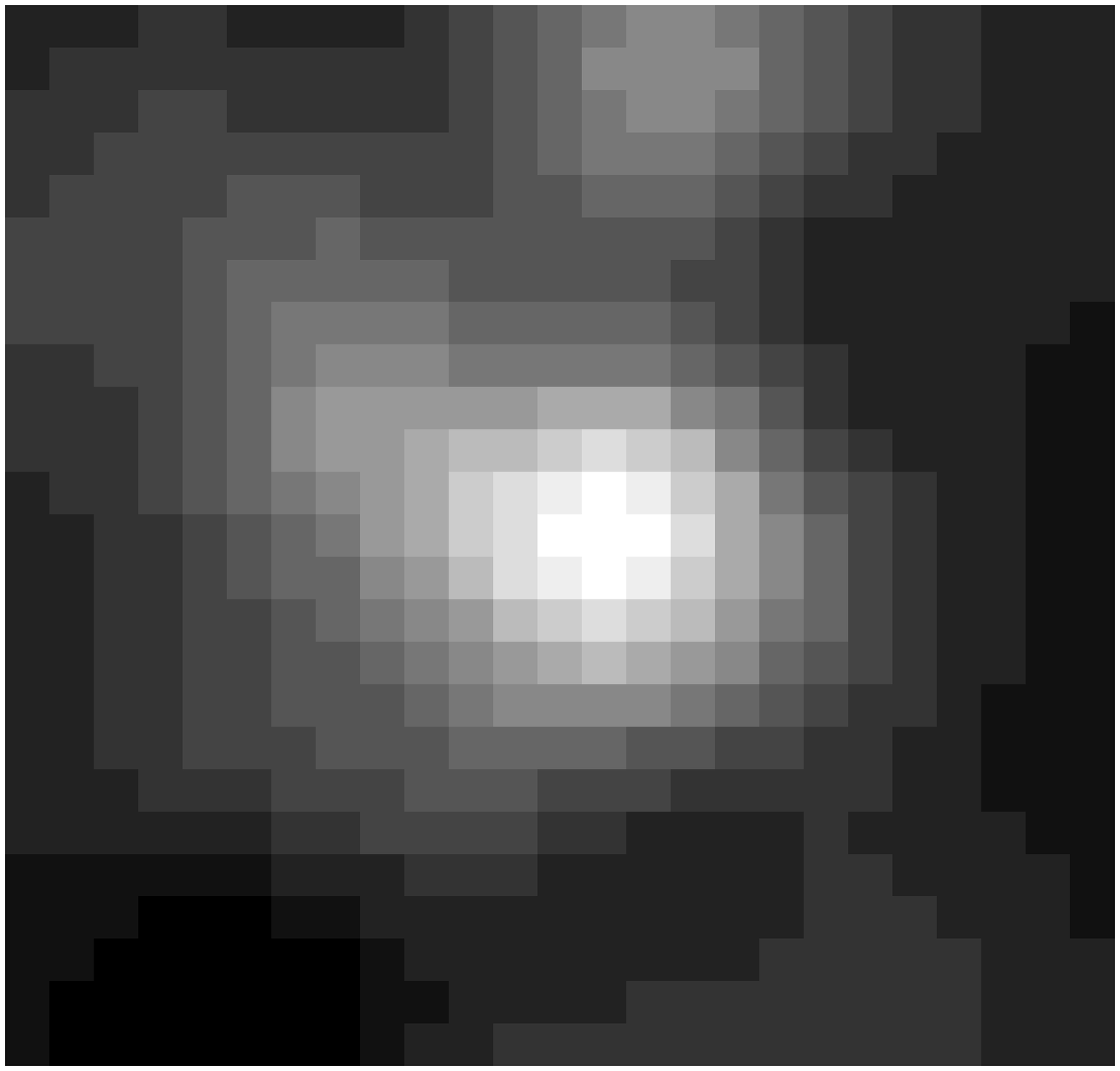,width=20mm}
\psfig{figure=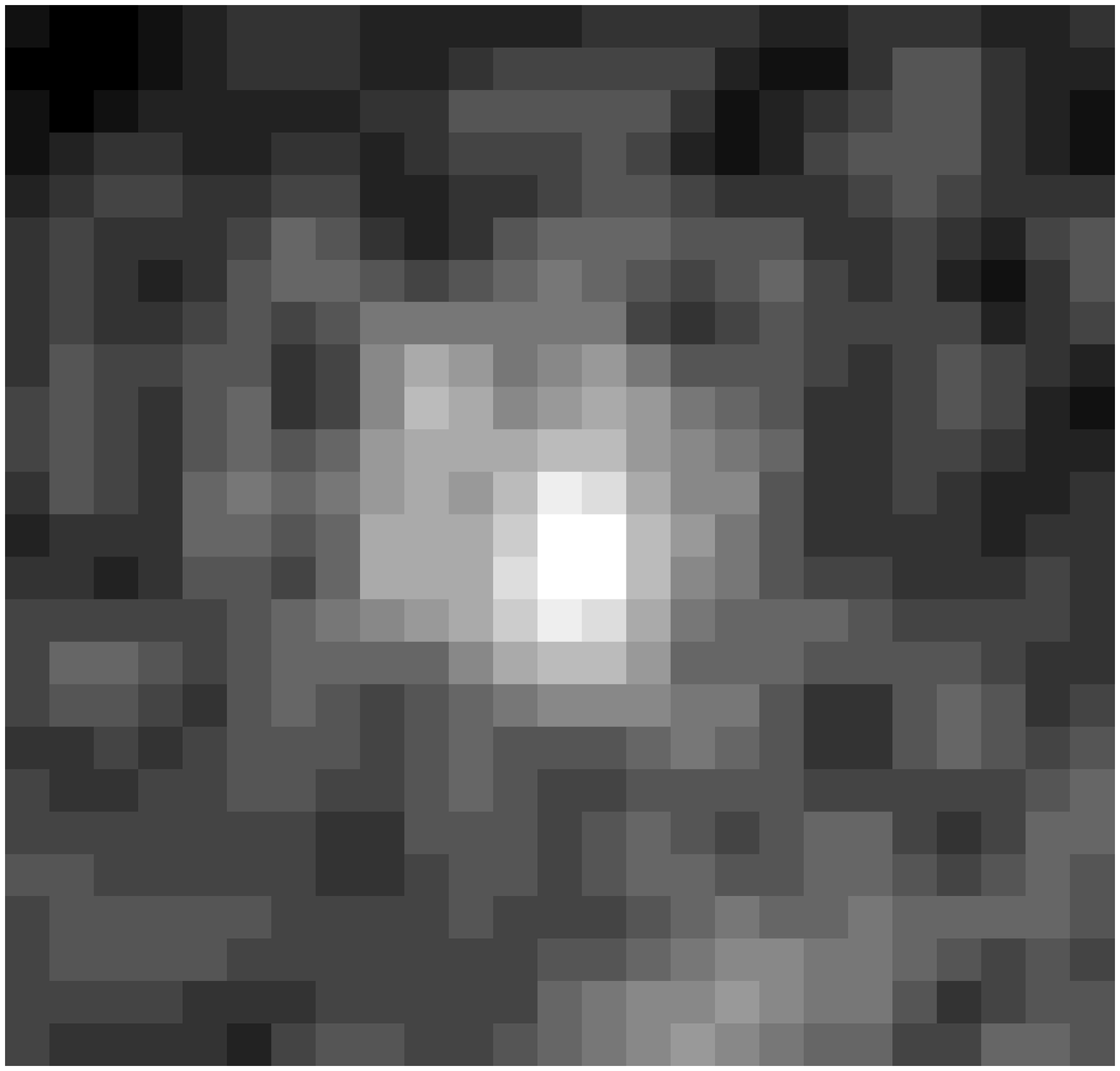,width=20mm}\psfig{figure=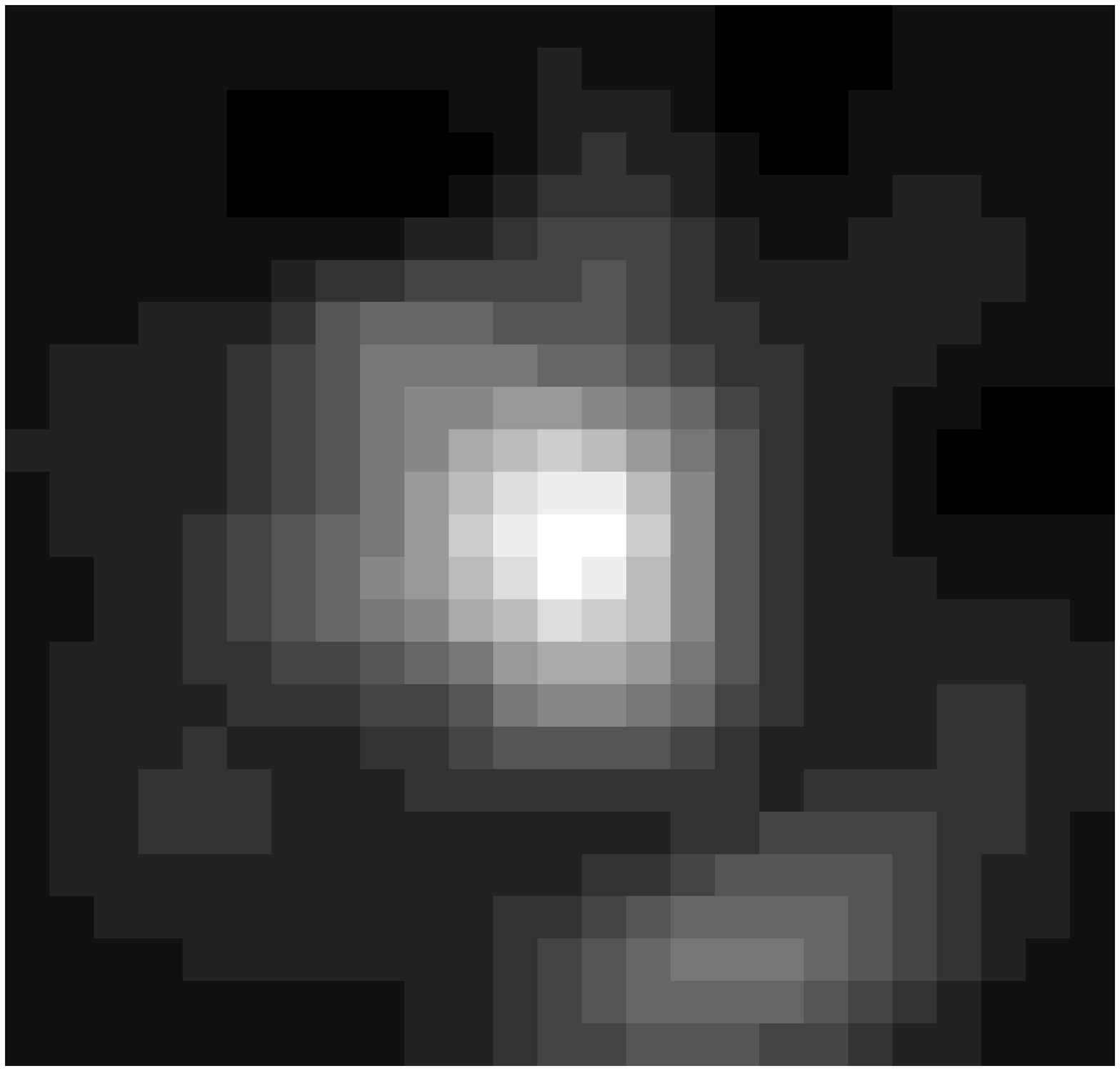,width=20mm}\hspace{2mm}
\psfig{figure=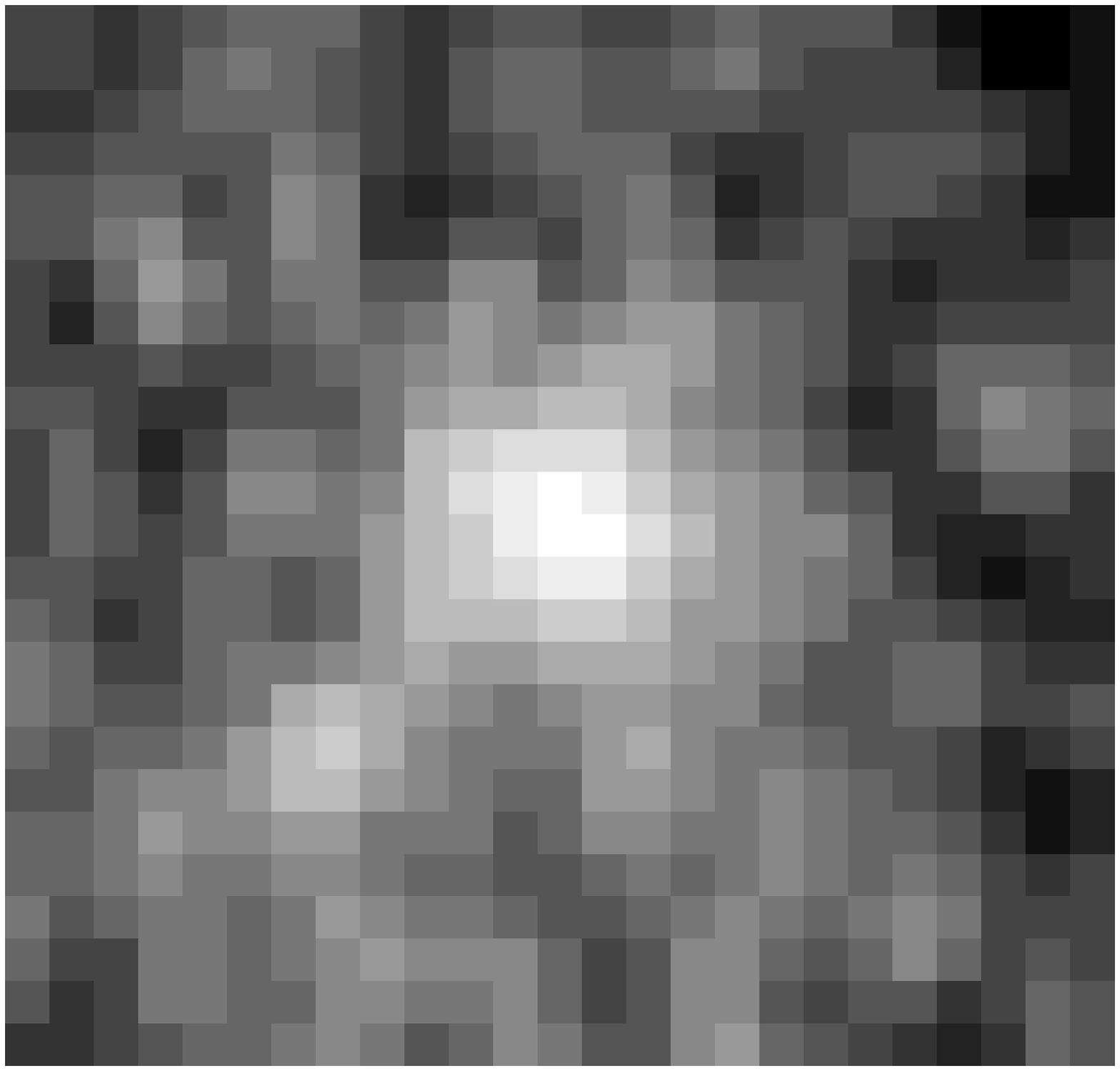,width=20mm}\psfig{figure=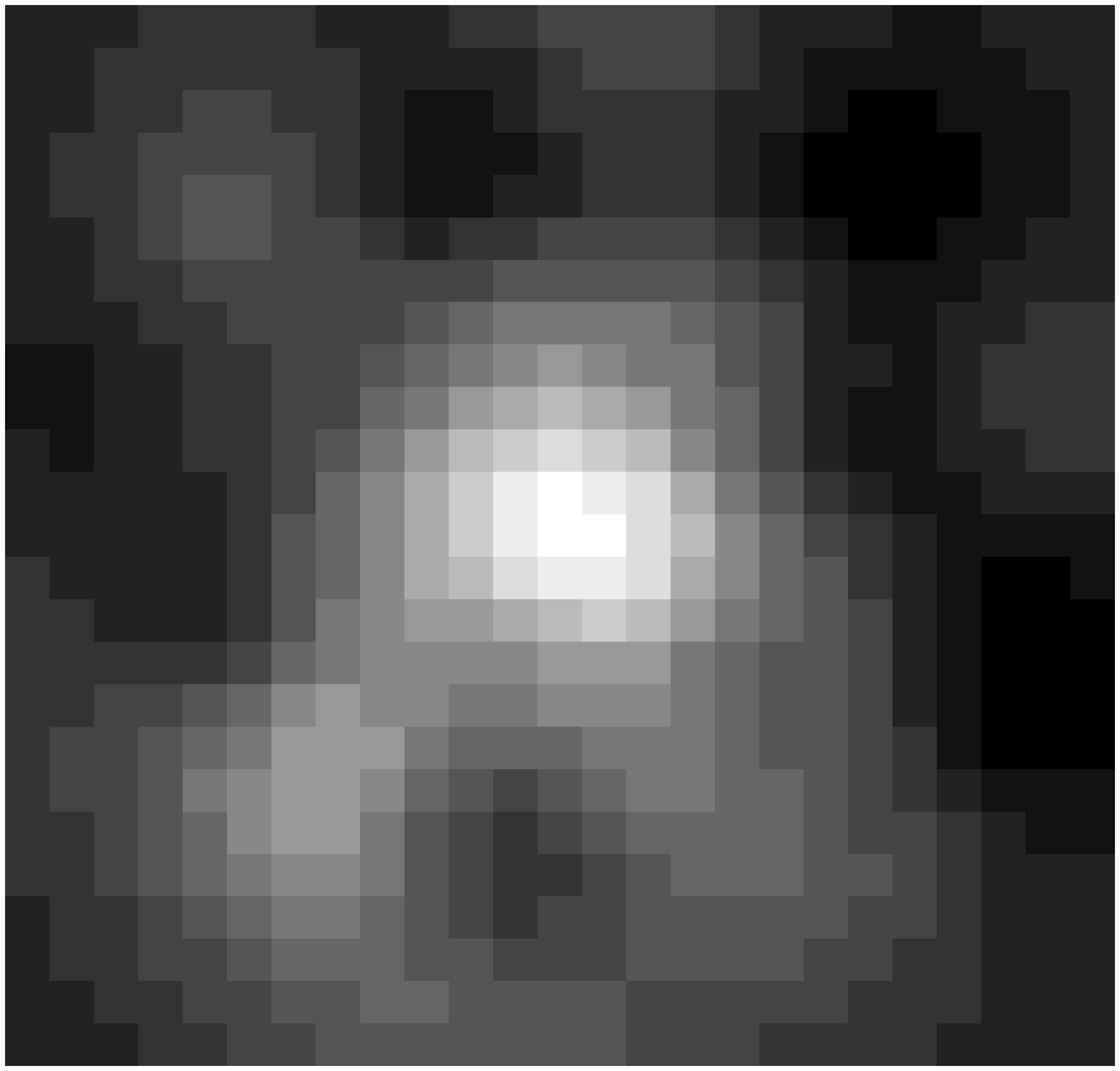,width=20mm}\hspace{2mm}
\psfig{figure=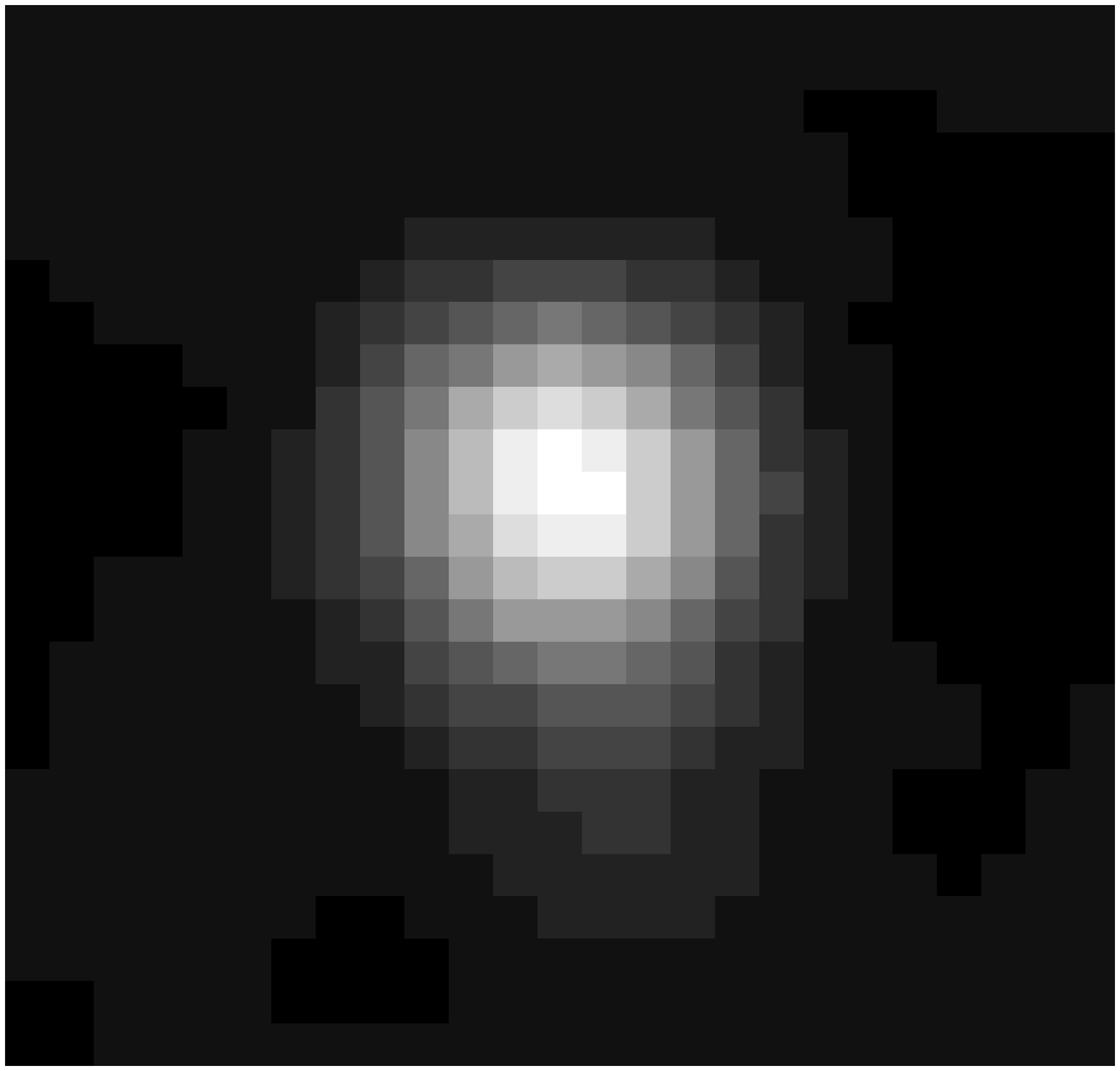,width=20mm}\psfig{figure=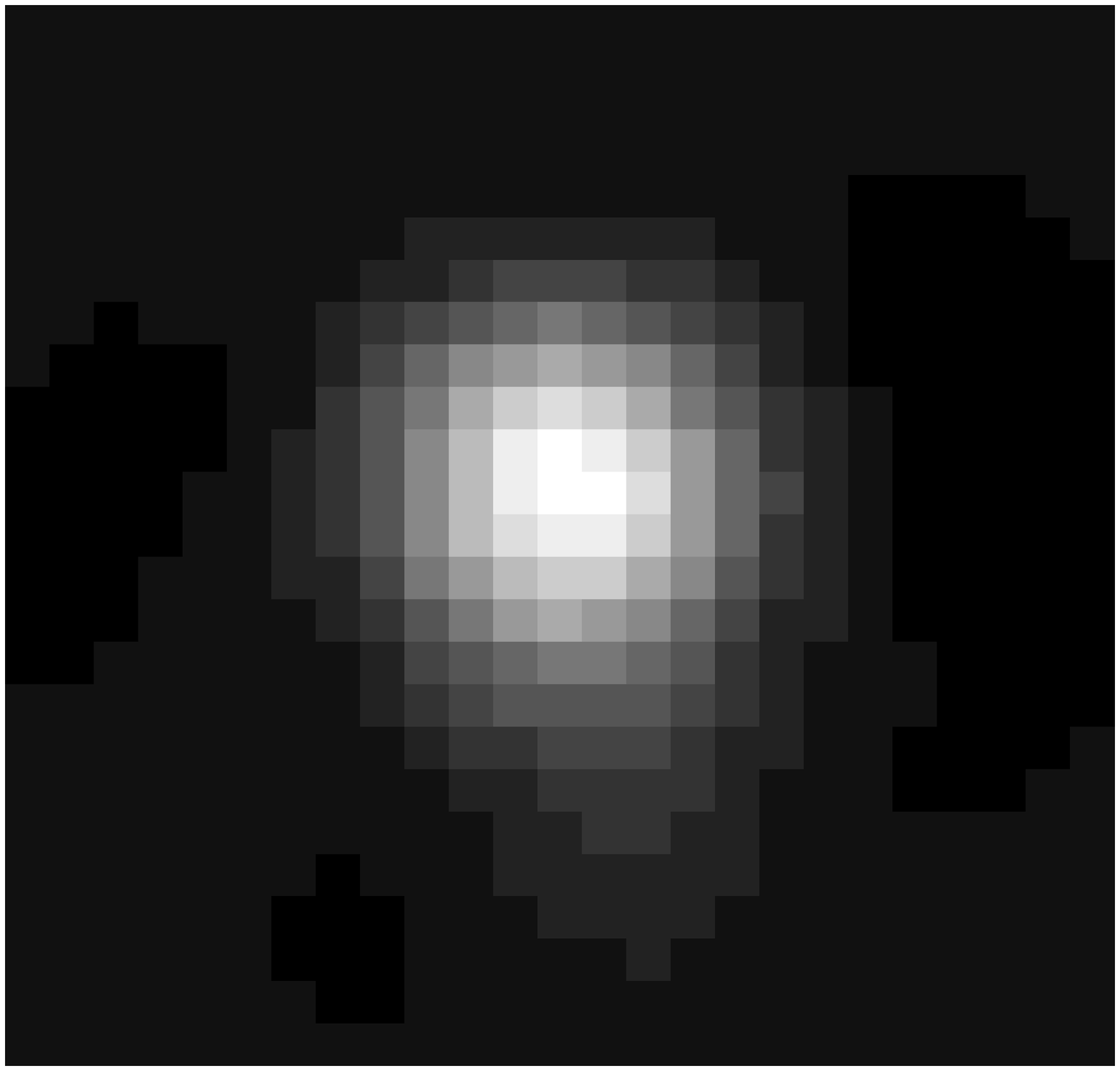,width=20mm}\hspace{2mm}
\psfig{figure=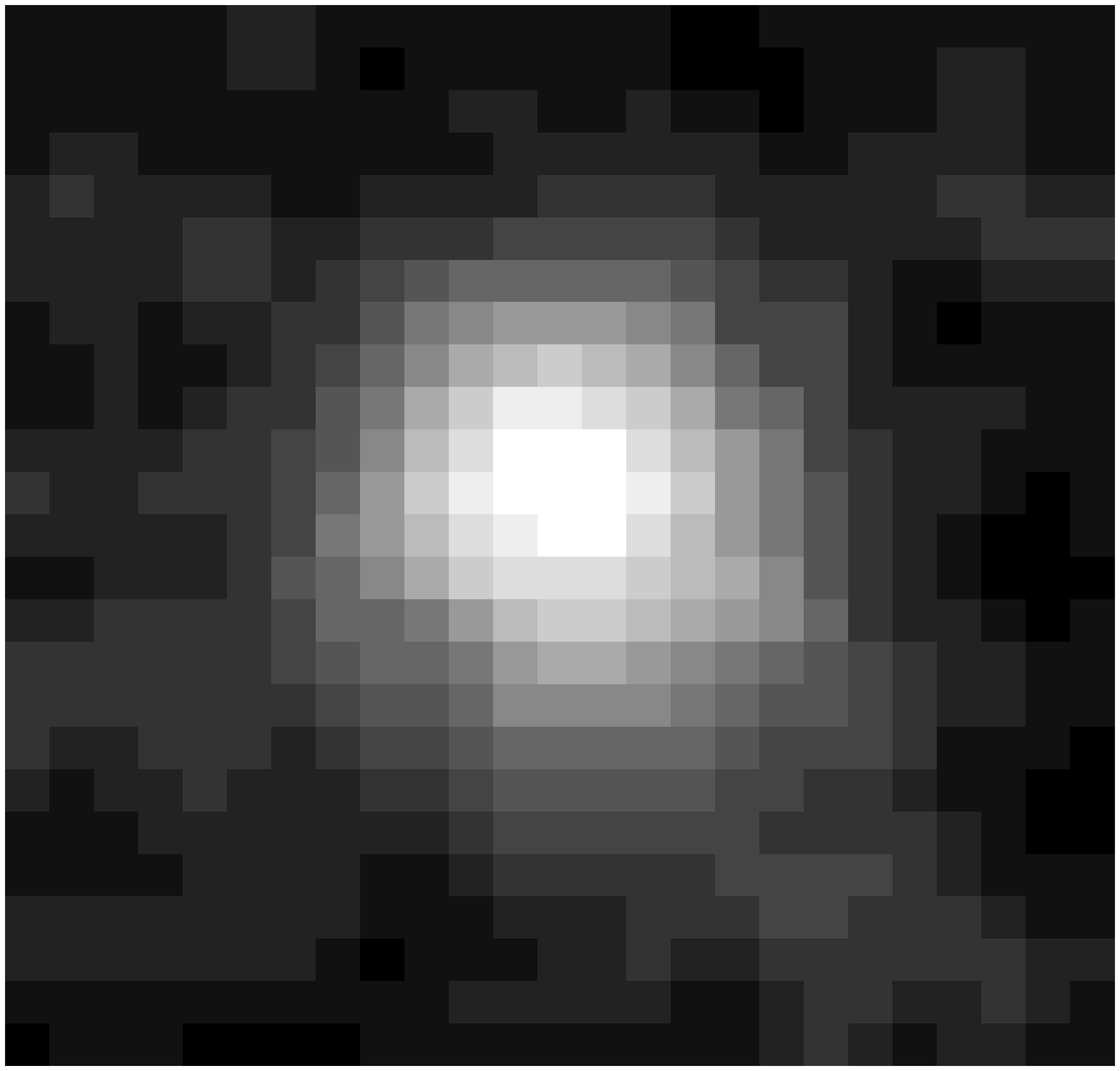,width=20mm}\psfig{figure=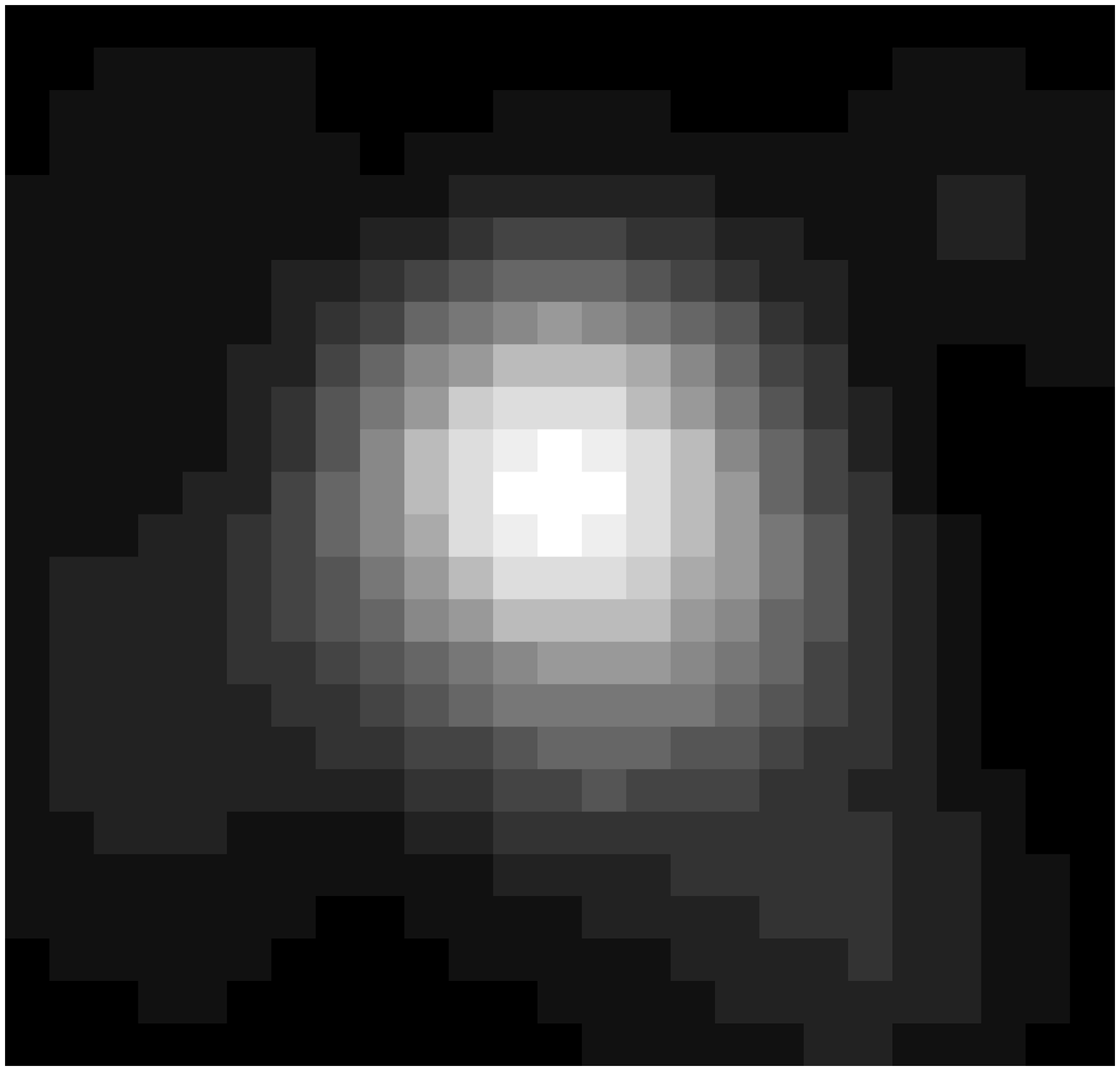,width=20mm}
\psfig{figure=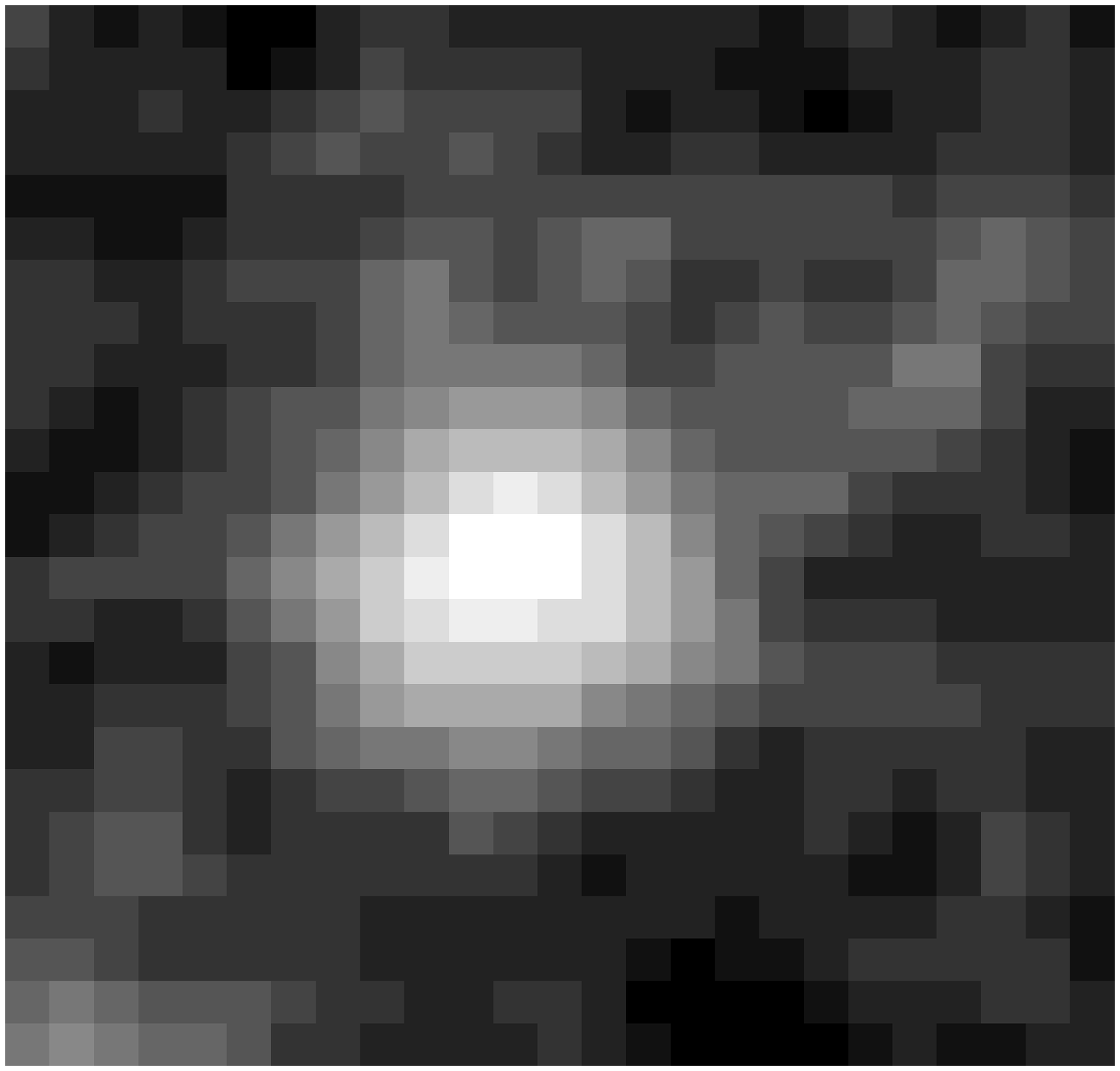,width=20mm}\psfig{figure=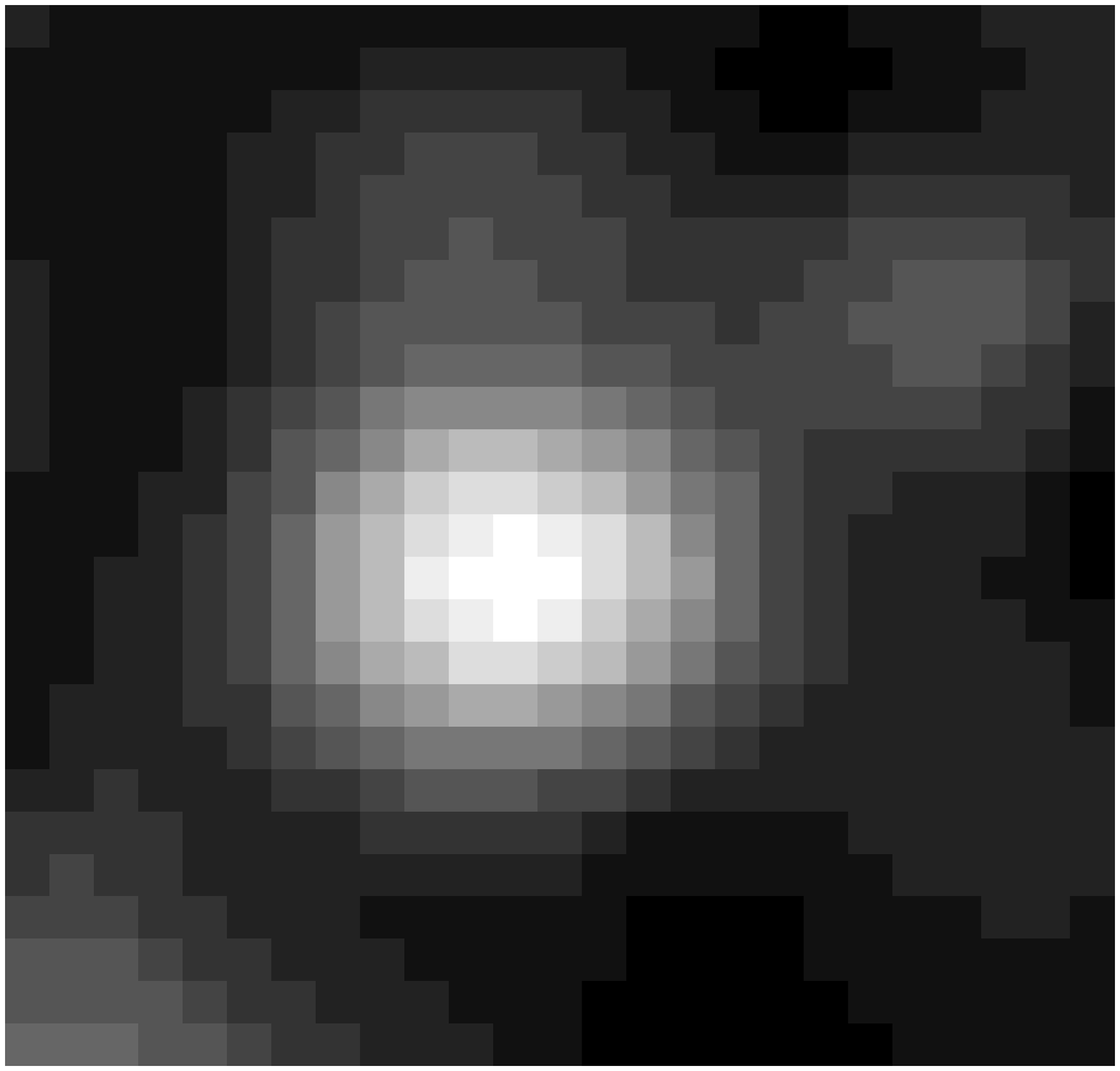,width=20mm}\hspace{2mm}
\psfig{figure=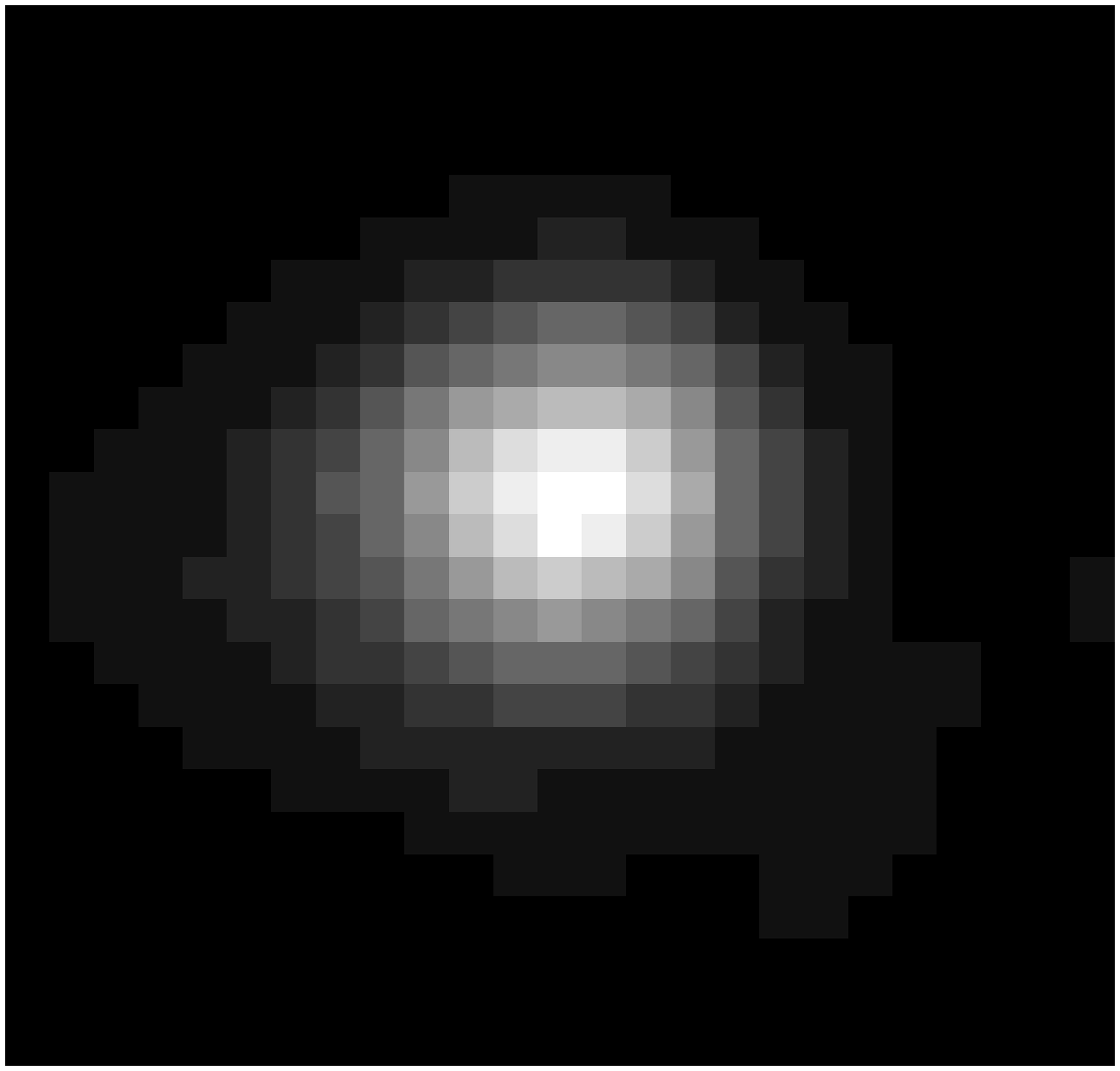,width=20mm}\psfig{figure=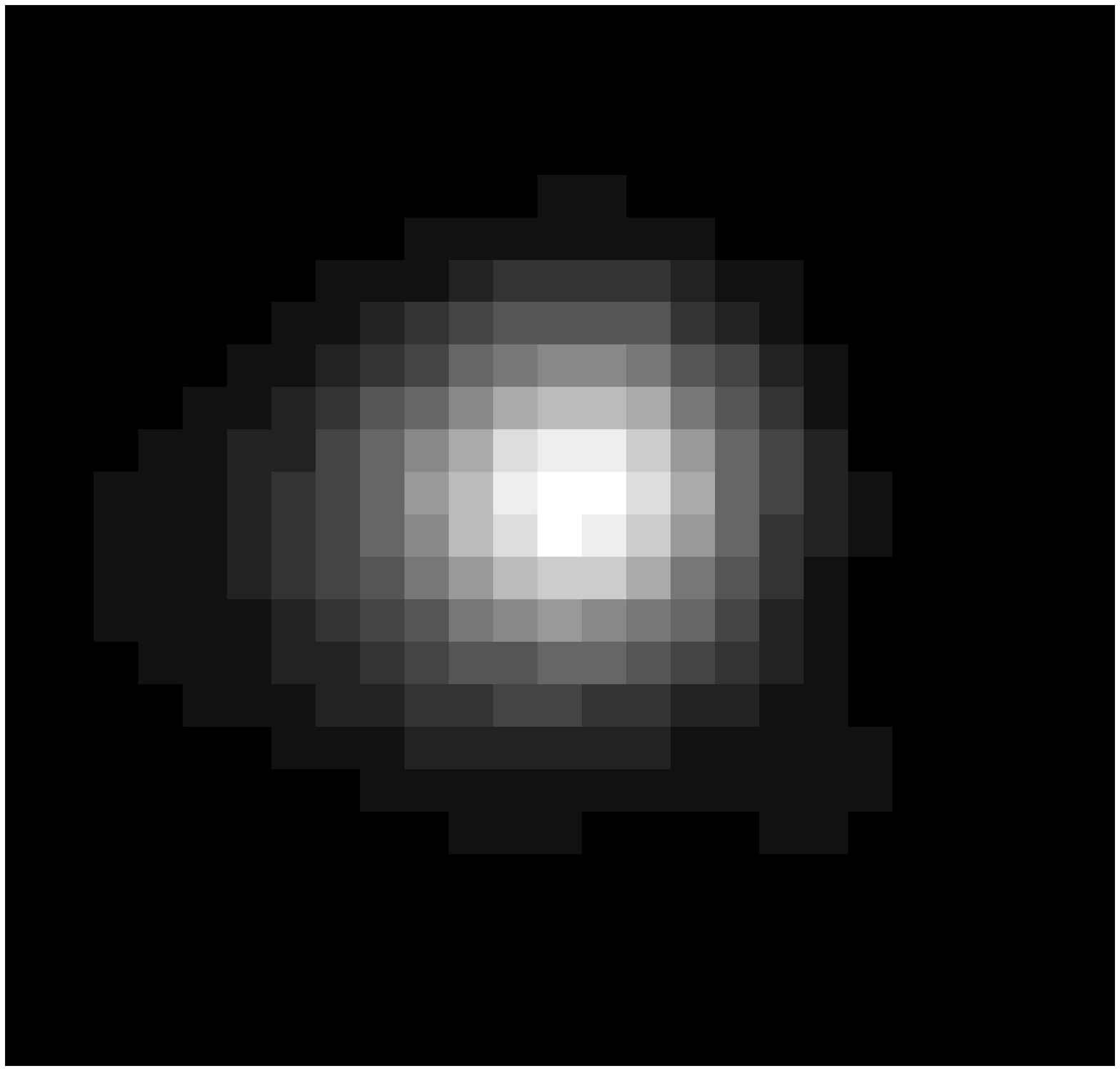,width=20mm}\hspace{2mm}
\psfig{figure=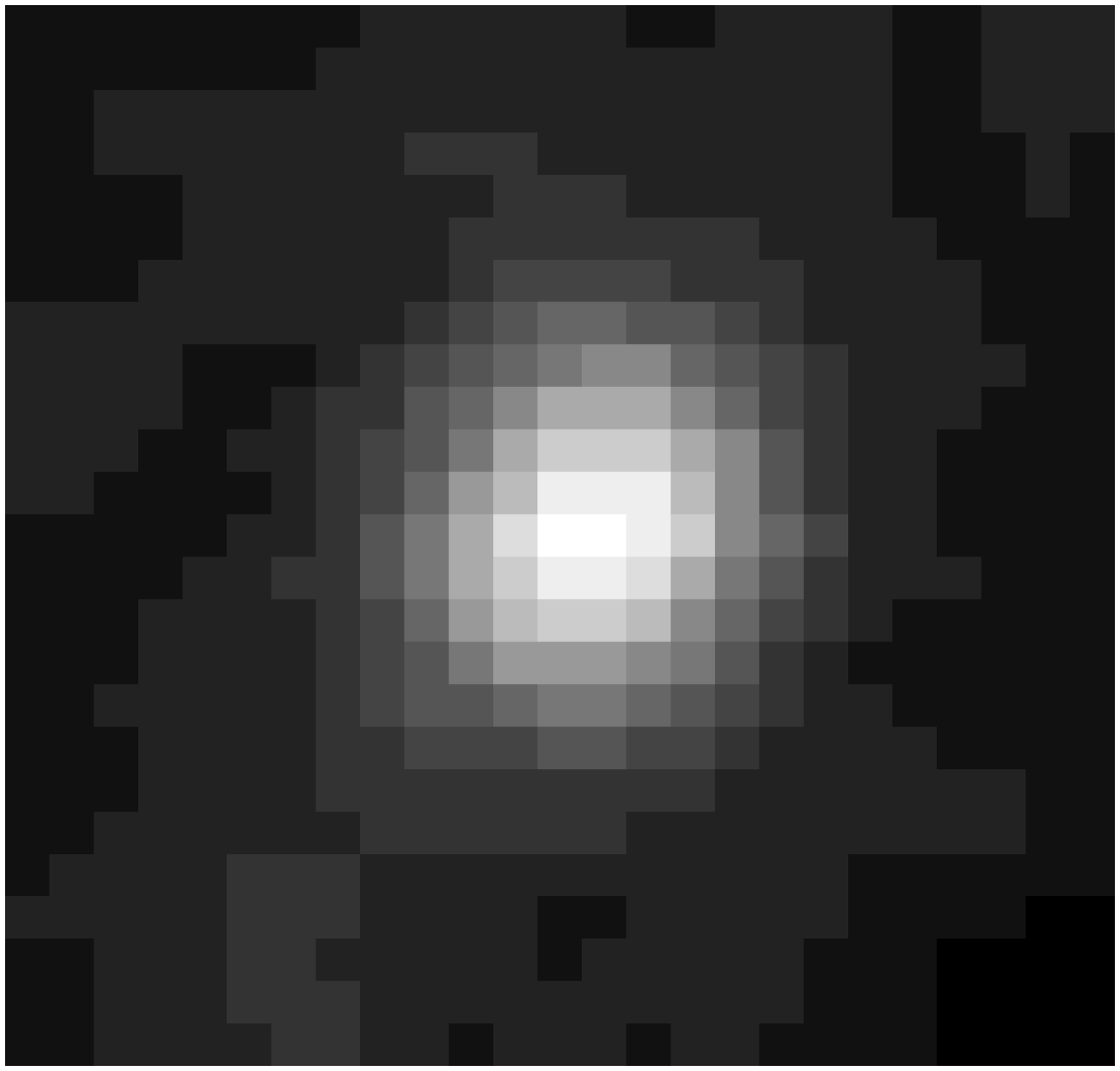,width=20mm}\psfig{figure=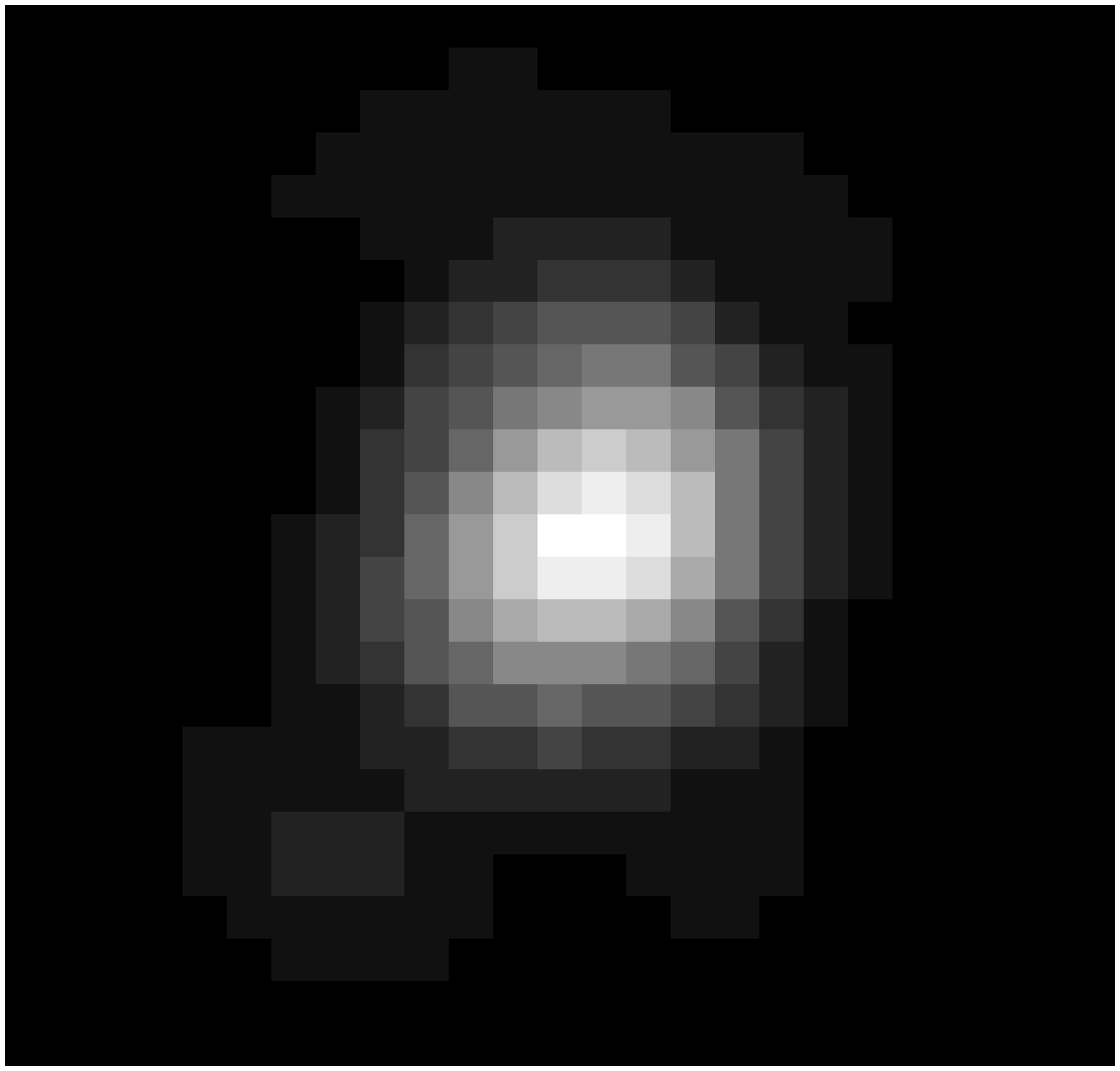,width=20mm}\hspace{2mm}
\psfig{figure=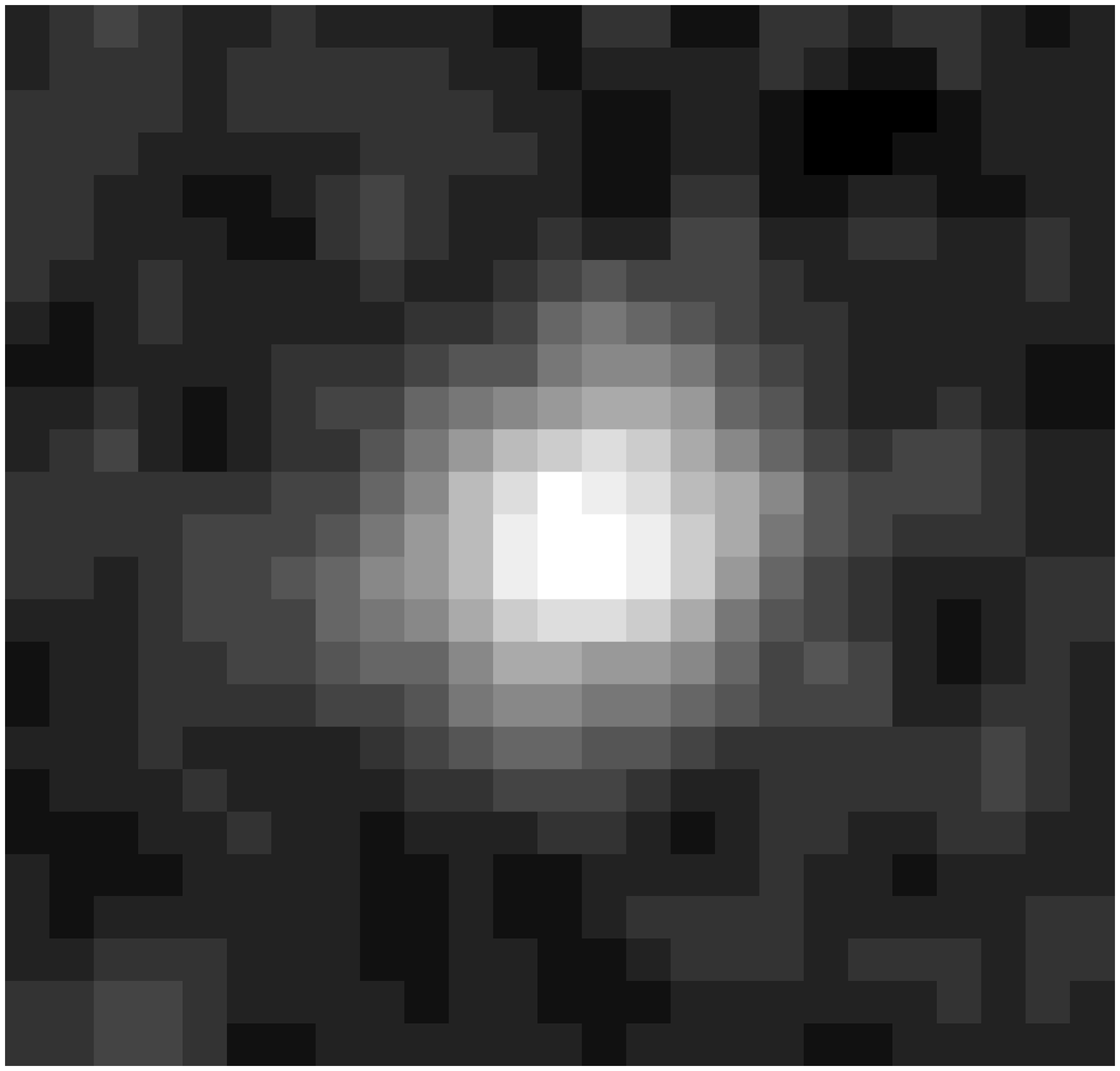,width=20mm}\psfig{figure=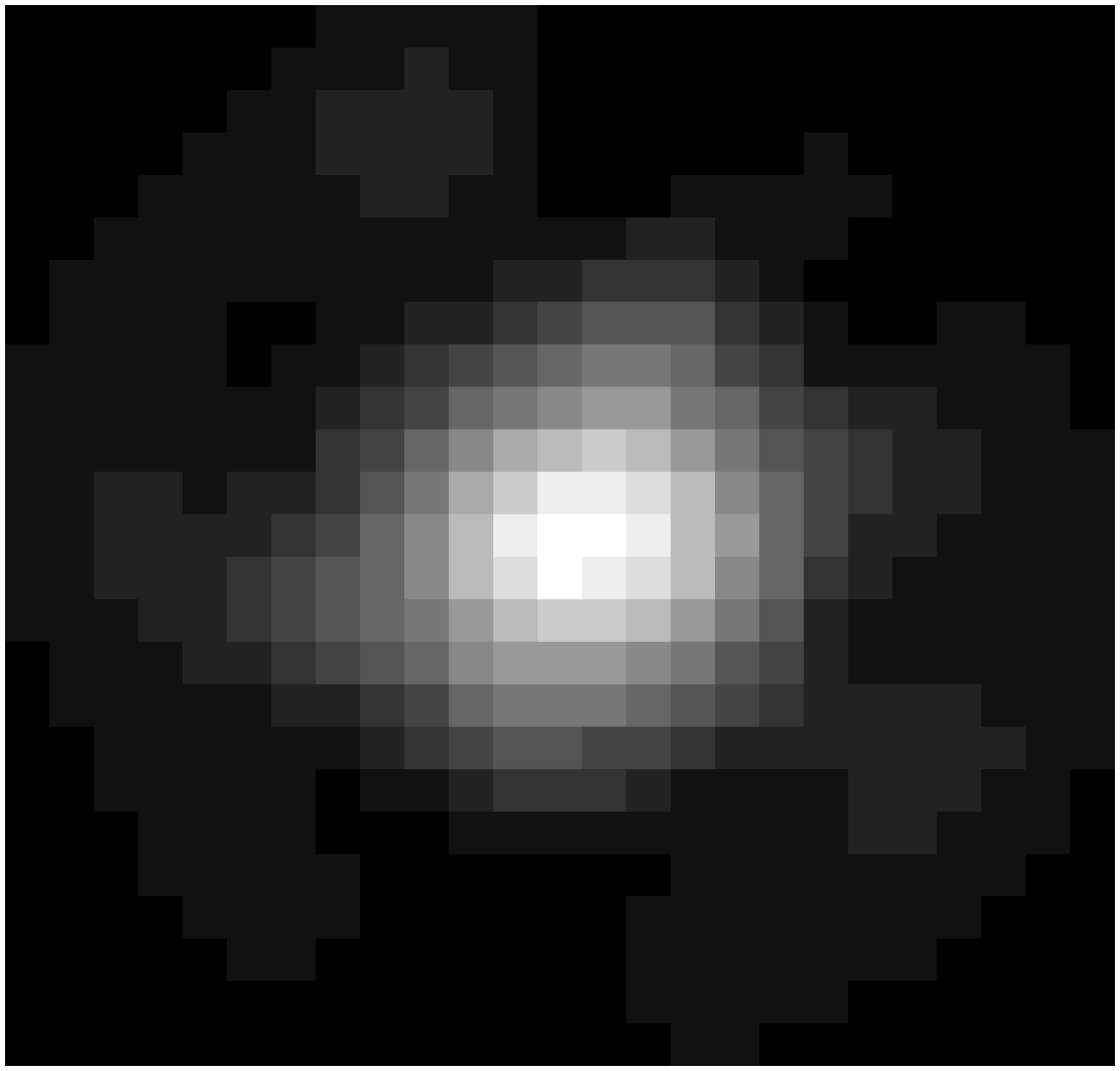,width=20mm}
\psfig{figure=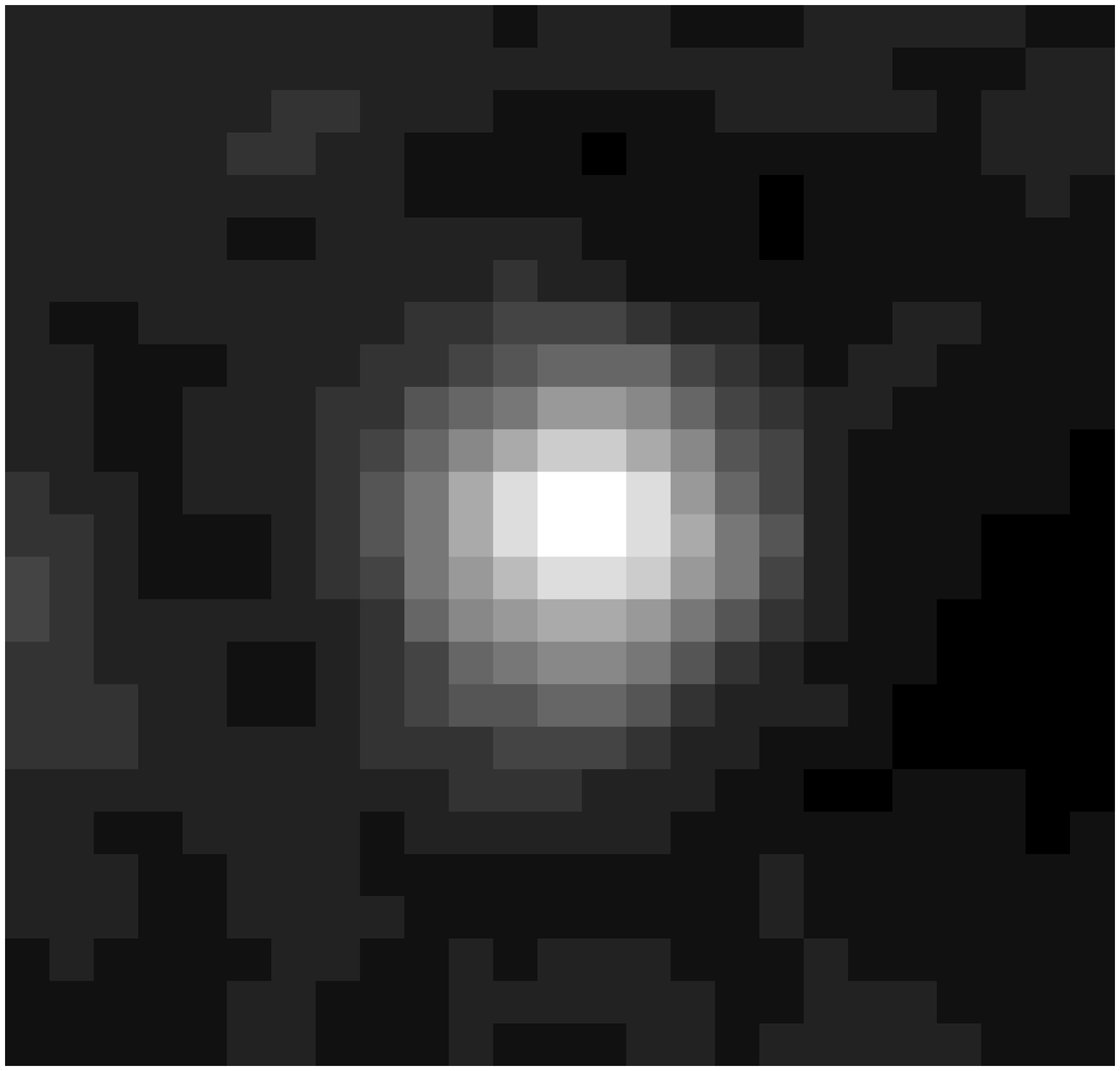,width=20mm}\psfig{figure=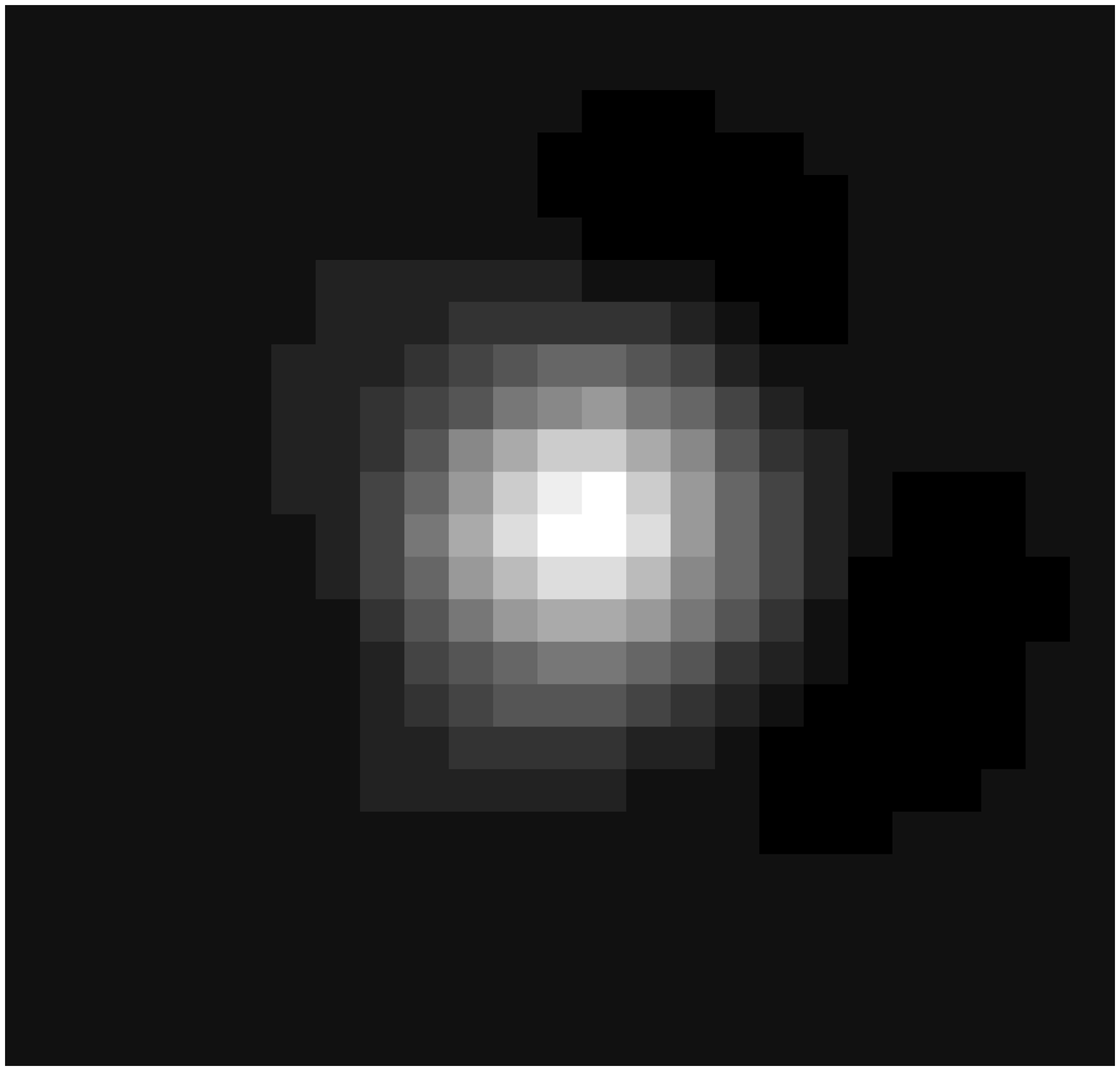,width=20mm}\hspace{2mm}
\psfig{figure=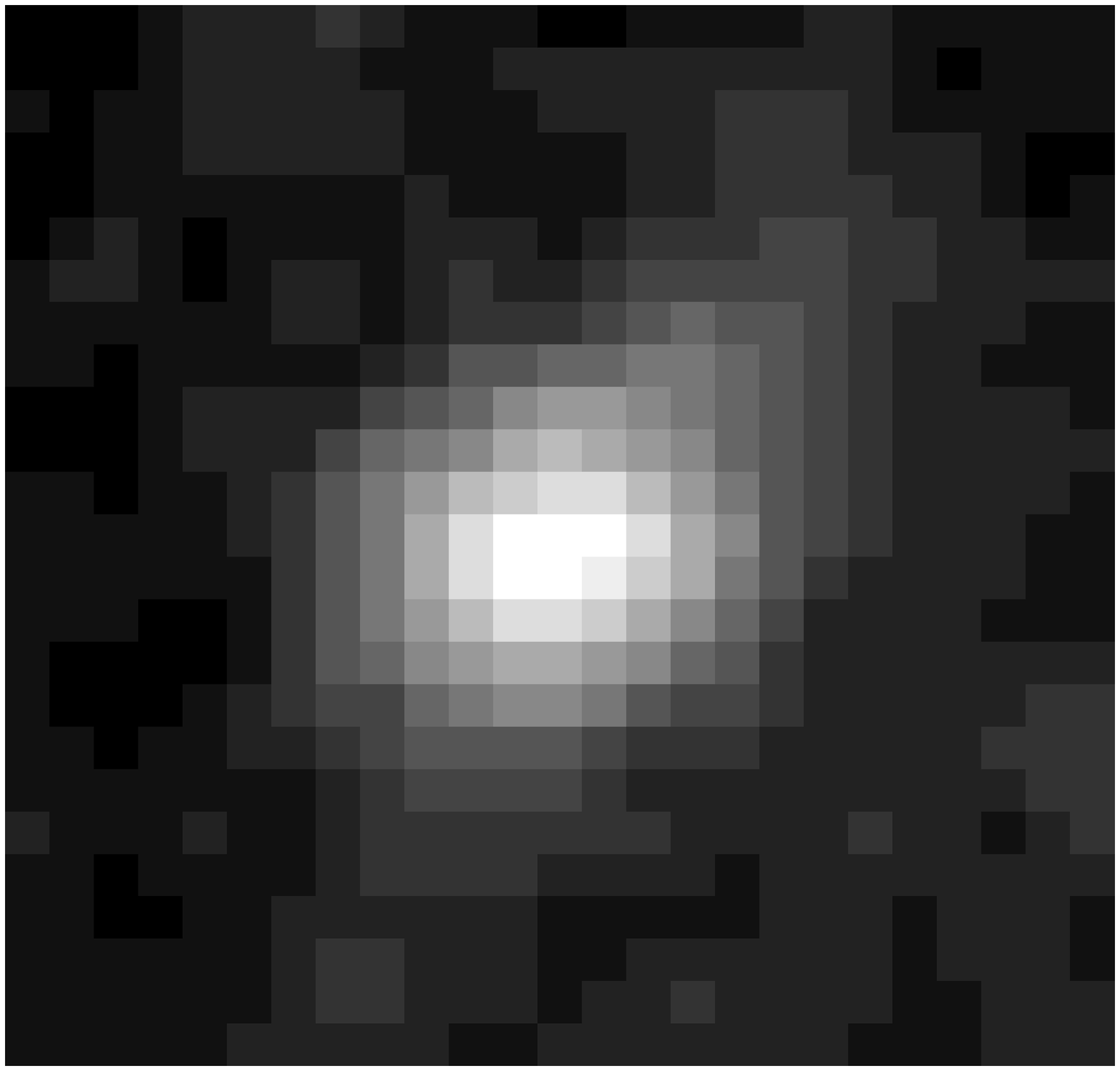,width=20mm}\psfig{figure=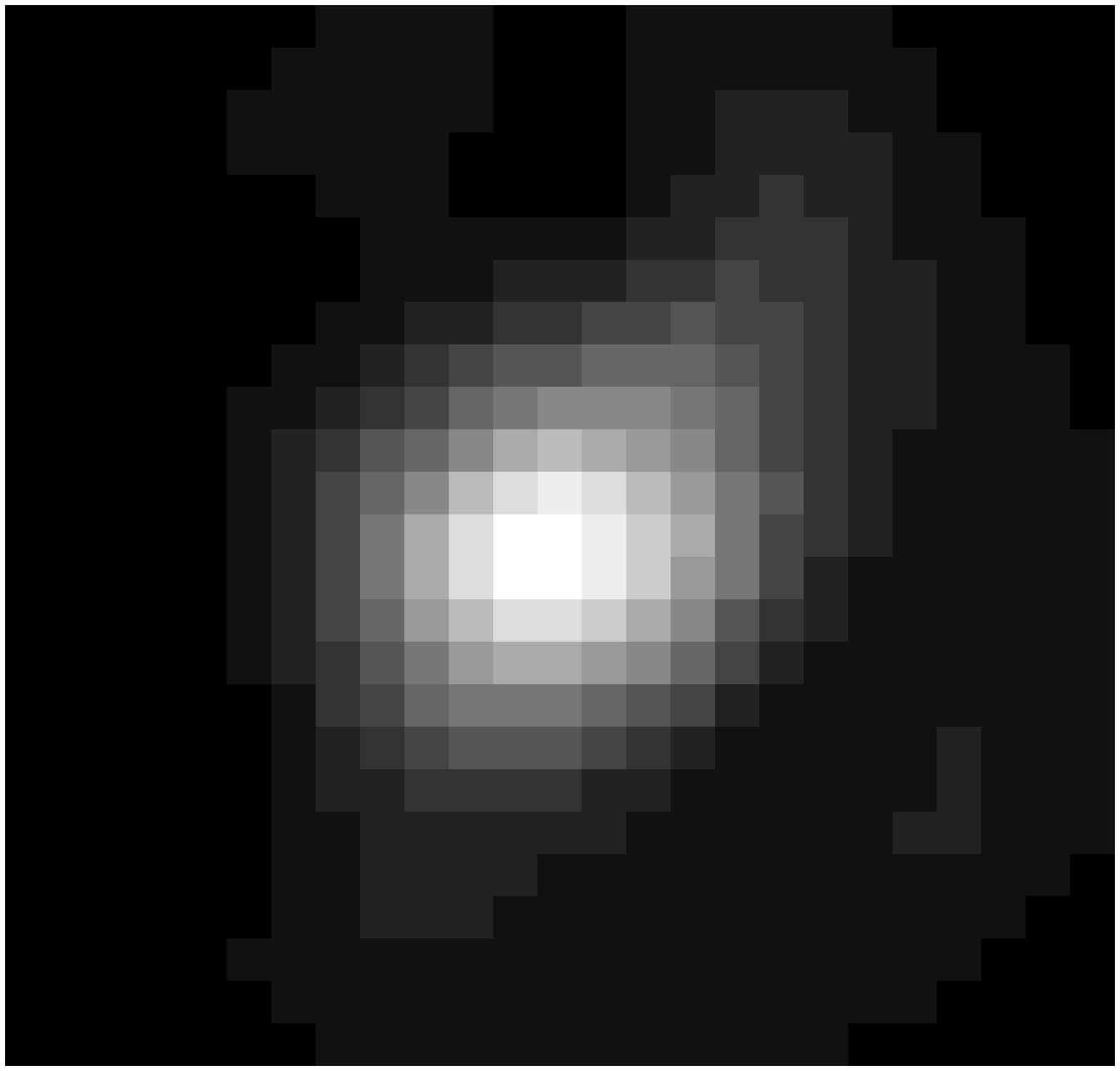,width=20mm}\hspace{2mm}
\psfig{figure=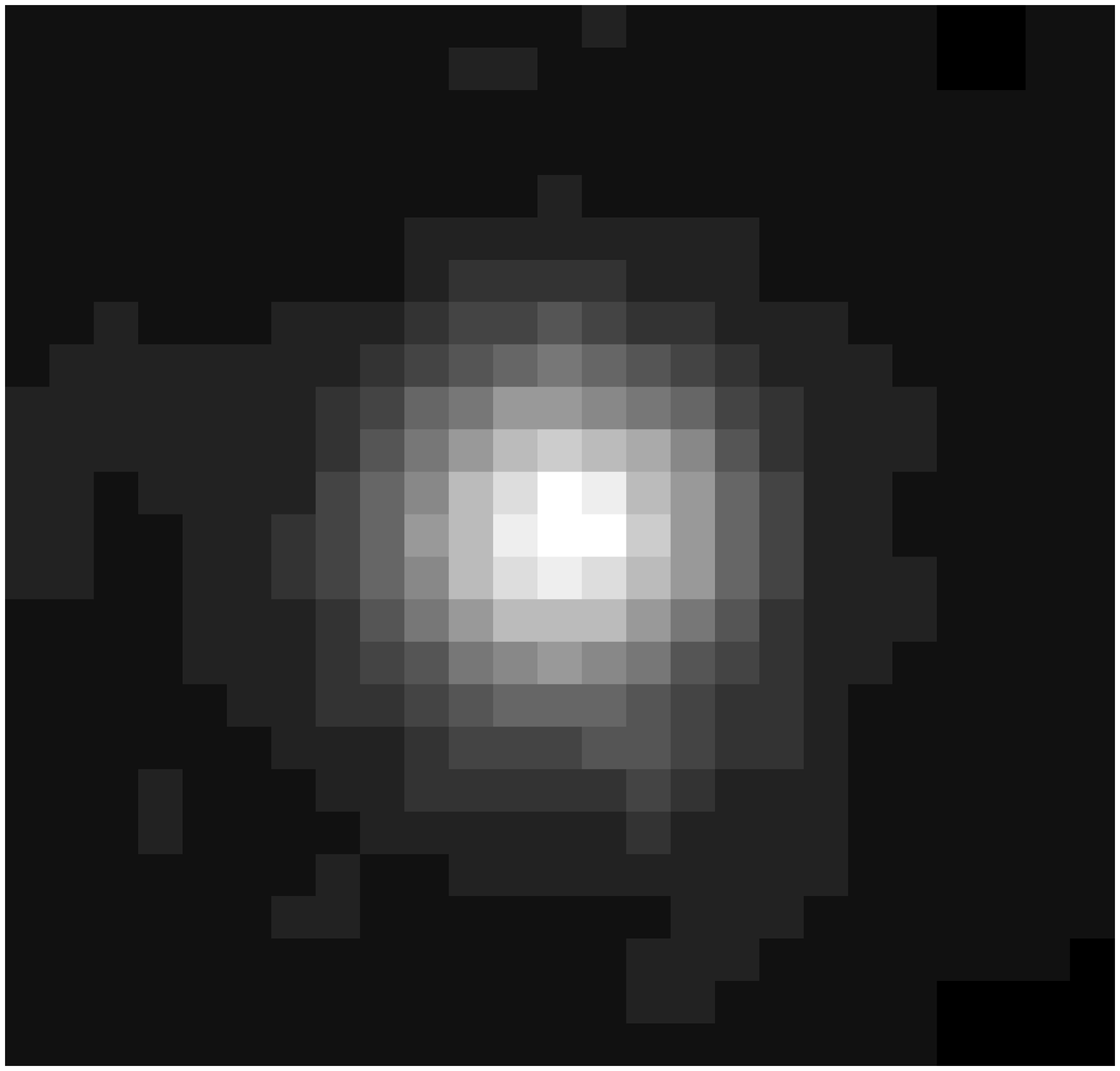,width=20mm}\psfig{figure=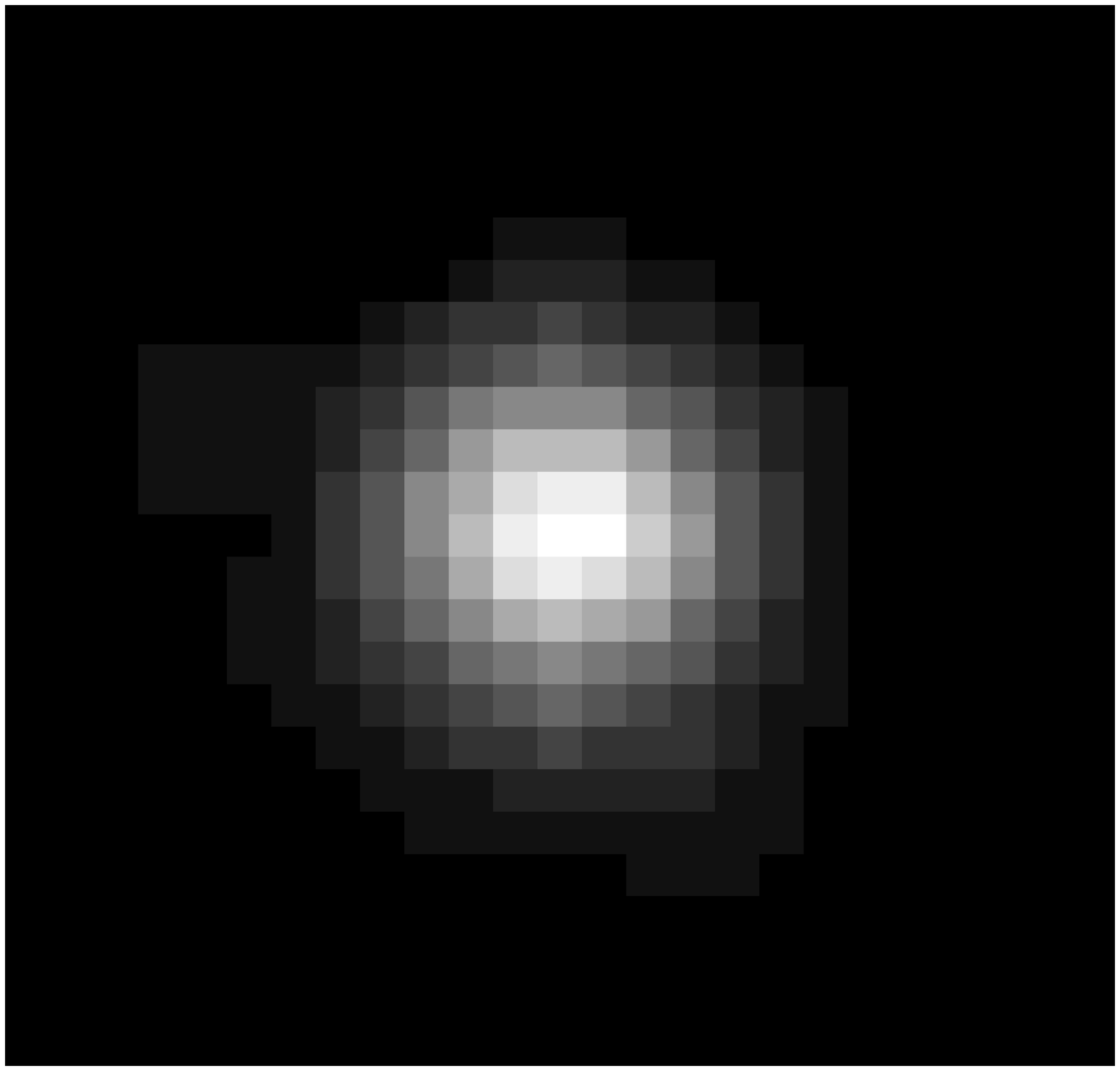,width=20mm}\hspace{2mm}
\psfig{figure=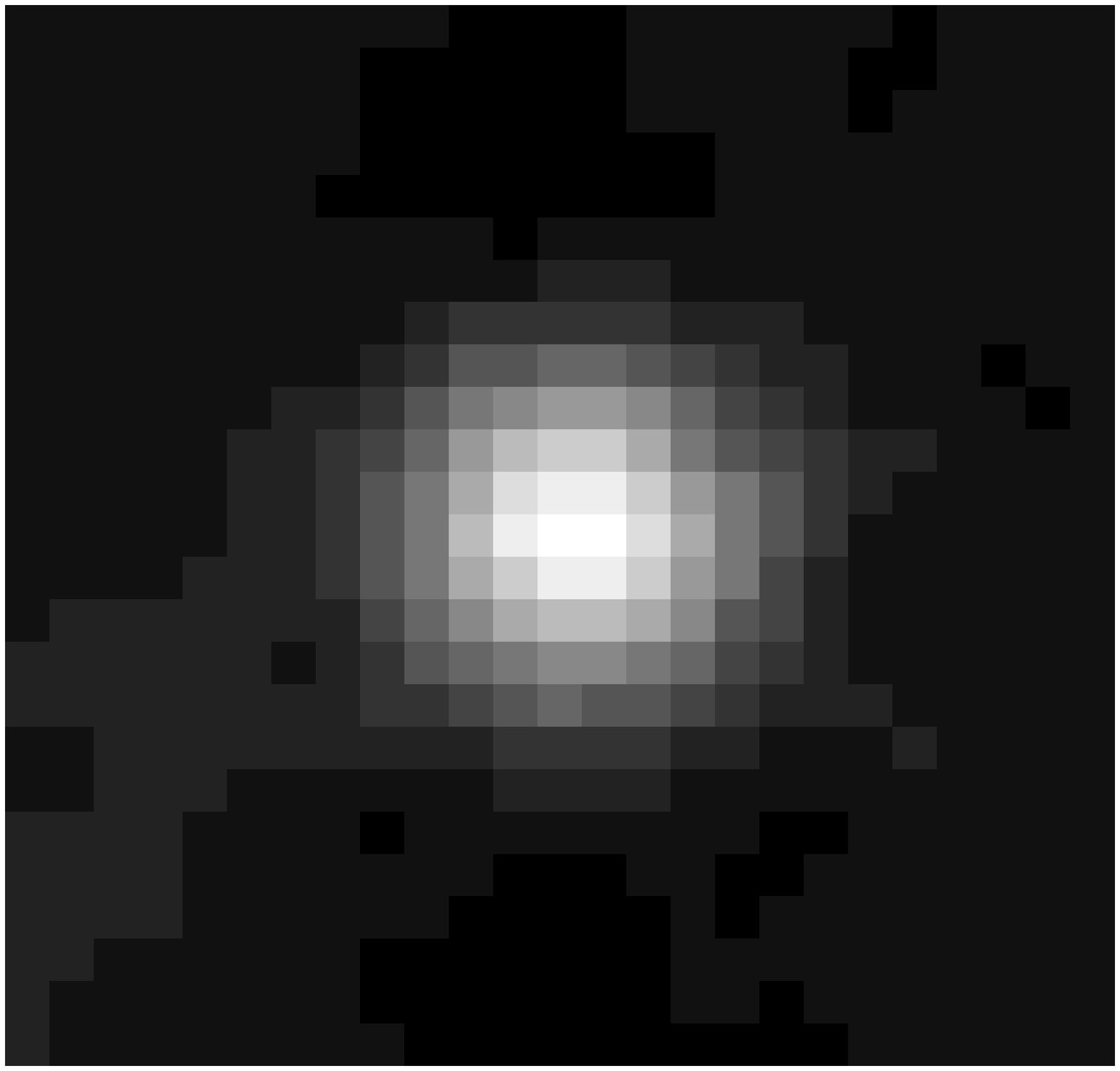,width=20mm}\psfig{figure=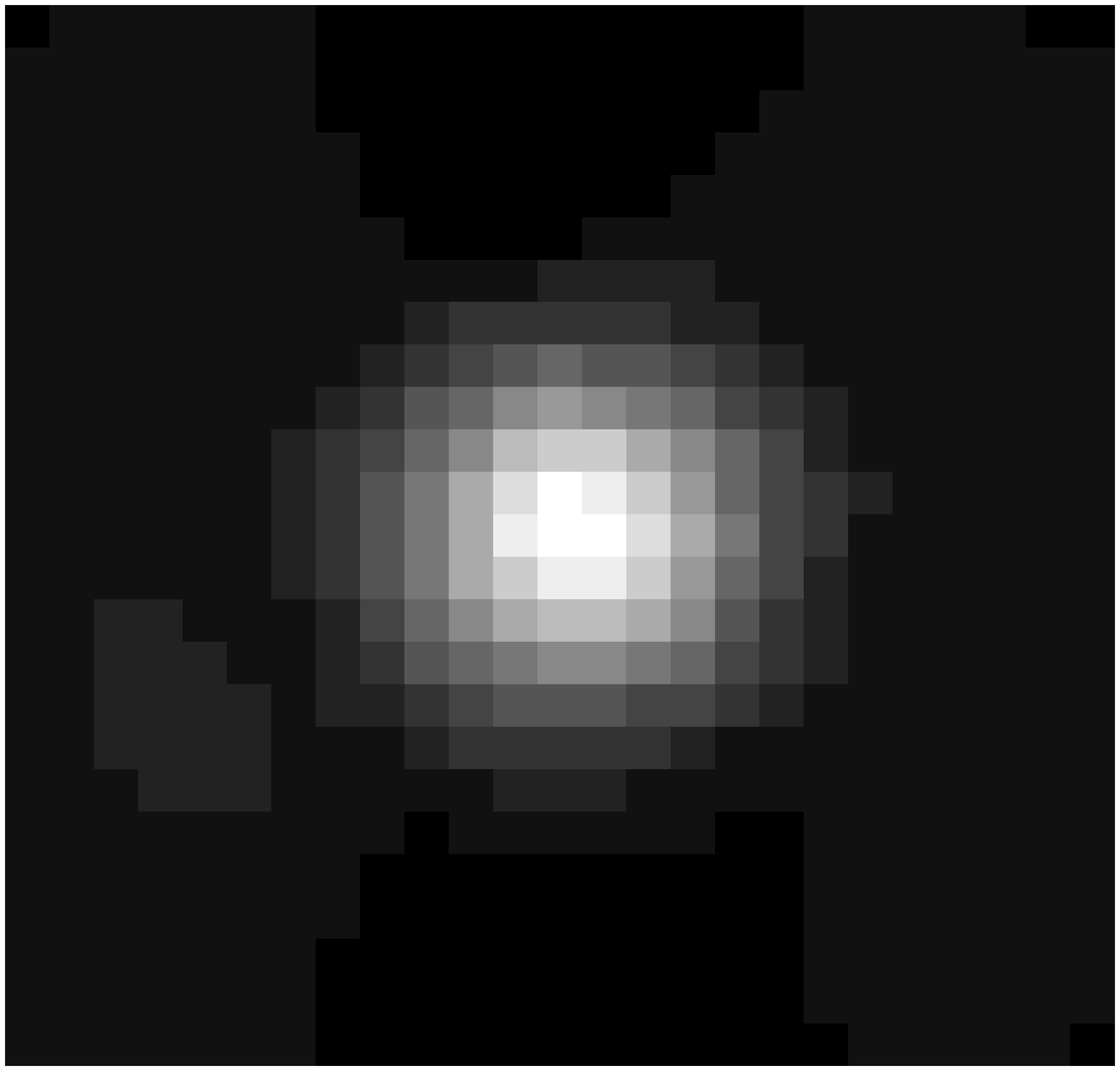,width=20mm}
\psfig{figure=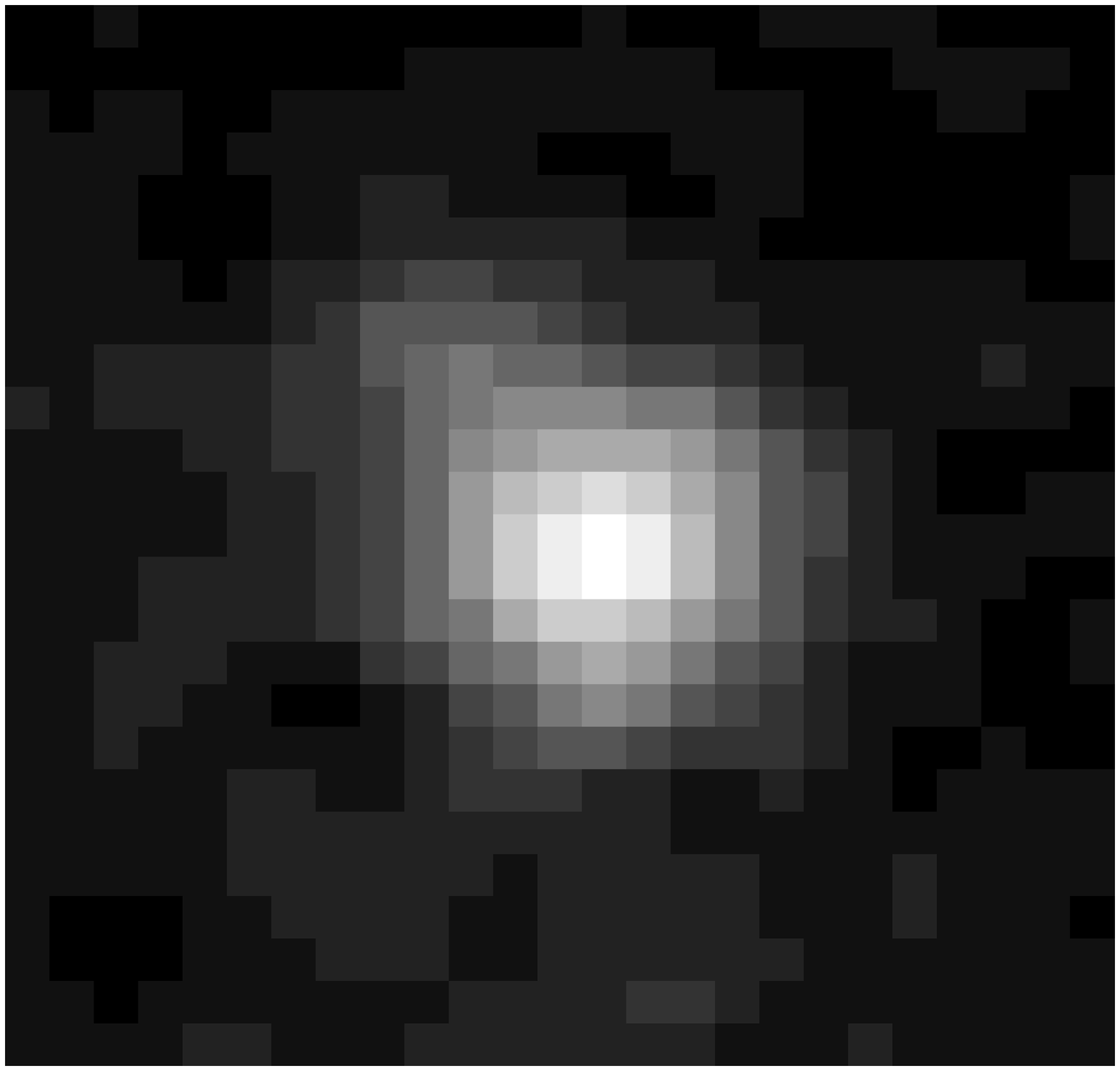,width=20mm}\psfig{figure=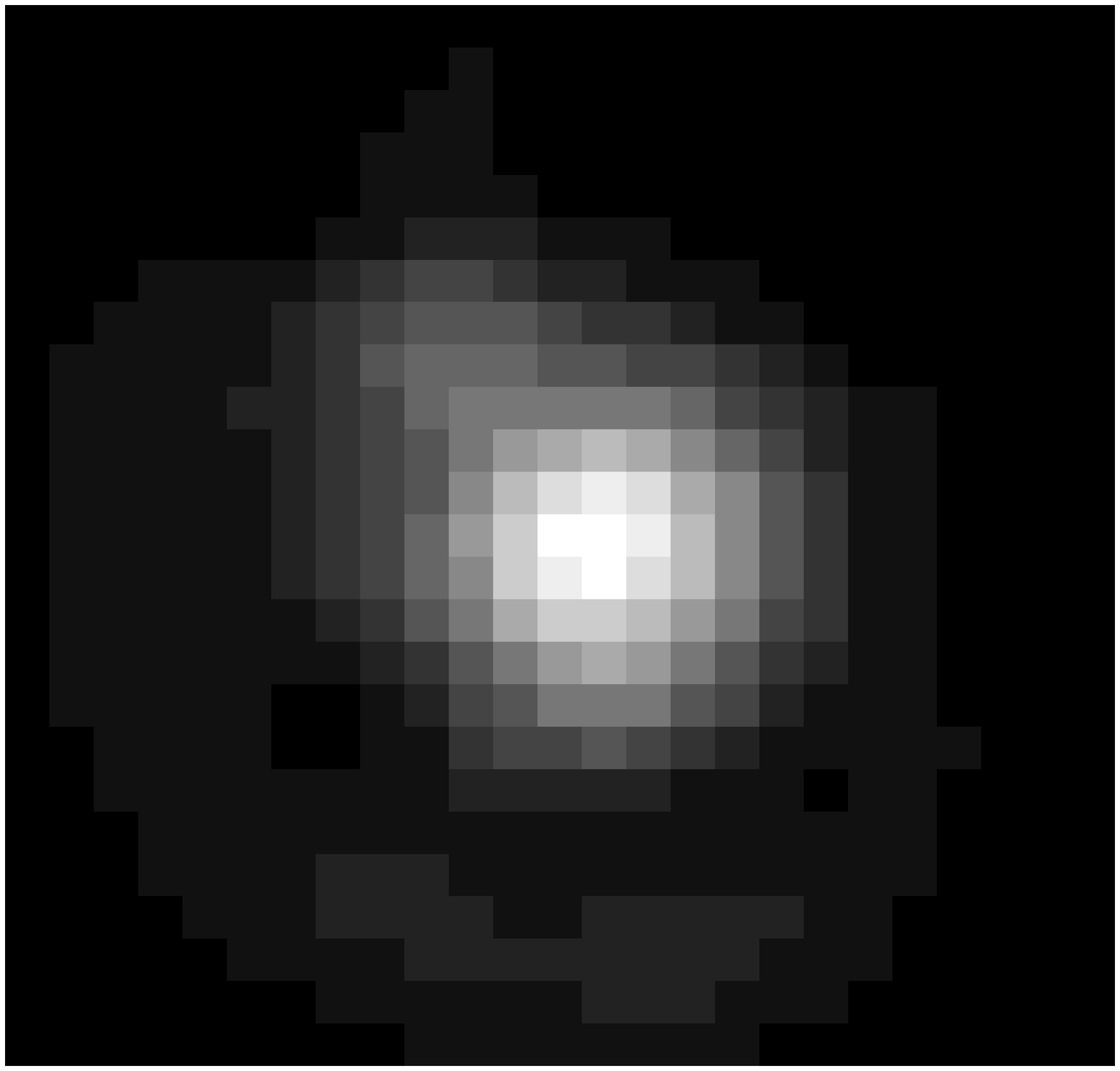,width=20mm}\hspace{2mm}
\psfig{figure=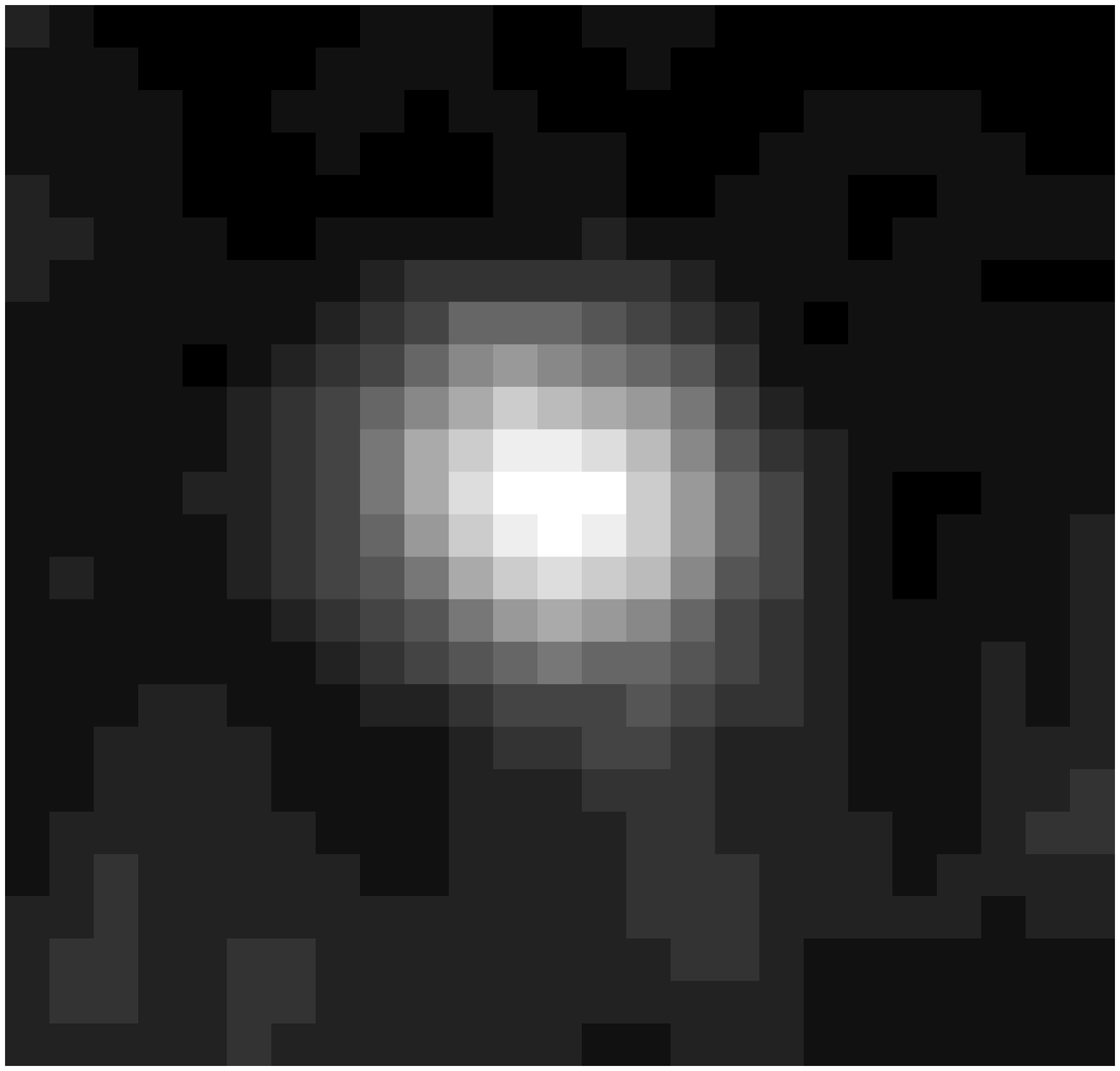,width=20mm}\psfig{figure=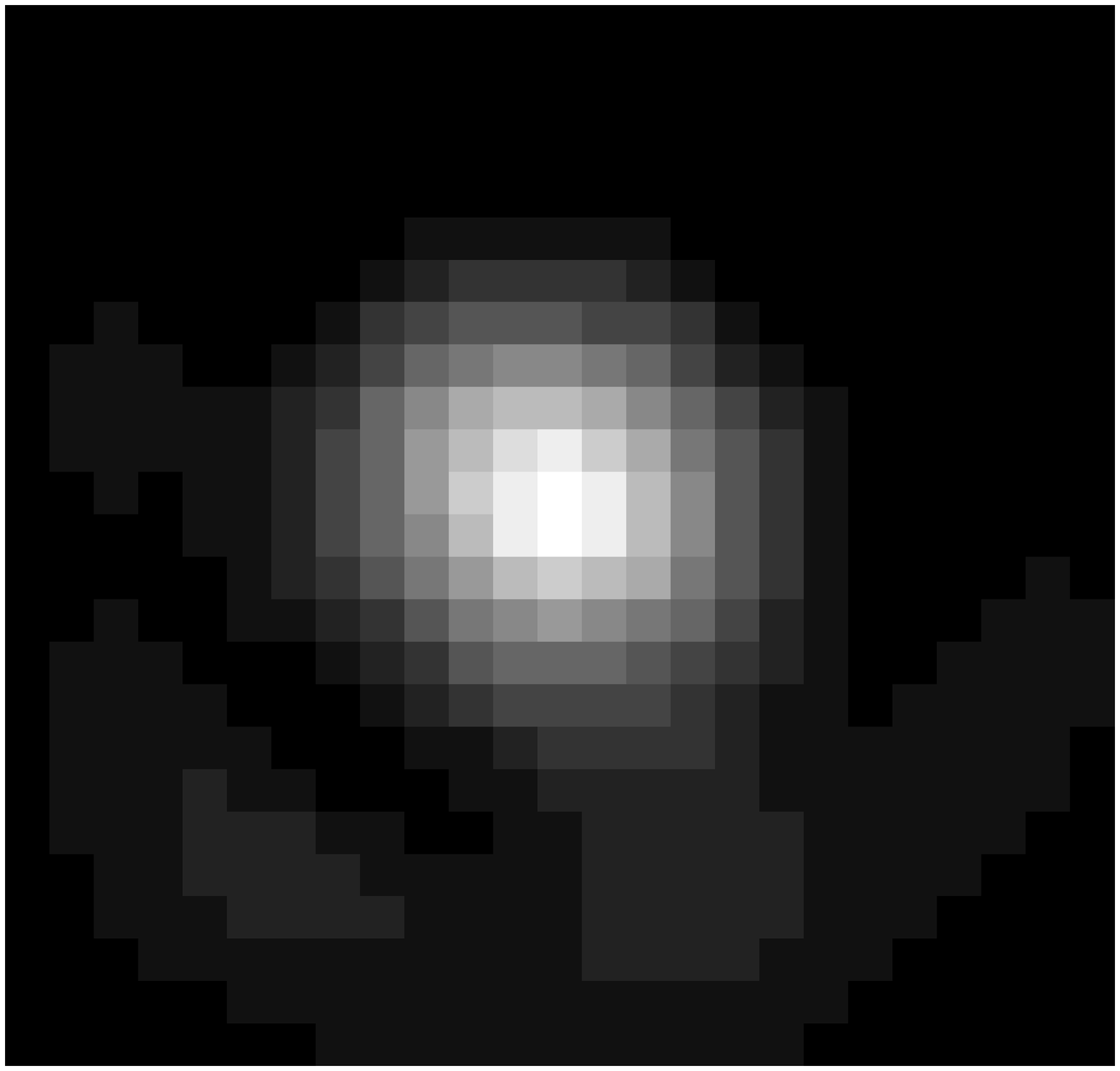,width=20mm}\hspace{2mm}
\psfig{figure=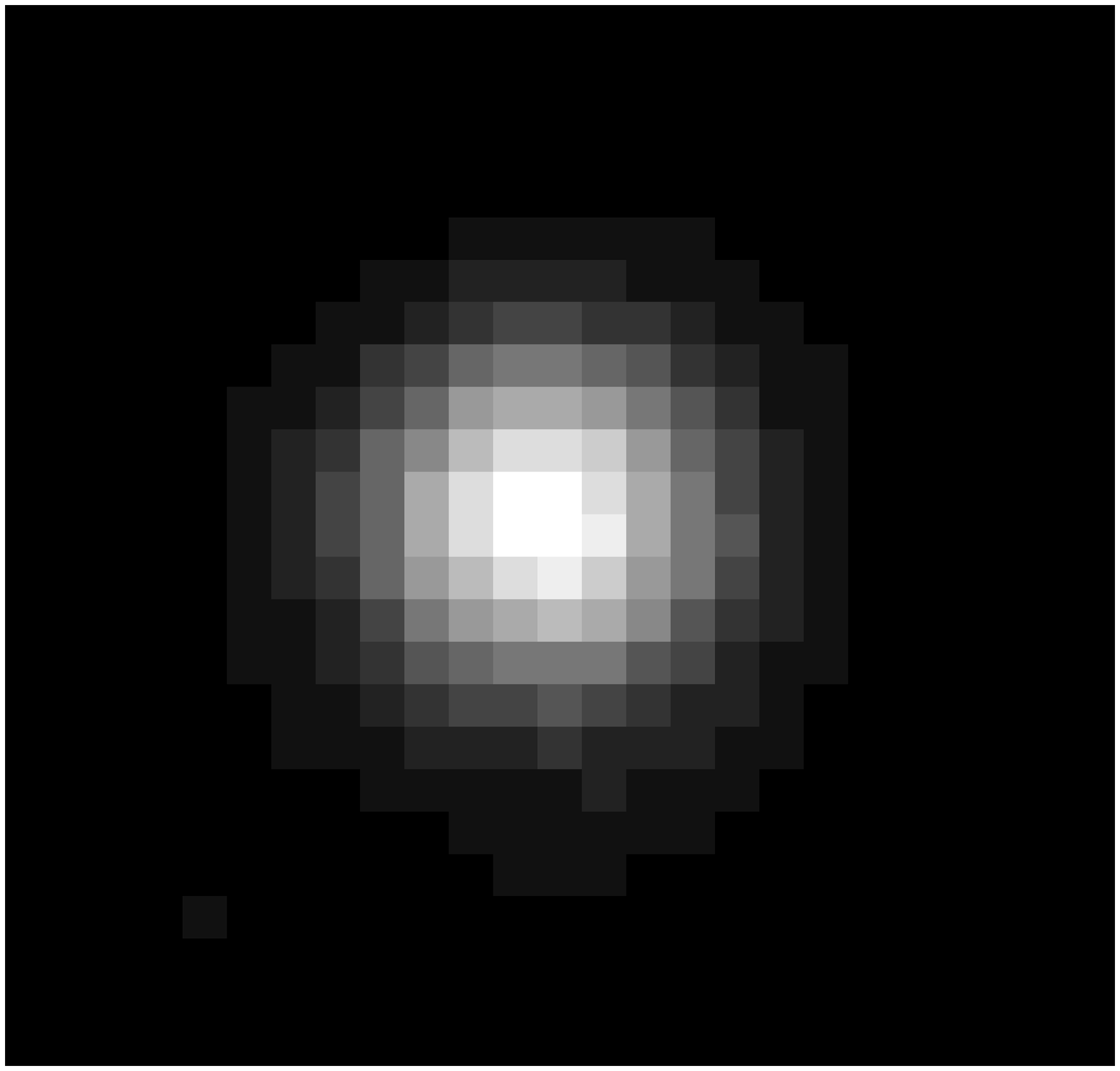,width=20mm}\psfig{figure=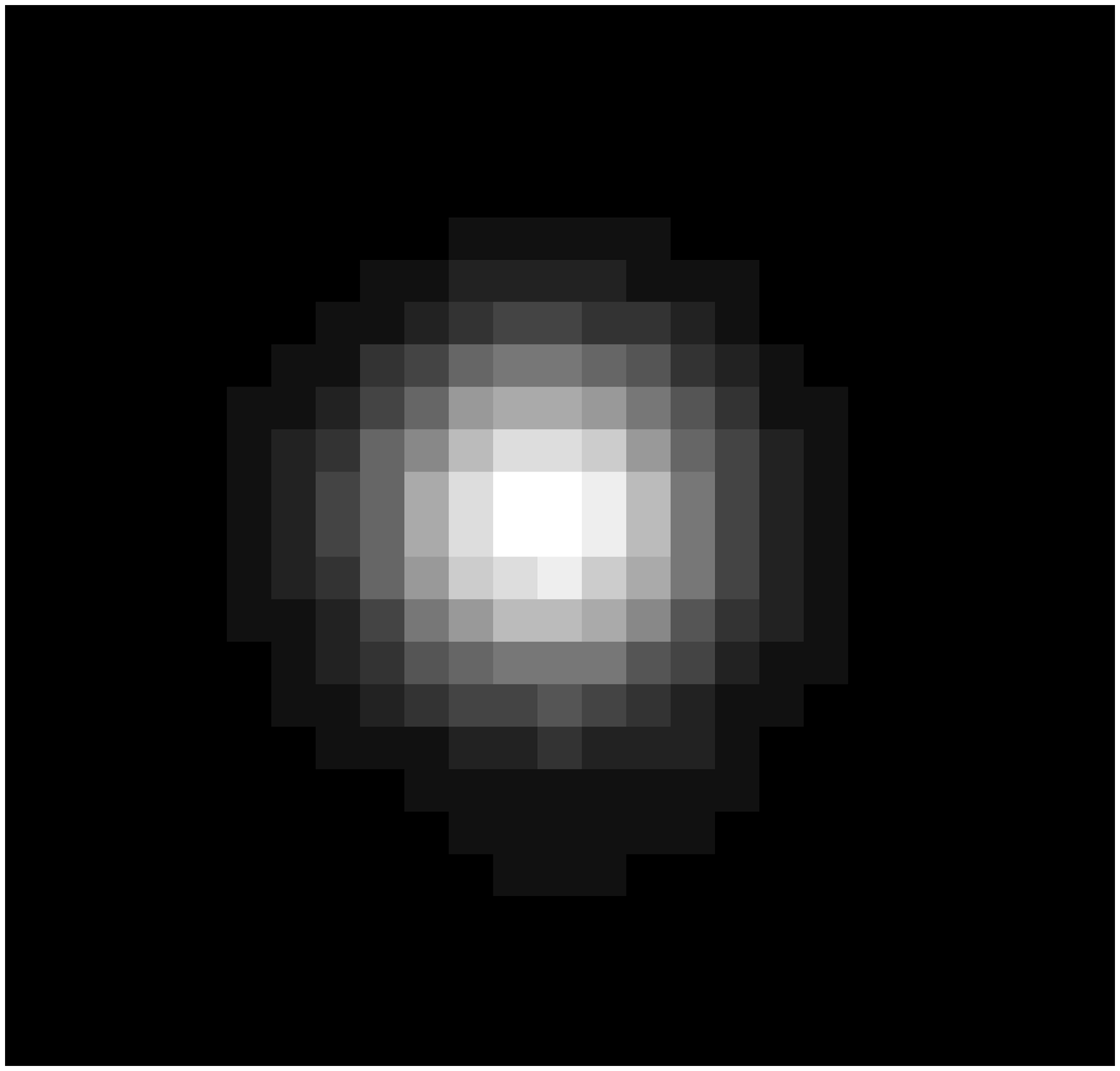,width=20mm}\hspace{2mm}
\psfig{figure=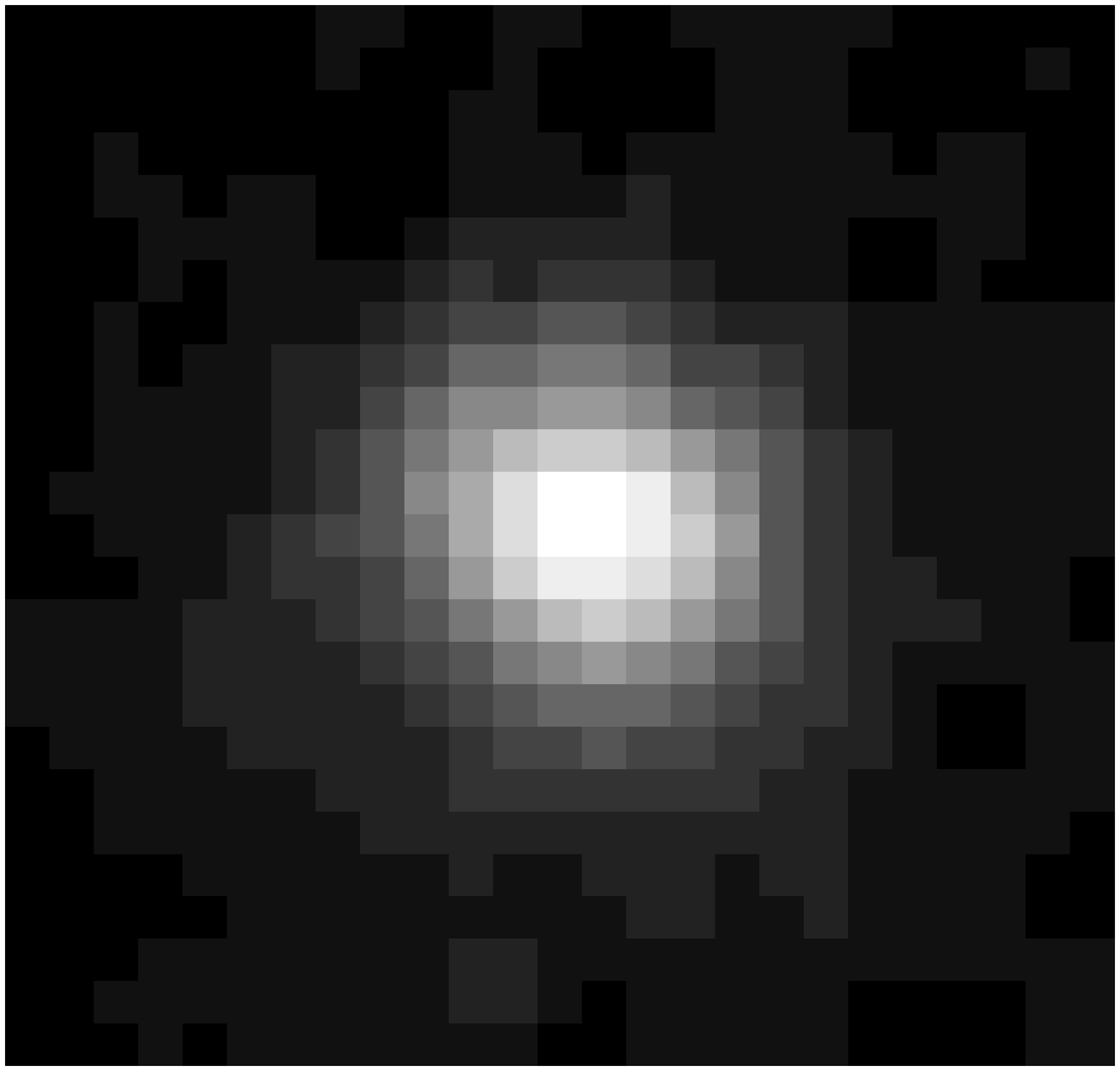,width=20mm}\psfig{figure=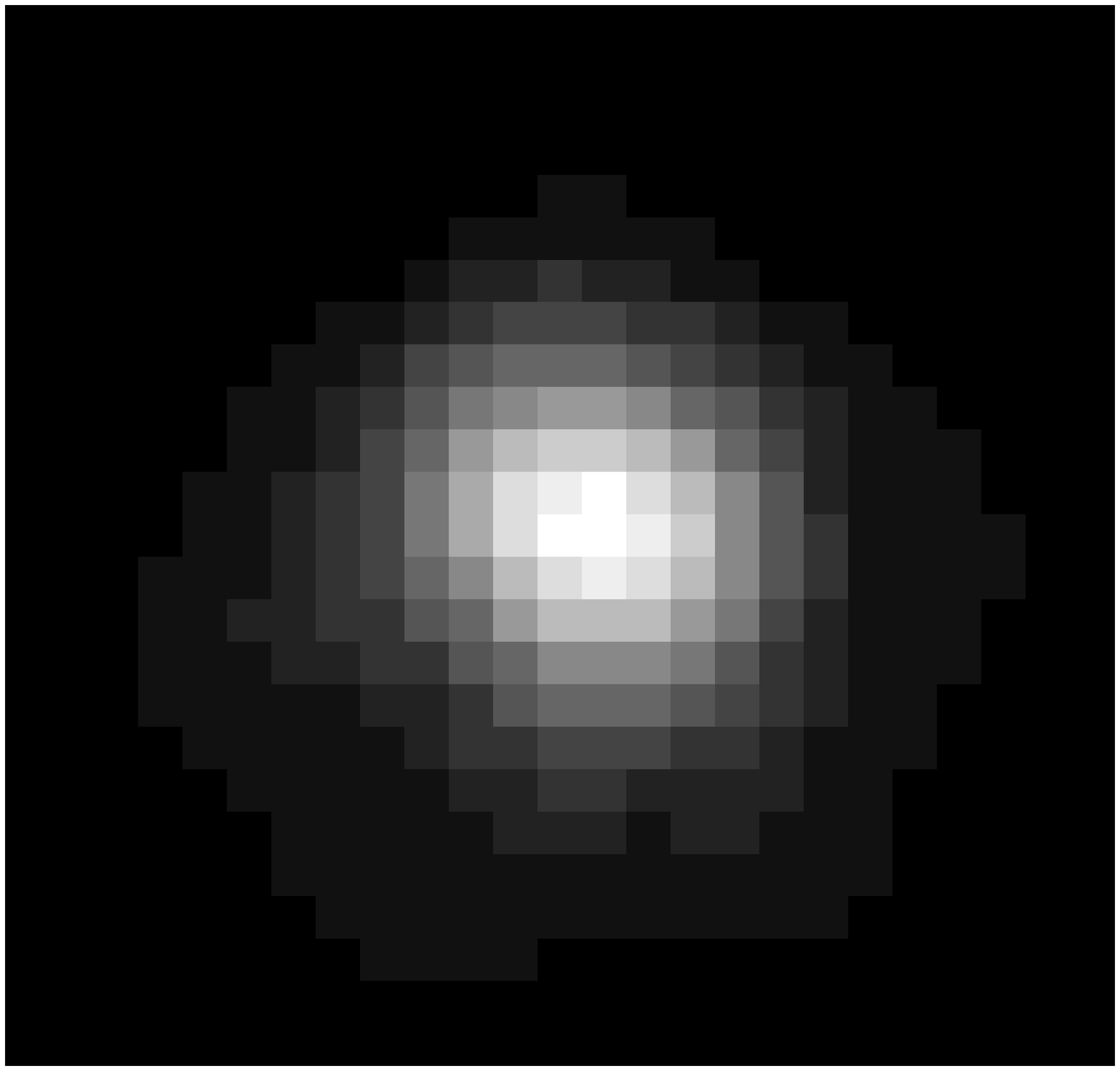,width=20mm}
\caption{Examples of radio sources in the HDF-N (left of each column) and their convolved shapelet reconstructions (right of each column). These decompositions have been carried out with fixed SExtractor centroid and $\beta$ proportional to the FWHM. \label{fig:shapeletsexamples}}
\end{figure*}
Here we summarize the shapelets method which is described more fully in \citet{2003MNRAS.338...35R}, \citet{2003MNRAS.338...48R} and \citet{2005MNRAS.363..197M}. In this approach, in the Cartesian formalism, the surface brightness $f(\mathbf{x})$ of a galaxy is decomposed into a series of localised orthonormal basis functions $B_{n_{1},n_{2}}$ called shapelets:
\begin{equation}
\label{eq:decomp}
f(\mathbf{x})=\sum_{n_{1},n_{2}}f_{n_{1},n_{2}}B_{n_{1},n_{2}}(\mathbf{x};\beta),
\end{equation} 
where
\begin{equation}
\label{eq:cartbasis}
B_{n_{1},n_{2}}({\bf x};\beta)=\frac{H_{n_{1}}\left(\frac{x_{1}}{\beta}\right)H_{n_{2}}\left(\frac{x_{2}}{\beta}\right)e^{-\frac{|x|^{2}}{2\beta^{2}}}}{\left(2^{n_{1}n_{2}}\beta^{2}\pi n_{1}!n_{2}! \right) ^{\frac{1}{2}}},
\end{equation}
and where $H_{m}(\eta)$ is the Hermite polynomial of order $m$, with the characteristic scale of the basis described by $\beta$. The series converges most quickly if the characteristic scale $\beta$ is chosen to be similar to the size of the galaxy, and the centroid of the object is located accurately. The sum of $n_{1}$ and $n_{2}$ is referred to as the order of the basis functions. In practice any decomposition has to be truncated at some order $n_{\rm max}$ such that the decomposition yields a sufficiently accurate model of the galaxy while also being computationally efficient, as the computation time of each object's decomposition is $\propto n_{\rm max}^4$. From orthonormality, we can find shapelet coefficients for a galaxy by calculating
\begin{equation}
\label{eq:overlap}
f_{n_{1},n_{2}}=\int\,f(\mathbf{x})B_{n_{1},n_{2}}({\bf x};\beta)\,d^{2}\mathbf{x}.
\end{equation}
We use the publicly available shapelets software package\footnote{http://www.astro.caltech.edu/$\sim$rjm/shapelets/code/} described in \citet{2005MNRAS.363..197M} in order to make shapelet decompositions for all our objects. This code is well tested using optical data (c.f. \citealp{2006MNRAS.368.1323H} and \citealp{2007MNRAS.376...13M}), and we seek to extend its applicability to radio data here.

The code usually fits convolved shapelet coefficients to a galaxy while also optimizing centroid $\mathbf{x}_{c}$, $\beta$ and $n_{\rm max}$ using a non-linear algorithm. For the radio objects we found that this led to a large number of failures, due to the incorrect estimation of $\beta$ or the centroid wandering off the edge of the postage stamp; in order to stabilise the behaviour, we fix the centroid position $\mathbf{x}_{c}$ to the SExtractor detection centroid. We also have the freedom to fix $\beta$ to 0.4 times the SExtractor FWHM which we find consistently leads to models with reasonably low $n_{\max}$. Figure \ref{fig:shapeletsexamples} shows some of the radio objects and their resulting shapelet models (still convolved with the beam). 

\subsection{PSF Deconvolution}
To deconvolve the beam/PSF from the radio/optical data, we require a shapelet model of the relevant kernel. In the radio image the restoring beam is exactly known, so the deconvolution process is relatively straightforward. As mentioned above, the restoring beam is a 0.4 arcsec circular Gaussian. \citet{2005MNRAS.363..197M} describe the ensuing deconvolution step in detail; briefly, the shapelets are convolved with the PSF model, and the resulting functions are least-squares fit to the data. The coefficients of the fit correspond to the deconvolved model.

To estimate the ACS PSF, we also use SExtractor to create a catalogue of stars using the SExtractor star/galaxy classifier index (see \citet{1996A&AS..117..393B} for details). We decompose each stellar image into shapelet coefficients; for each shapelet mode, the mean coefficient is used for the PSF model. Again, the methodology of \citet{2005MNRAS.363..197M} is used to deconvolve all galaxy images. 

\subsection{Shear Catalogue}
Before arriving at a final shear catalogue, various necessary cuts were applied to the datasets. Galaxies with failures in the shapelet modelling, due to poor $\chi^2$ fits, have been removed. In the radio case we also removed all objects that were not resolved, i.e. FWHM$<0.4''$, and also applied a flux cut of $S_{1.4}>54\mu\mbox{Jy}$ to remove low level noise peaks. In the optical case we applied a magnitude cut of $m_{z}<25$ in order to only work with objects with S/N$>6$; this cut also removed all unresolved objects.

We now need to combine the shapelet coefficients of each object to estimate the weak shear they have experienced. We use the simple Gaussian-weighted shear estimator given by \citet{2007MNRAS.380..229M},
\begin{equation}
\mathbf{\tilde{\gamma}=\frac{\sqrt{2}f_{2,2}}{\langle f_{0,0}-f_{4,0}\rangle}.}
\end{equation}
Note that the average on the denominator is taken over the objects once the cuts described above have been made. In this fashion we calculate a two-component shear estimator for each useable galaxy in the survey. We are now ready to examine the properties of weak lensing in the radio at current flux limits.


\section{Weak Lensing Results}
\label{results}
\subsection{Useable number density}
In our final shear catalogue constructed as described above, we obtain
number densities of $n=0.75\mbox{ arcmin}^{-2}$ for the gold radio
dataset, $n=3.76 \mbox{ arcmin}^{-2}$ for the silver radio dataset, and $n=40.66\mbox{ arcmin}^{-2}$ for the optical data. It can immediately be seen that
currently, radio number densities are substantially lower than those
available at optical wavelengths. However, it should also be noted how
much $n$ has increased in relation to \citet{2004ApJ...617..794C}, where there were only $\simeq20$ objects
per square degree. With the imminent arrival of {\it e}-MERLIN and LOFAR, radio number densities will begin to compare well with optical number densities; this is crucial for weak lensing 2-point statistics, where the noise is inversely proportional to the number density.

\subsection{Shear Statistics}
\begin{figure}
\centering
\psfig{figure=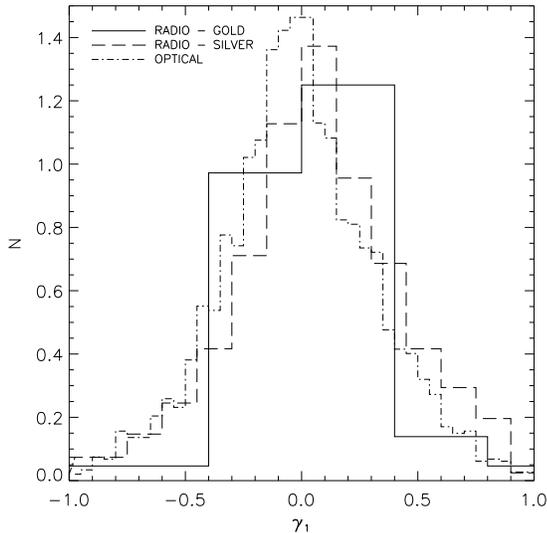,width=80mm}
\caption{Normalised distribution of shear estimators for the deconvolved gold (solid line) and silver (dashed lines) radio objects and the optical (dot-dashed line) objects.\label{fig:shearhists}}
\end{figure}

In addition, the noise on weak lensing 2-point statistics is proportional to the shear estimator variance, $\sigma_\gamma^2$, so this is an important item to quantify, together with the overall distribution of shear estimators. A histogram showing the distribution of the shear estimators for the gold and silver radio sets are shown in Figure \ref{fig:shearhists}, together with the distribution of the optical shear estimators. We see that the bin size is necessarily large for the gold set, on account of the small number of objects; nevertheless, we can immediately see that the variance of both radio sets and the optical set are comparable. Focussing on the silver and optical sets where there is more detail, we can see qualitatively that the distributions are similar in shape, with a peak and wings deviating from Gaussian form.

More quantitatively, for the radio samples we find that $\sigma_{\gamma_{1}}=0.29\pm0.03, \sigma_{\gamma_{2}}=0.30\pm0.03$ for the gold set, and $\sigma_{\gamma_{1}}=0.35\pm0.02, \sigma_{\gamma_{2}}=0.41\pm0.02$ for the silver set. This can be compared with $\sigma_{\gamma_{1}}=0.326\pm0.004, \sigma_{\gamma_{2}}= 0.328\pm0.004$ for the optical objects.

This is encouraging; it shows that radio shear estimators are not much more noisy than optical shear estimators at this flux limit. The combined effects of intrinsic ellipticity and measurement noise give comparable shear variance in both parts of the spectrum.

\begin{figure}
\centering
\psfig{figure=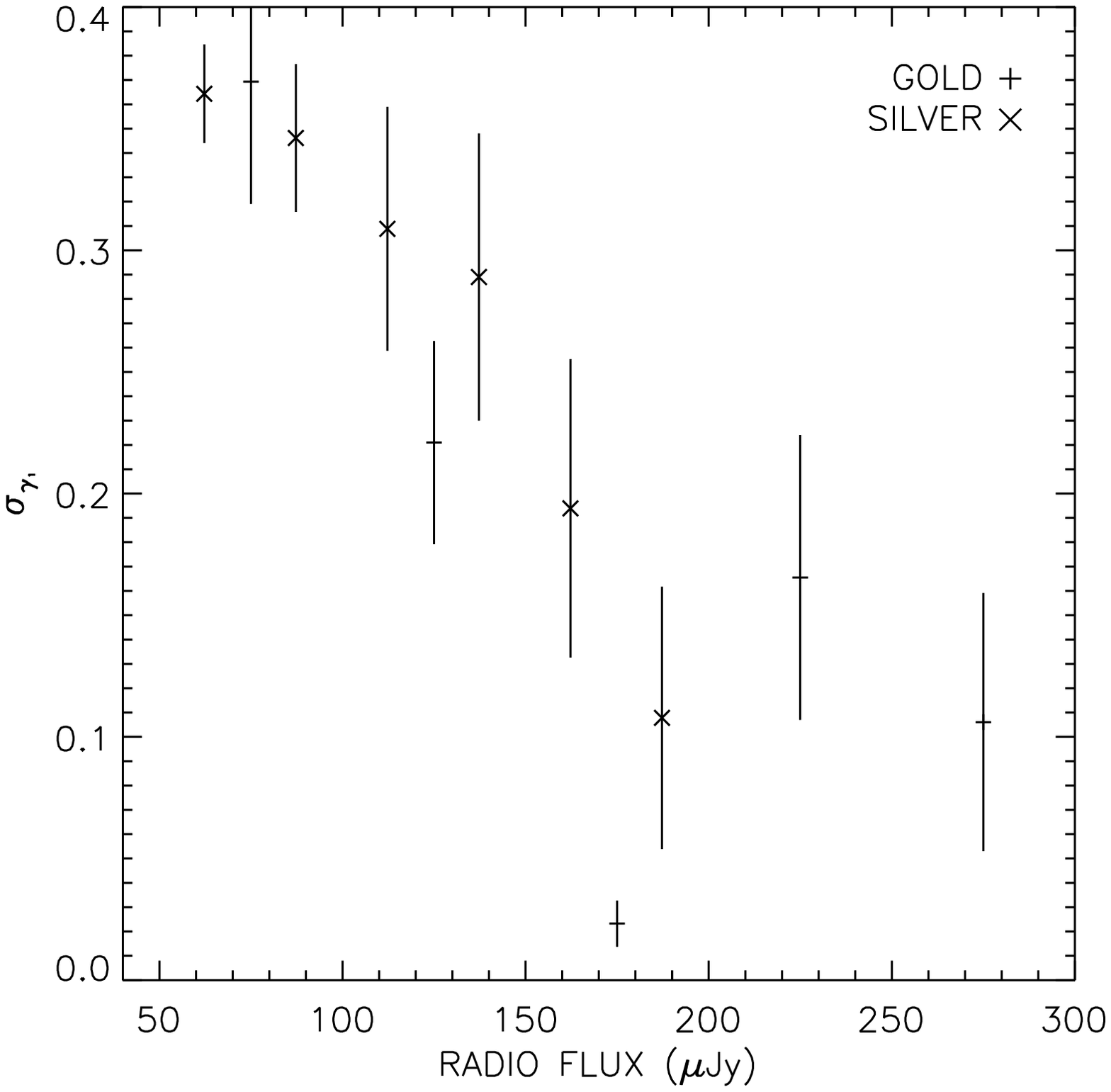,width=80mm}
\psfig{figure=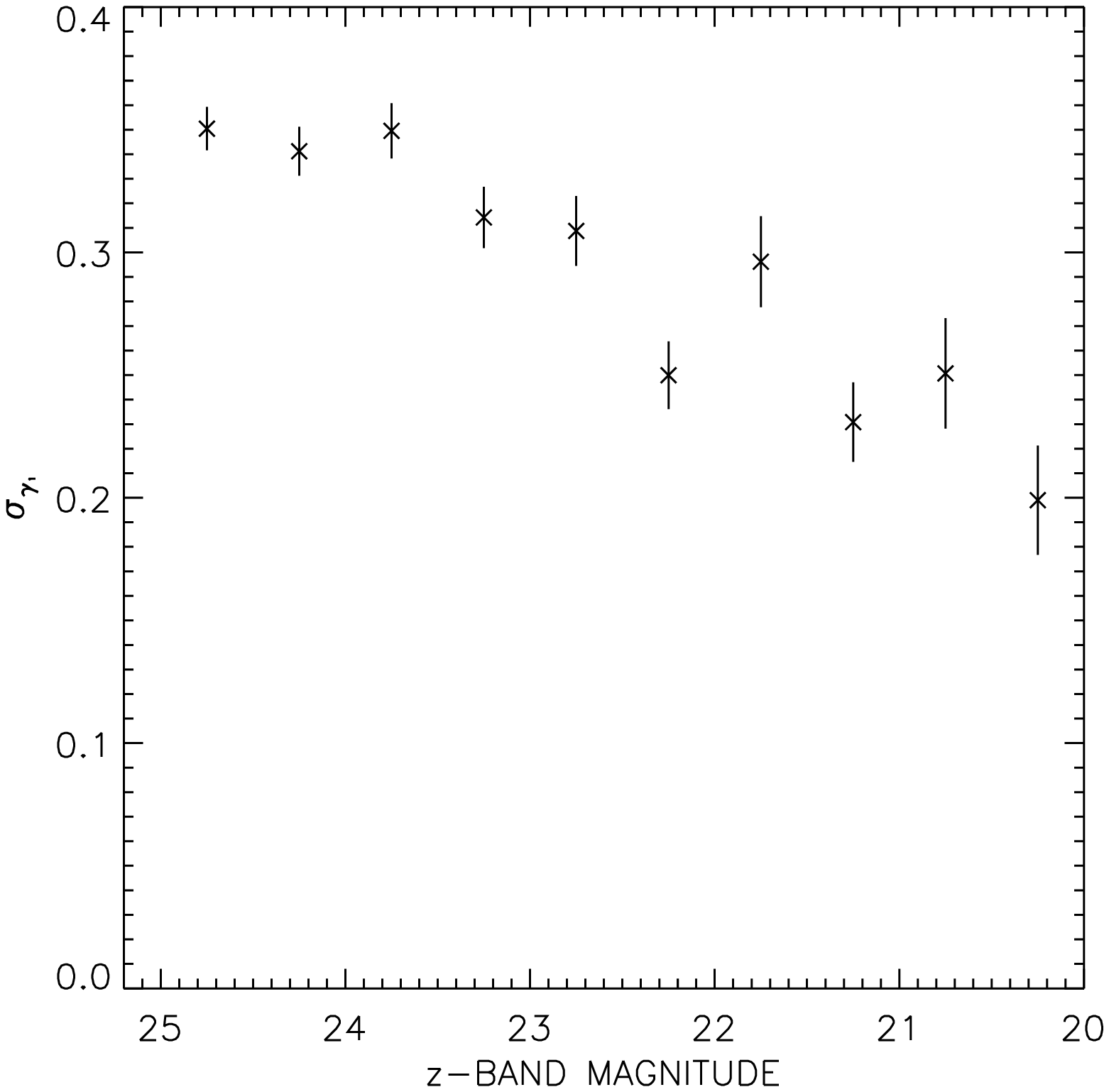,width=80mm}
\caption{\textit{Top panel}: Shear estimator dispersion $\sigma_{\gamma_{1}}$ for the deconvolved radio objects as a function of flux. \textit{Lower panel}: $\sigma_{\gamma_1}$ for optical data, as a function of z-band magnitude (with axis flipped for comparison with top plot).\label{fig:sig}}
\end{figure}

In Figure \ref{fig:sig} we show how the quantity $\sigma_{\gamma_{1}}$ varies with flux in the radio case (top panel) and with the $z$-band magnitude in the optical case (lower panel). For both the gold and silver objects, we find that the fainter objects have a larger scatter compared to the brighter ones, due to the reduced signal-to-noise, and the greater noise associated with deconvolution for small objects. The optical objects show a less pronounced trend, with bright objects having a little less scatter than the fainter ones.


\subsection{Systematic Errors}
\label{systematics}
We now wish to quantify the level of systematics present in the processed radio shear data. This paper seeks to assess the level of the problem, and to take a straightforward first step towards its amelioration in section \ref{cross}.

Firstly, we calculate the average shear for the entire catalogues. For the gold radio set,  we find $\langle\gamma_{1}\rangle=0.035\pm0.040$ and $\langle\gamma_{2}\rangle=-0.007\pm0.040$, consistent with no overall systematic offset. For the silver set, we find $\langle\gamma_{1}\rangle=0.072\pm0.021$ and $\langle\gamma_{2}\rangle=0.015\pm0.025$, indicating a systematic effect afflicting the estimators at this lower signal-to-noise. For the optical estimators we find $\langle\gamma_{1}\rangle=-0.014\pm0.006$ and $\langle\gamma_{2}\rangle=-0.003\pm0.006$, again showing an uncorrected small systematic due to the basic PSF correction, and/or a cosmic shear signal.

\begin{figure}
\centering
\psfig{figure=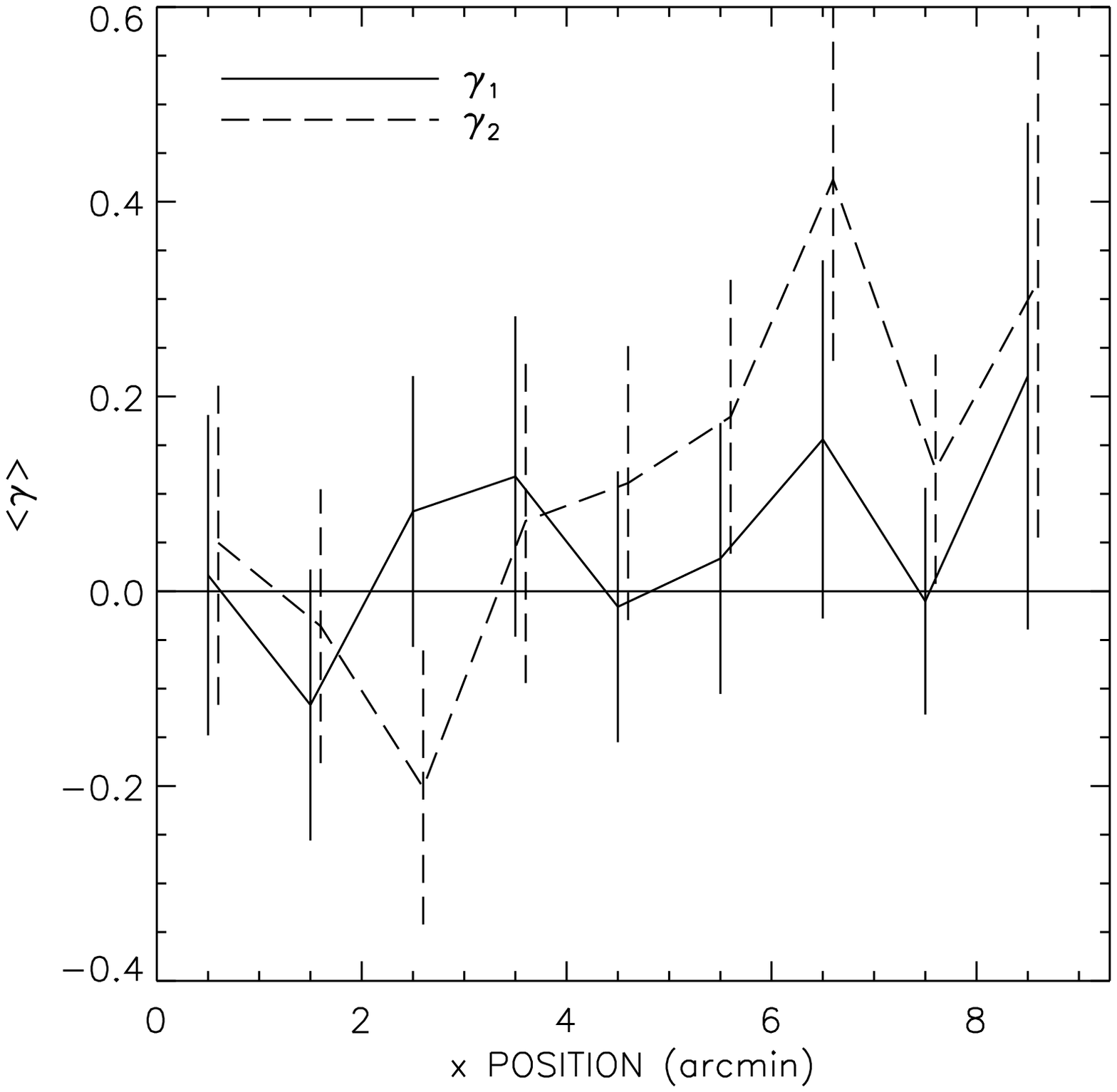,width=40mm}\psfig{figure=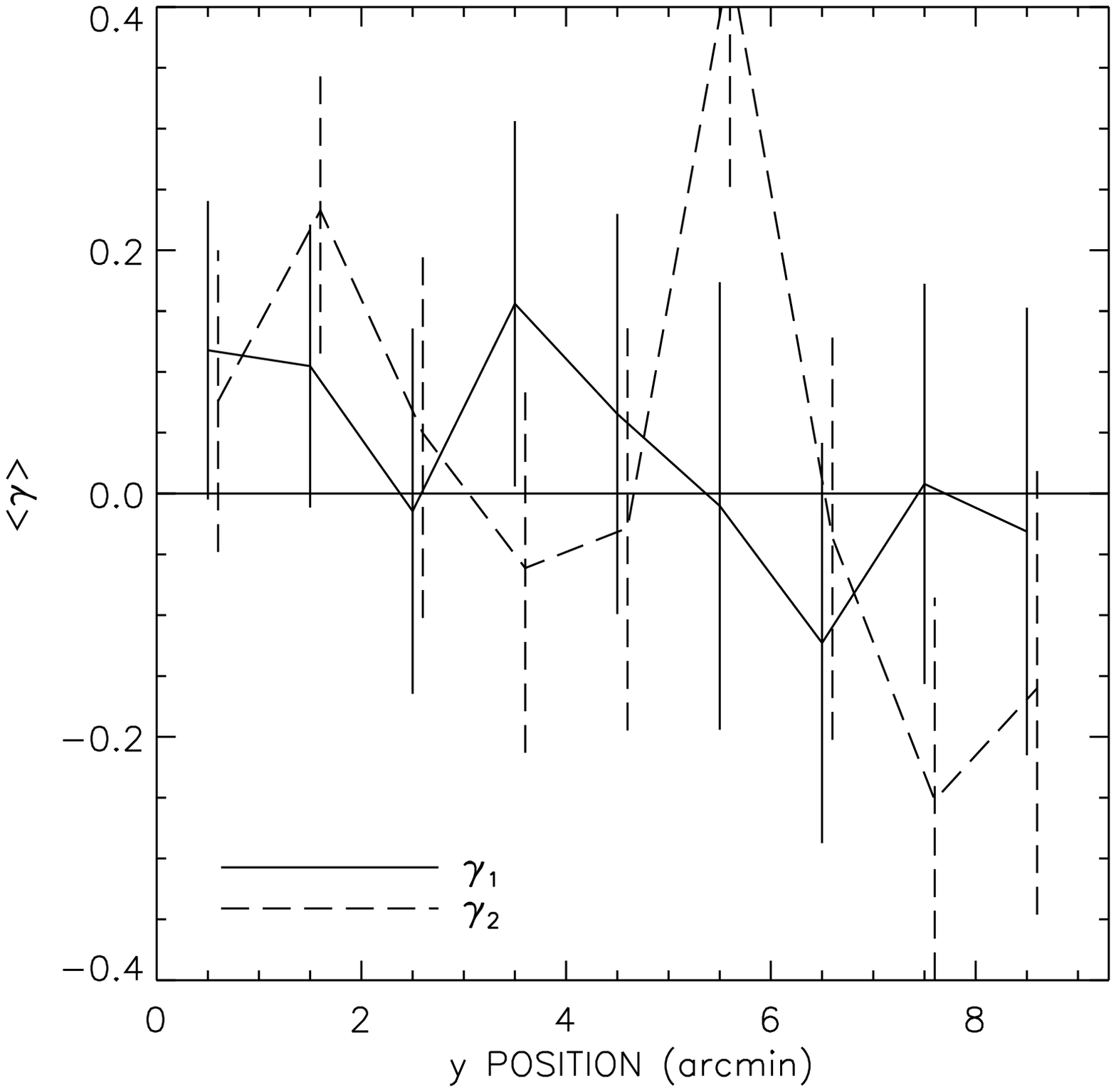,width=40mm} \\
\psfig{figure=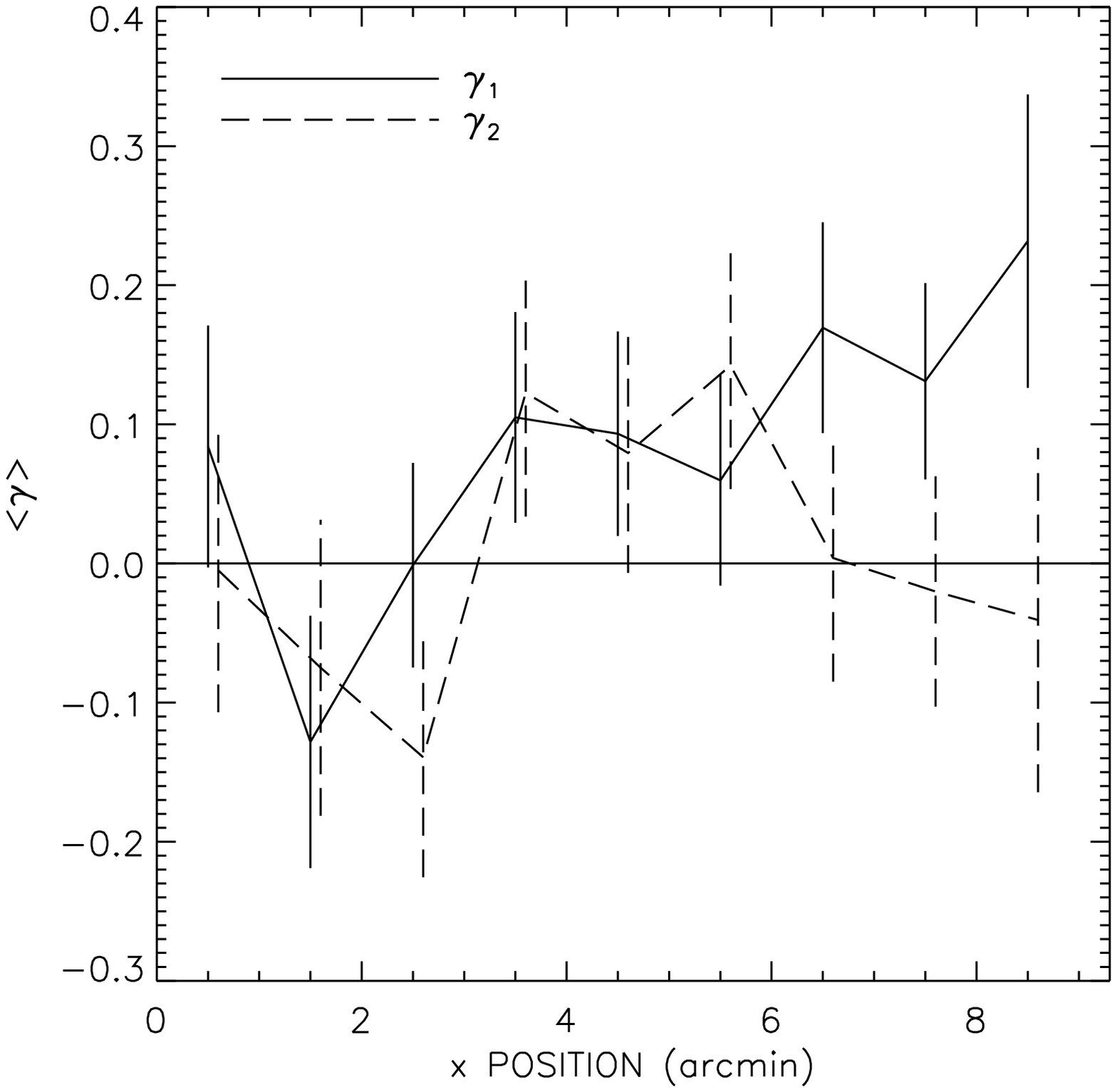,width=40mm}\psfig{figure=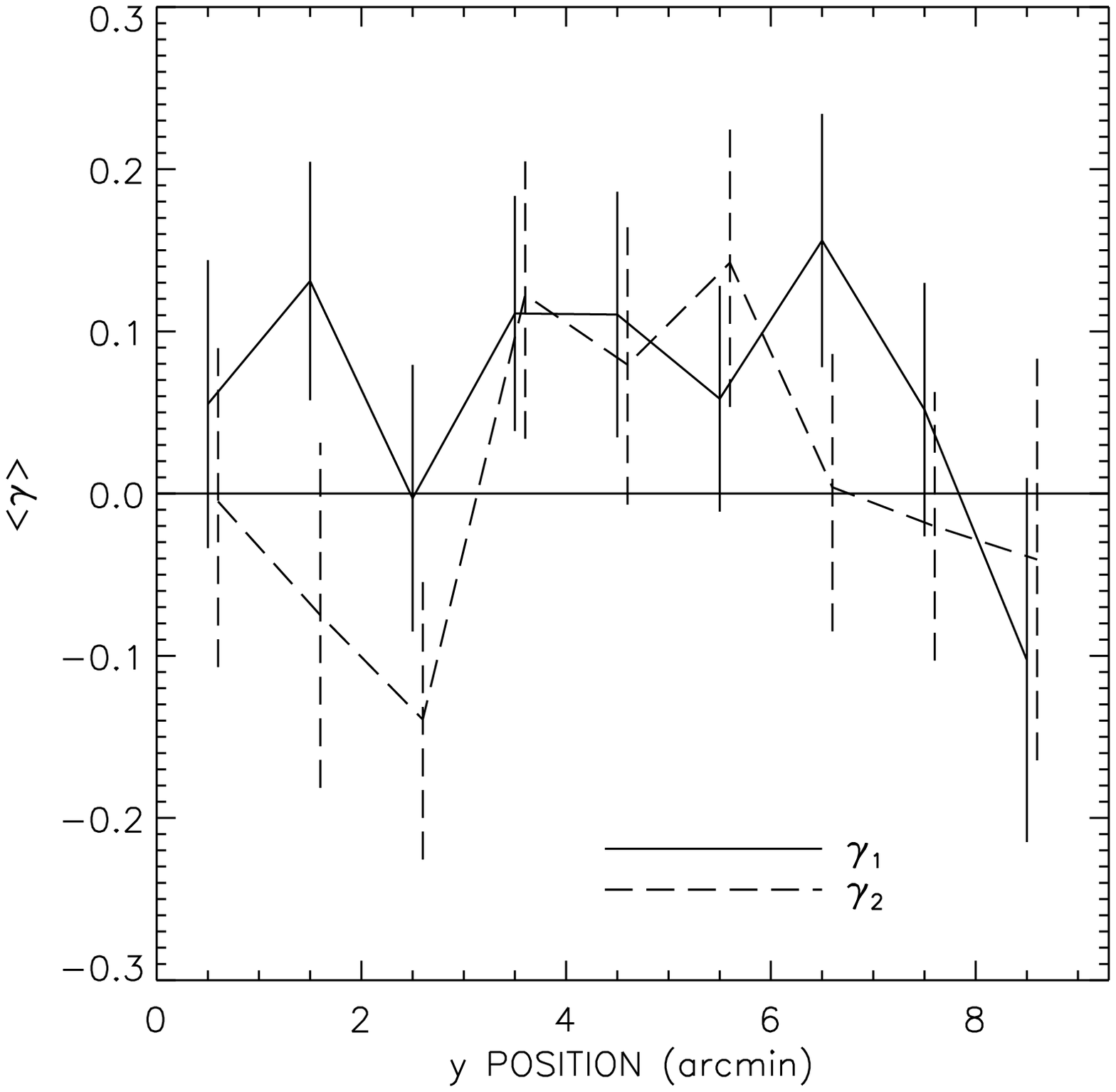,width=40mm} \\
\psfig{figure=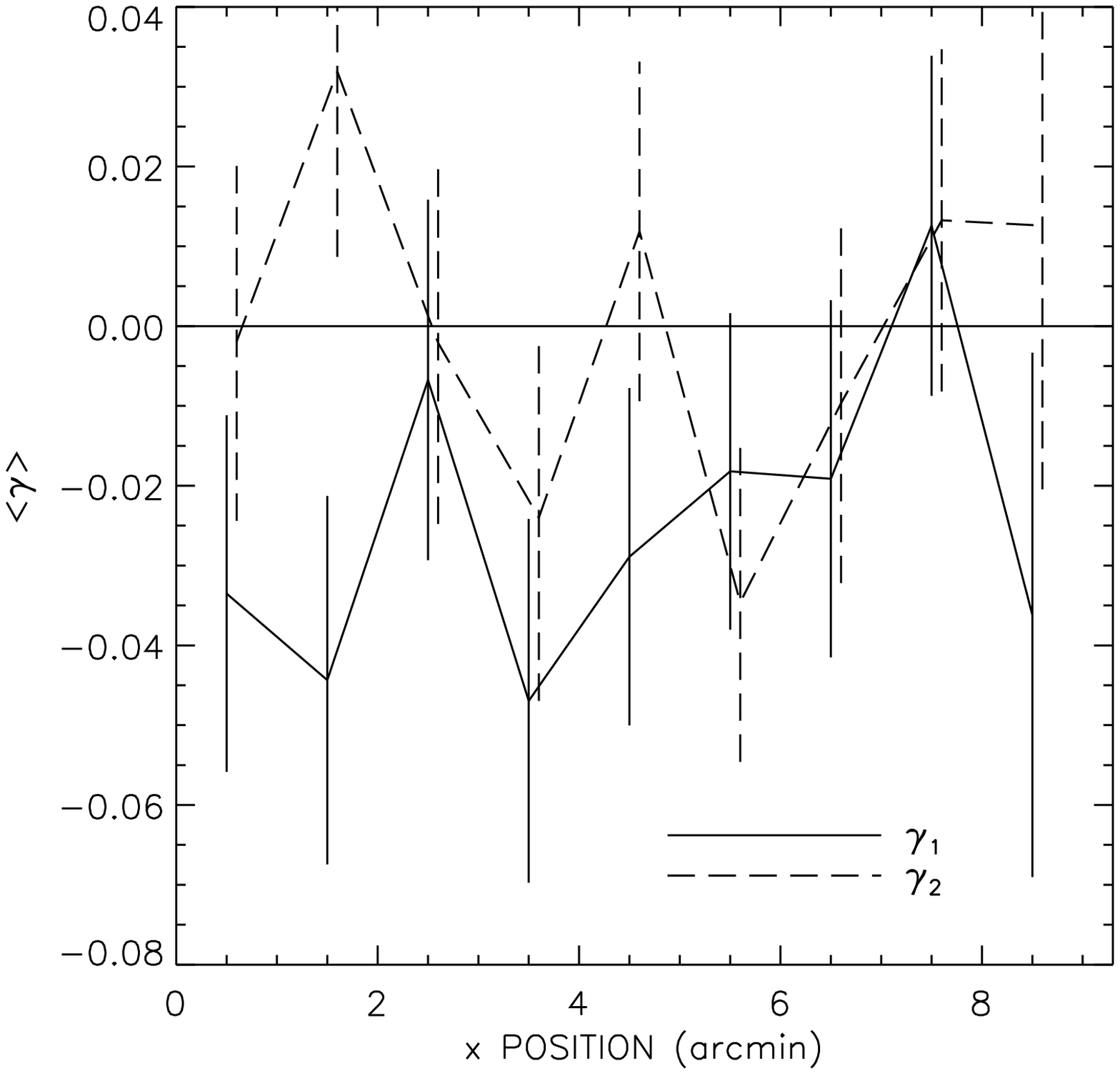,width=40mm}\psfig{figure=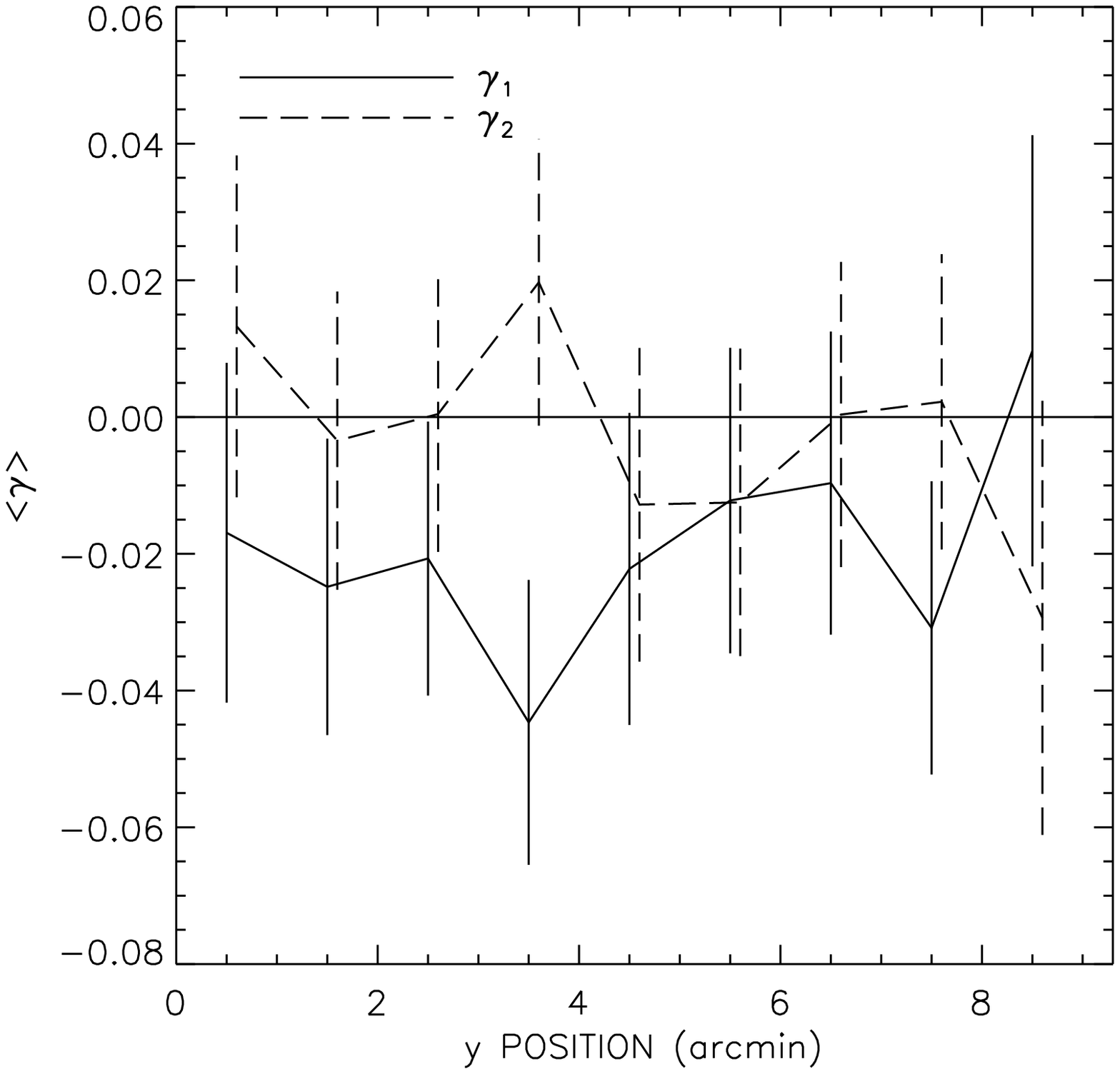,width=40mm}
\caption{Median shear estimators in $x$ and $y$ position bins. The top, middel and bottom panels show results for the gold radio, silver radio and optical objects respectively. 
\label{fig:gbarpos}}
\end{figure}

\begin{figure}
\centering
\psfig{figure=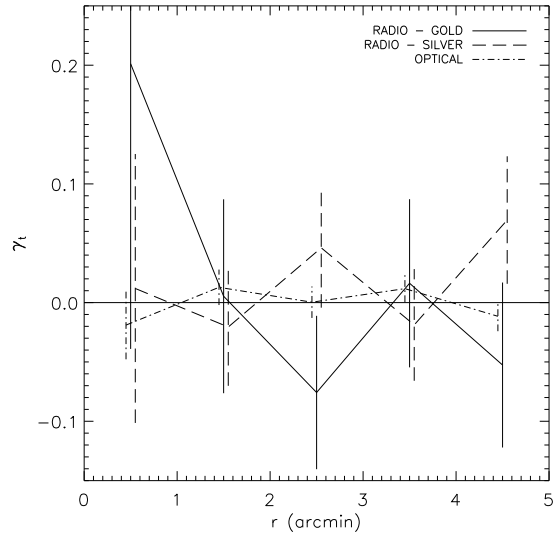,width=80mm}
\caption{Tangential shear as a function of radial distance from the centre of the image. \label{fig:gtr}}
\end{figure}

As a more detailed test, we median averaged our shear estimates in 1 arcminute bins in $x$ and $y$ position on the image (Figure \ref{fig:gbarpos}). For the gold radio objects (top panel) we see that $\gamma_{1}$ and $\gamma_{2}$ are consistent with zero in the $x$ direction, with the exception of one bin for $\gamma_2$. In the $y$ direction the results are again consistent with no systematic. For the silver set (middle panel) we find $\gamma_{1}$ to be contaminated in both the $x$ and $y$ direction, with $\gamma_{2}$ showing no evidence of systematics. For the optical objects (bottom panel) we see that there is a per cent level contamination in $\gamma_{1}$, with no evidence of systematics in $\gamma_2$. We also show in Figure \ref{fig:gtr} how the tangential component of the shear $\gamma_{t}=\gamma_{1}\cos(2\theta)-\gamma_{2}\sin(2\theta)$ varies with radial distance from the centre of the image. We see that in both the radio cases as well as the optical case this is consistent with zero even on the outskirts of the images.

\begin{figure}
\psfig{figure=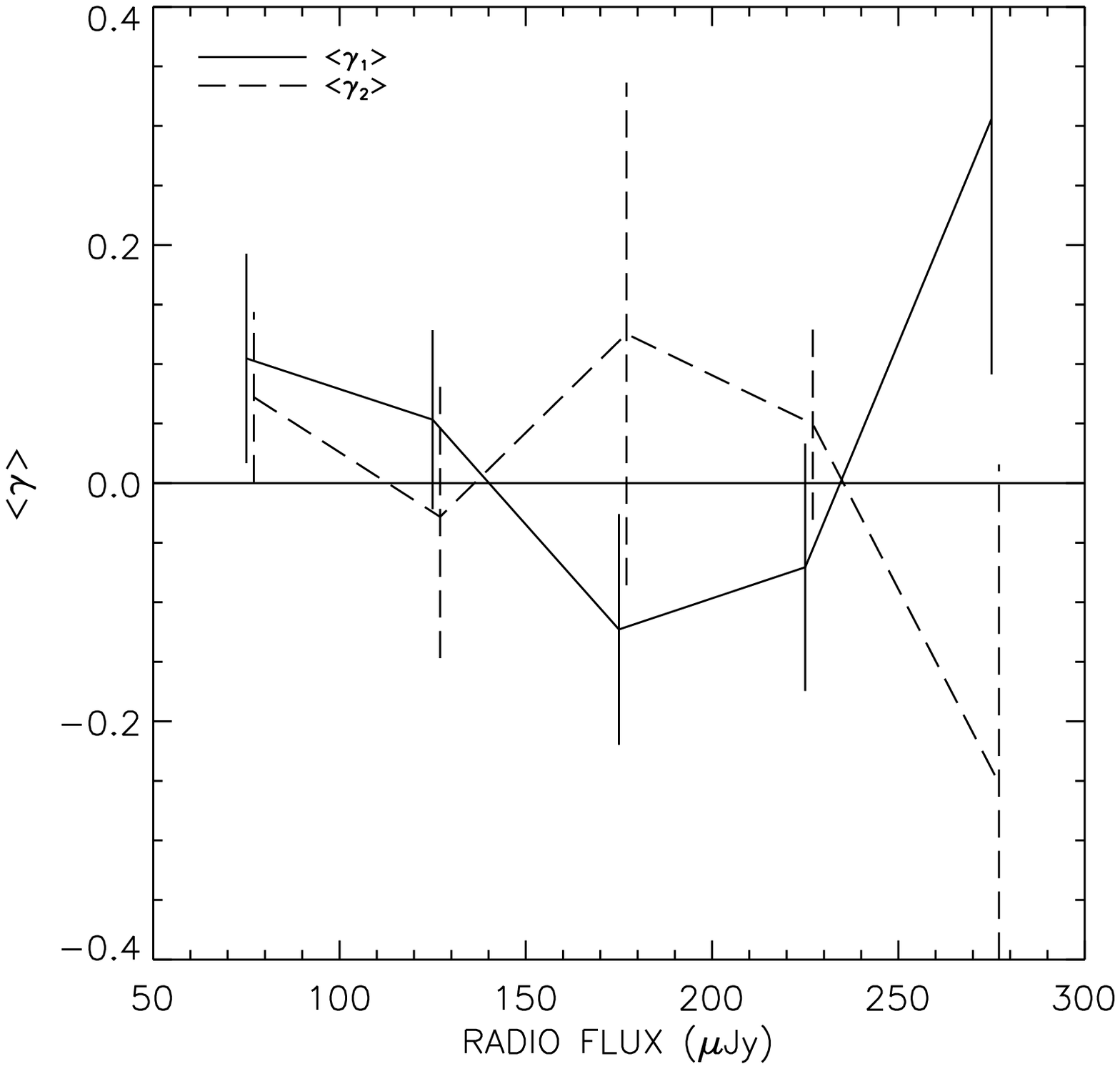,width=80mm}\\
\psfig{figure=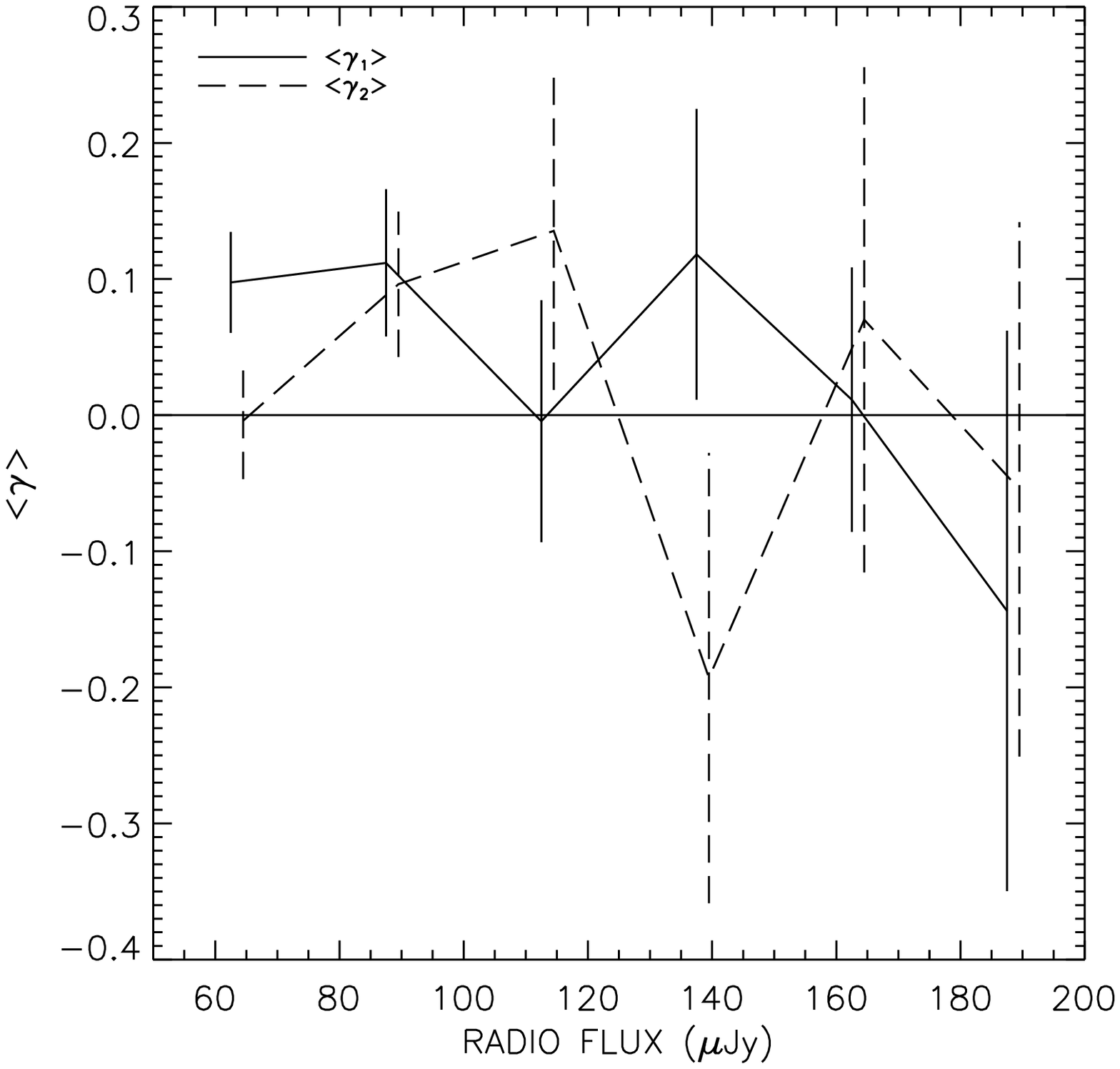,width=80mm}\\ 
\psfig{figure=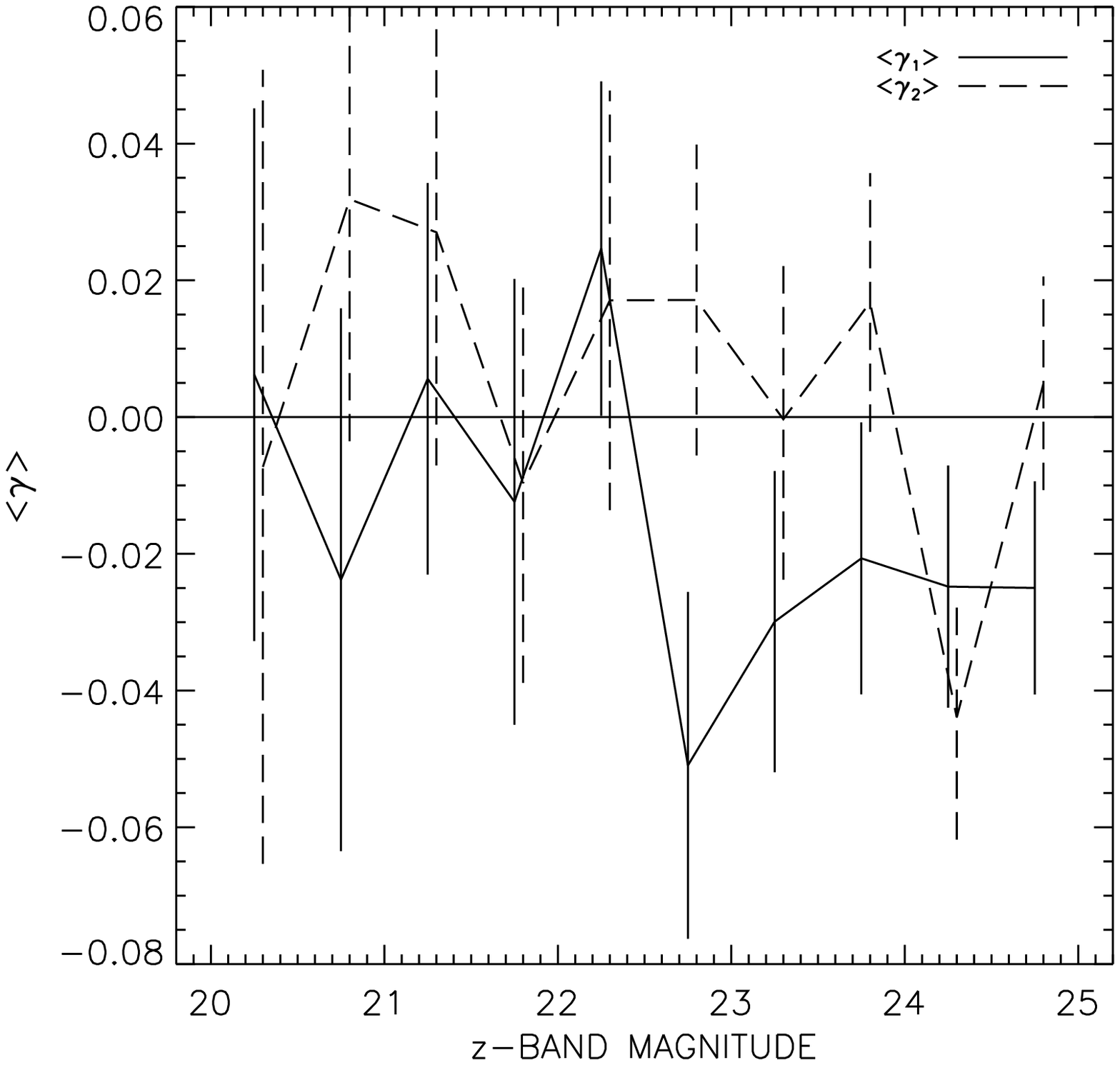,width=80mm}
\caption{$\langle\gamma\rangle$ in radio flux bins for the gold and silver radio objects (top and middle panels) and in $z$-band magnitude bins for the optical objects (lower panel). \label{fig:gbarfwhmmag}}
\end{figure}

\begin{figure}
\centering
\psfig{figure=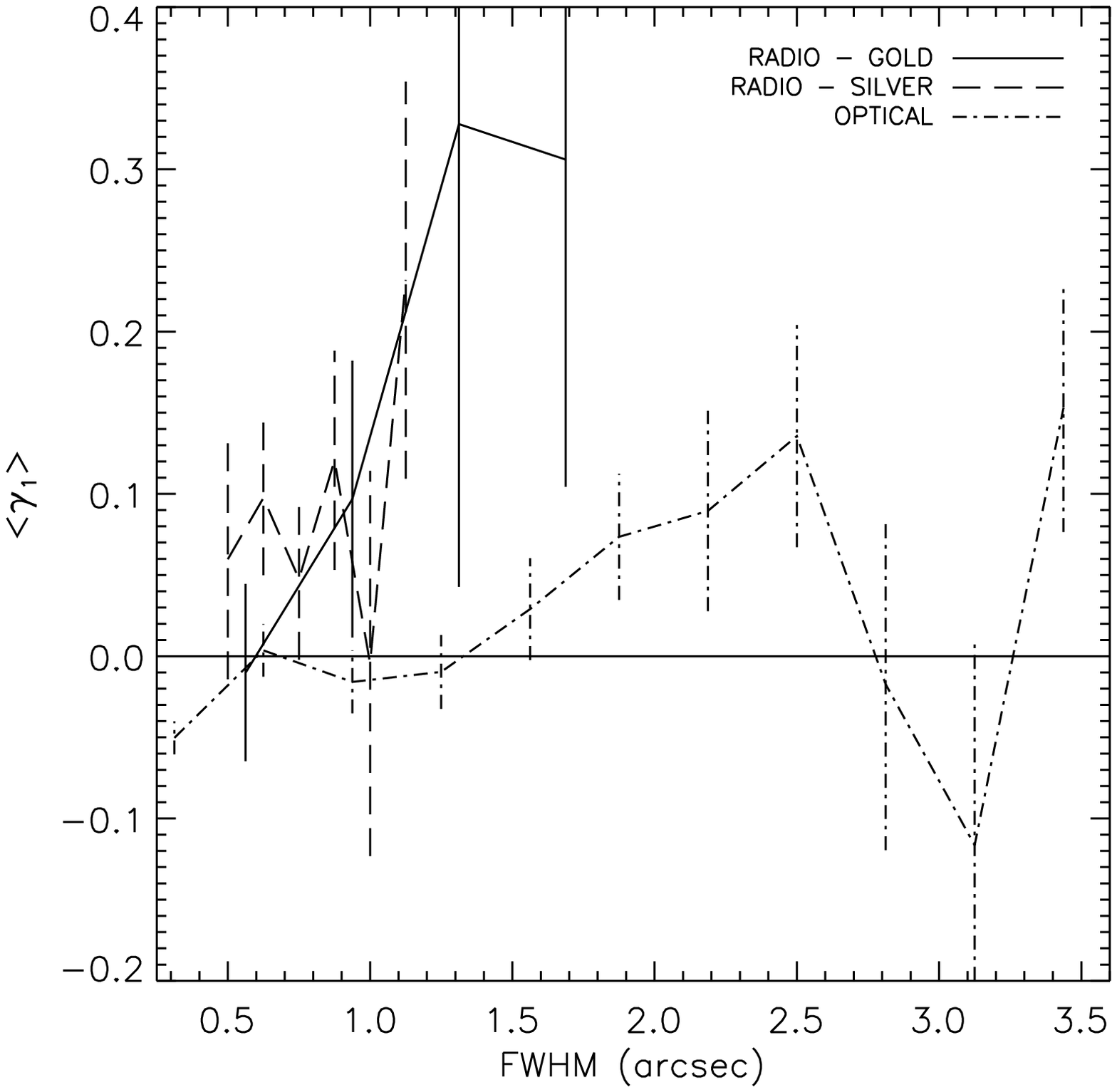,width=80mm}\\
\psfig{figure=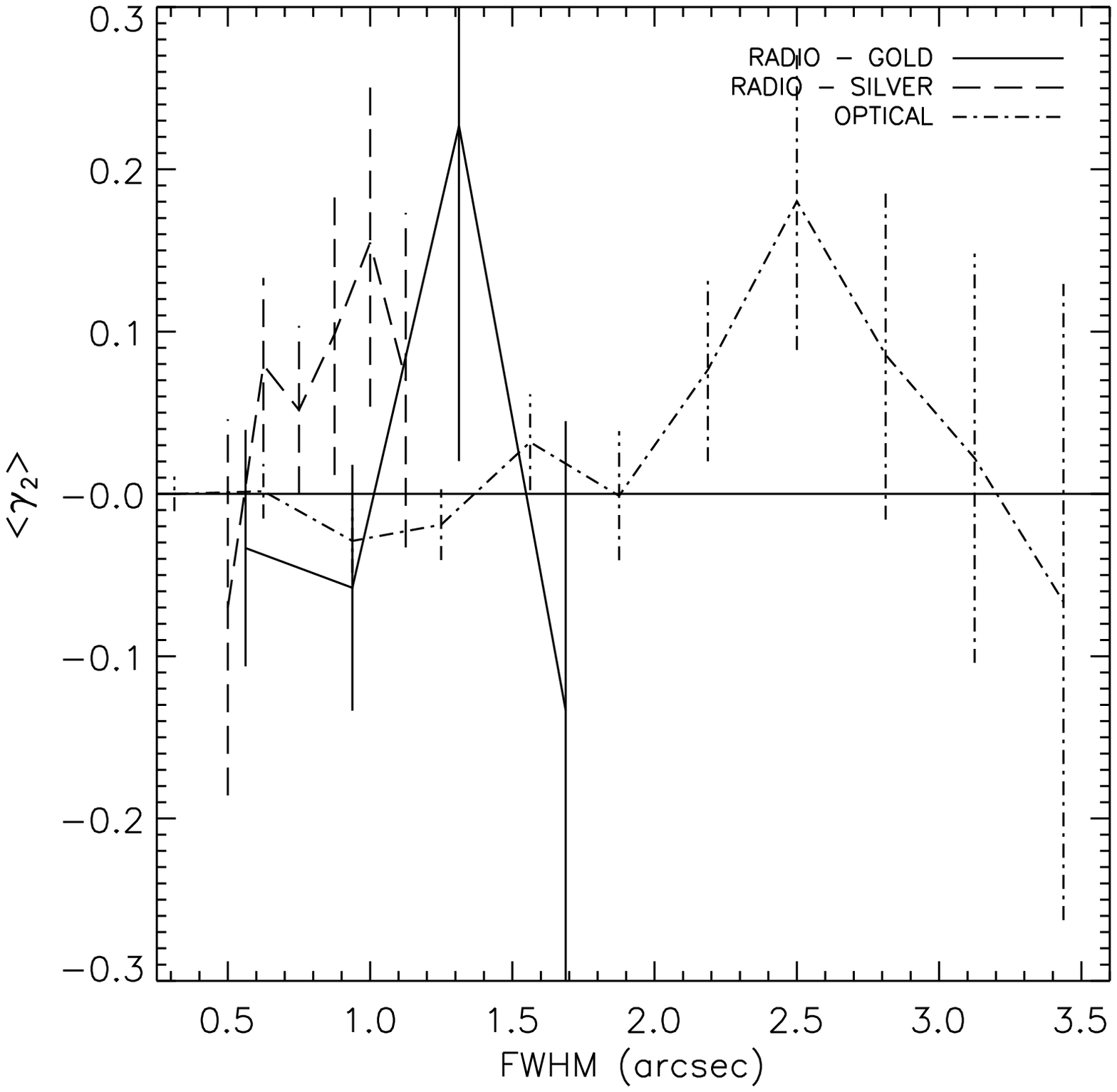,width=80mm}
\caption{$\langle\gamma\rangle$ in FWHM bins for the gold, silver and optical objects, for $\gamma_1$ (top) and $\gamma_2$ (bottom). \label{fig:gfwhm}}
\end{figure}

In order to isolate the source of the systematics present in the silver and optical data sets, we examine the median average shear in radio flux bins or optical $z$-magnitude bins (Figure \ref{fig:gbarfwhmmag}) and also as a function of object size (Figure \ref{fig:gfwhm}).

Firstly considering Figure \ref{fig:gbarfwhmmag}, we see that for the gold radio objects $\langle\gamma\rangle$ is consistent with zero as a function of flux. In the silver radio case we see that the fainter objects exhibit non-zero systematic shear. This may be due to anisotropic noise in the radio image which has not been removed by our deconvolution process. In the optical case we see some variation of $\langle\gamma\rangle$ with magnitude, with the strongest deviation from zero seen at the fainter magnitudes. This can be understood as uncorrected PSF systematics due to our basic correction method.

When we plot $\langle\gamma\rangle$ as a function of FWHM (Figure \ref{fig:gfwhm}), we again see that there is deviation from zero shear in the radio objects. At large FWHM, a few objects with large ellipticity dominate the signal. More troublesome is the non-zero shear for the silver at small FWHM; this is a further manifestation of the possible anisotropic noise problem raised above. The optical case is mainly consistent with zero.

\begin{figure}
\centering
\psfig{figure=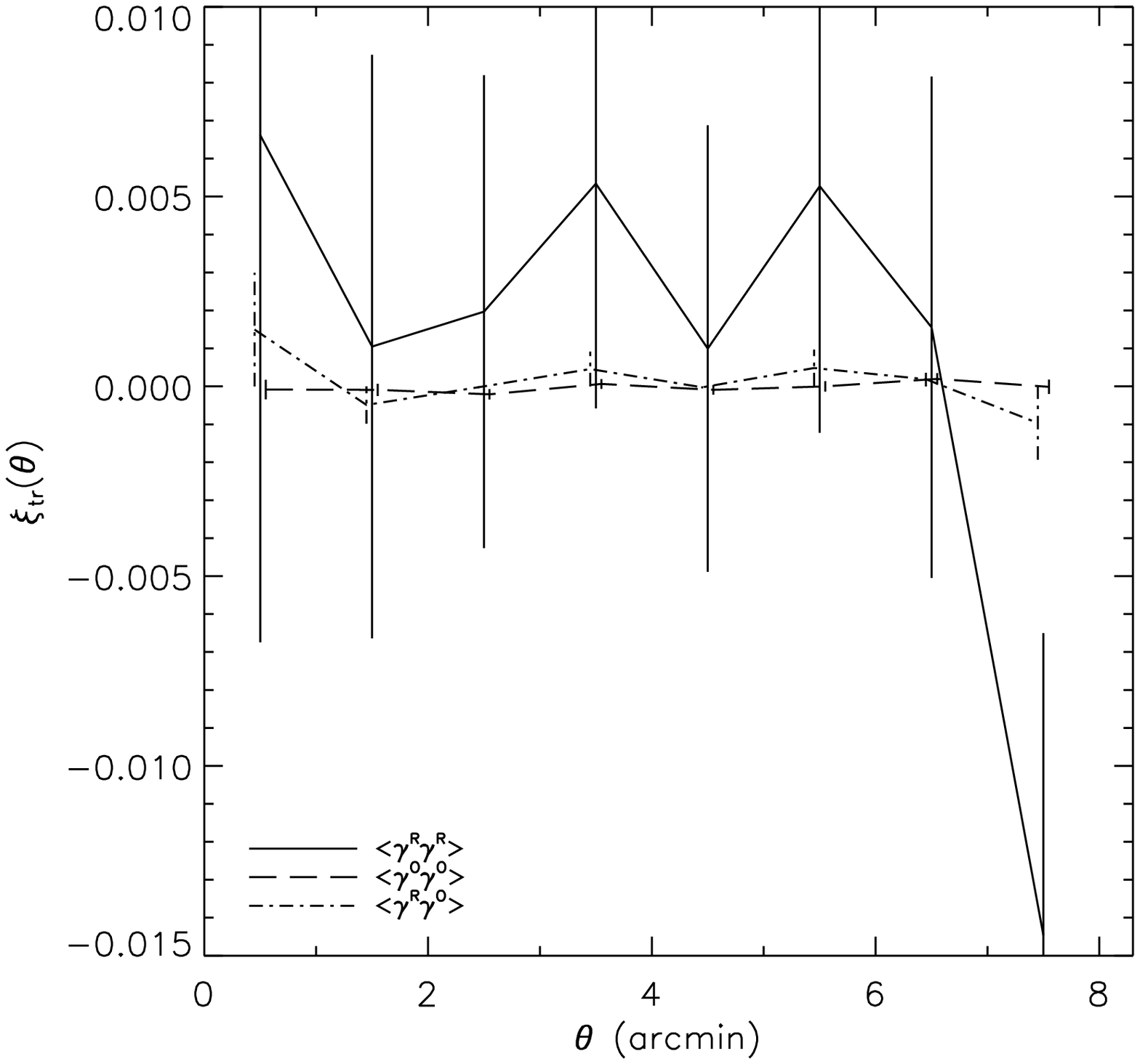,width=80mm}\\\psfig{figure=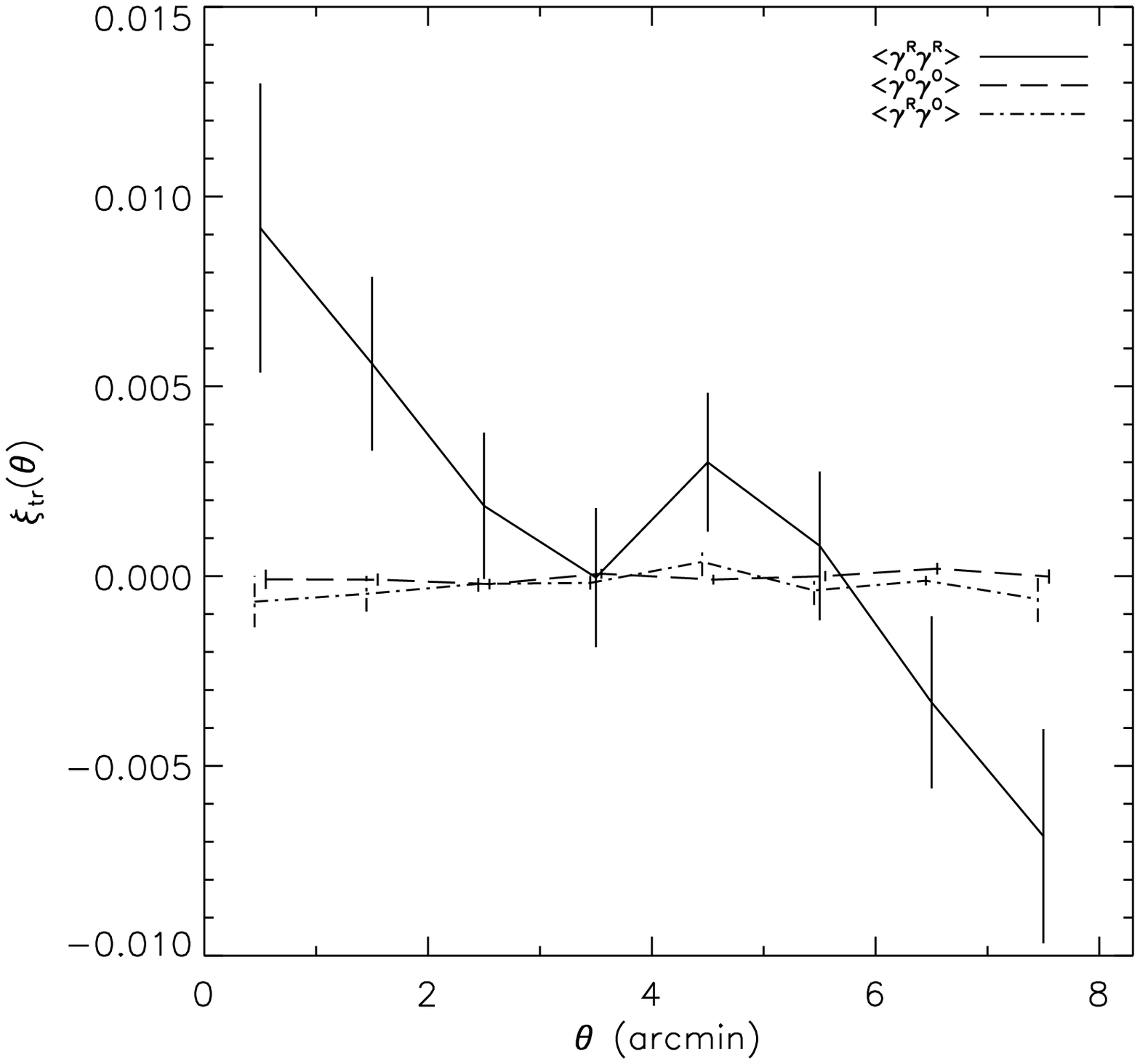,width=80mm}
\caption{\textit{Top panel}: $\xi_{tr}$ correlation functions for the gold radio objects overlaid with the optical and the gold-optical cross correlation functions. \textit{Lower panel}: same plot for the silver set, together with optical and optical-silver cross-correlation.\label{fig:c12}}
\end{figure}

A final test makes use of the correlation function $\xi_{tr}(\theta)$ that should be zero in the absence of systematics. In Figure \ref{fig:c12} we show our gold and silver radio $\xi_{tr}(\theta)$ correlation functions, along with the optical data and optical-radio cross-correlation.

We see that, while in the optical and gold cases, the results are consistent with zero, in the silver case there is clear evidence of systematic contamination on scales $\le 2'$ and $\ge 7'$. This can be fully removed by using the optical-radio cross-correlation, as will be discussed further in \S\ref{cross-wl}; in this cross-correlation case, the results are again consistent with zero as shown on the plot.

In summary, then, we see that there are substantial residual systematics in our silver set which are not evident in the gold set; these may be due to anisotropic noise, and pose a challenge which radio weak lensing studies need to solve. However, systematics of this size do not defeat the analysis of this paper, as we shall now see.

\subsection{Cosmic Shear Constraints}
\label{cosmic}

\begin{figure}
\psfig{figure=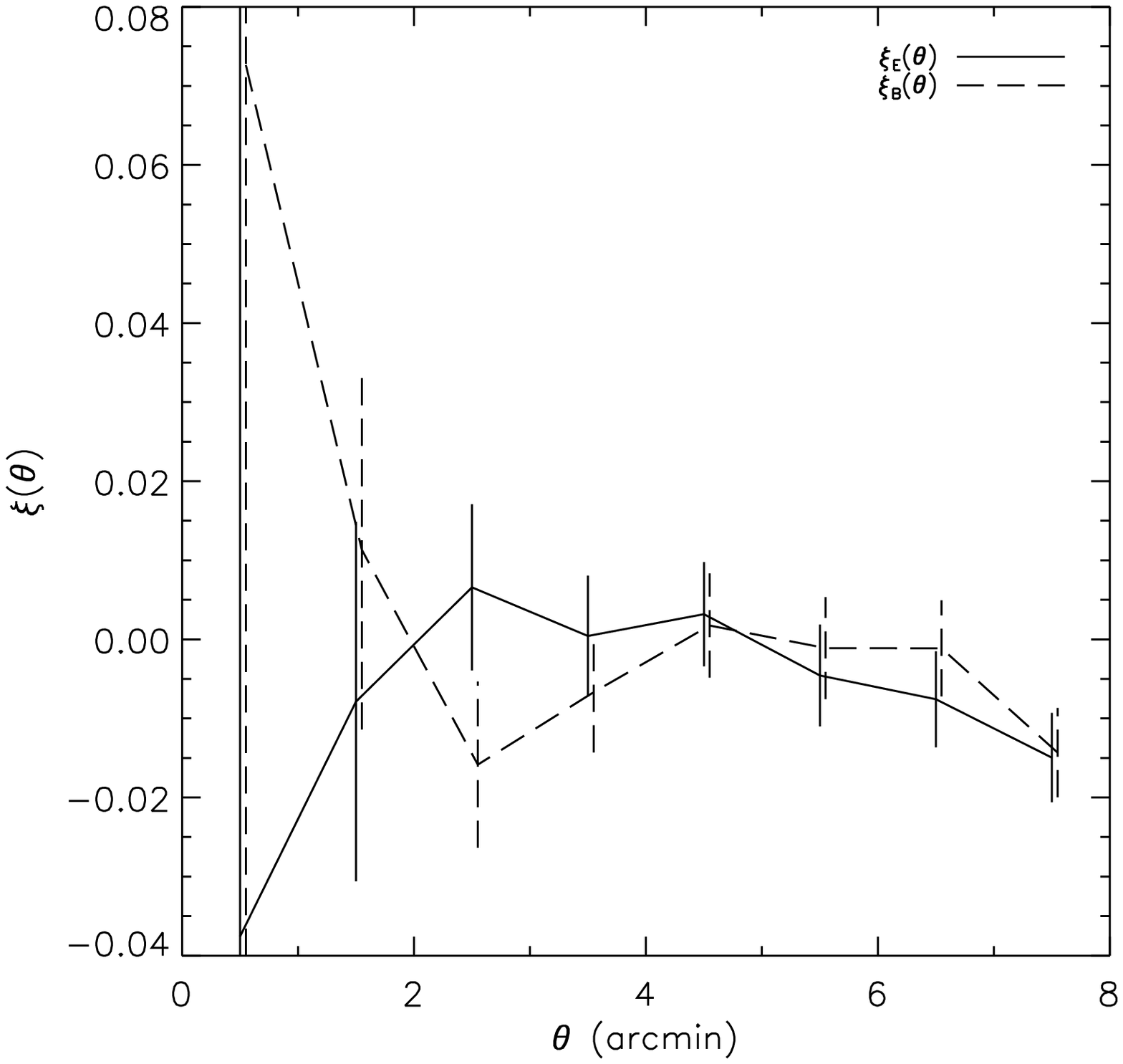,width=80mm}\\\vspace{-.5cm}
\psfig{figure=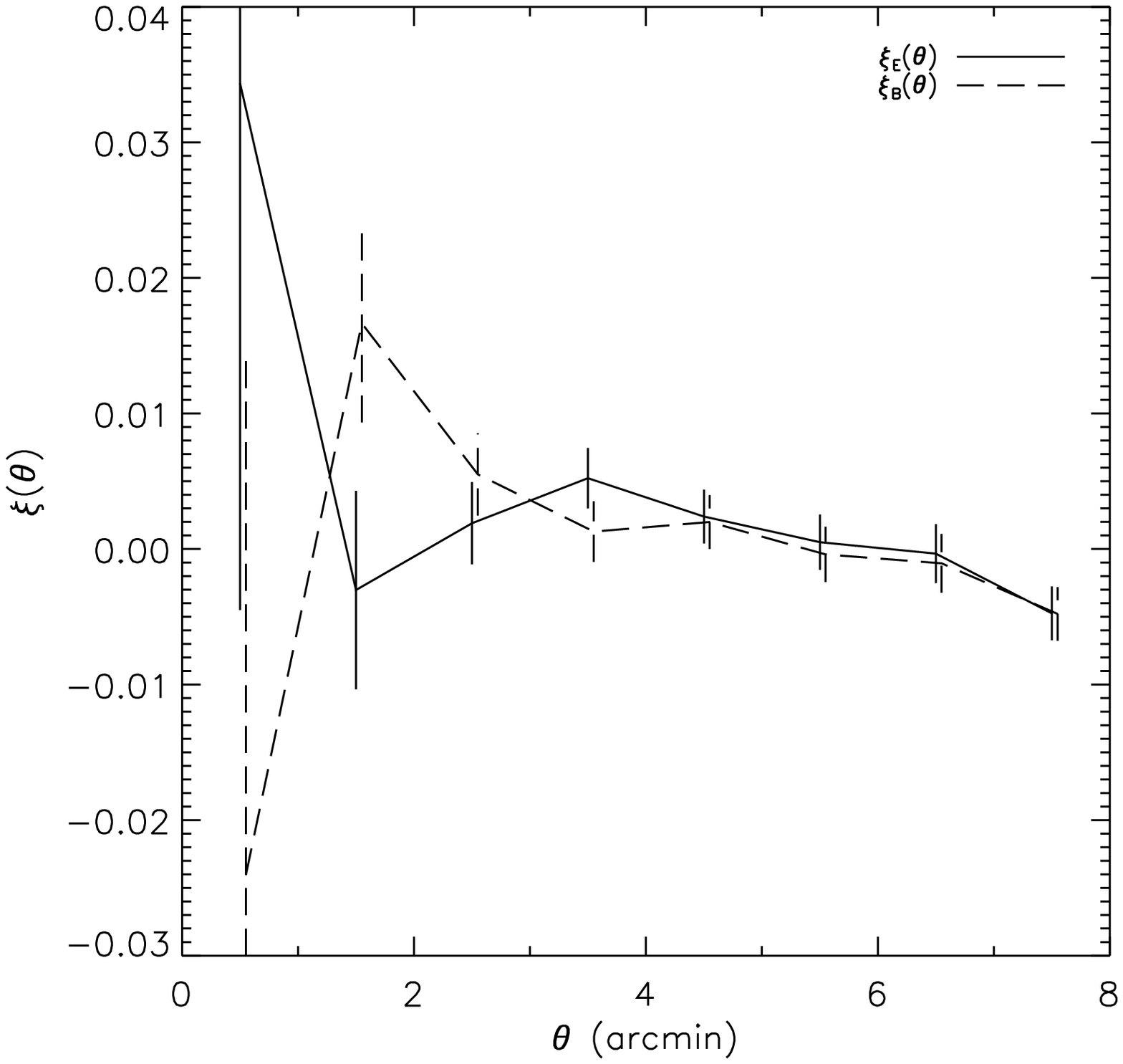,width=80mm}\\\vspace{-.5cm}
\psfig{figure=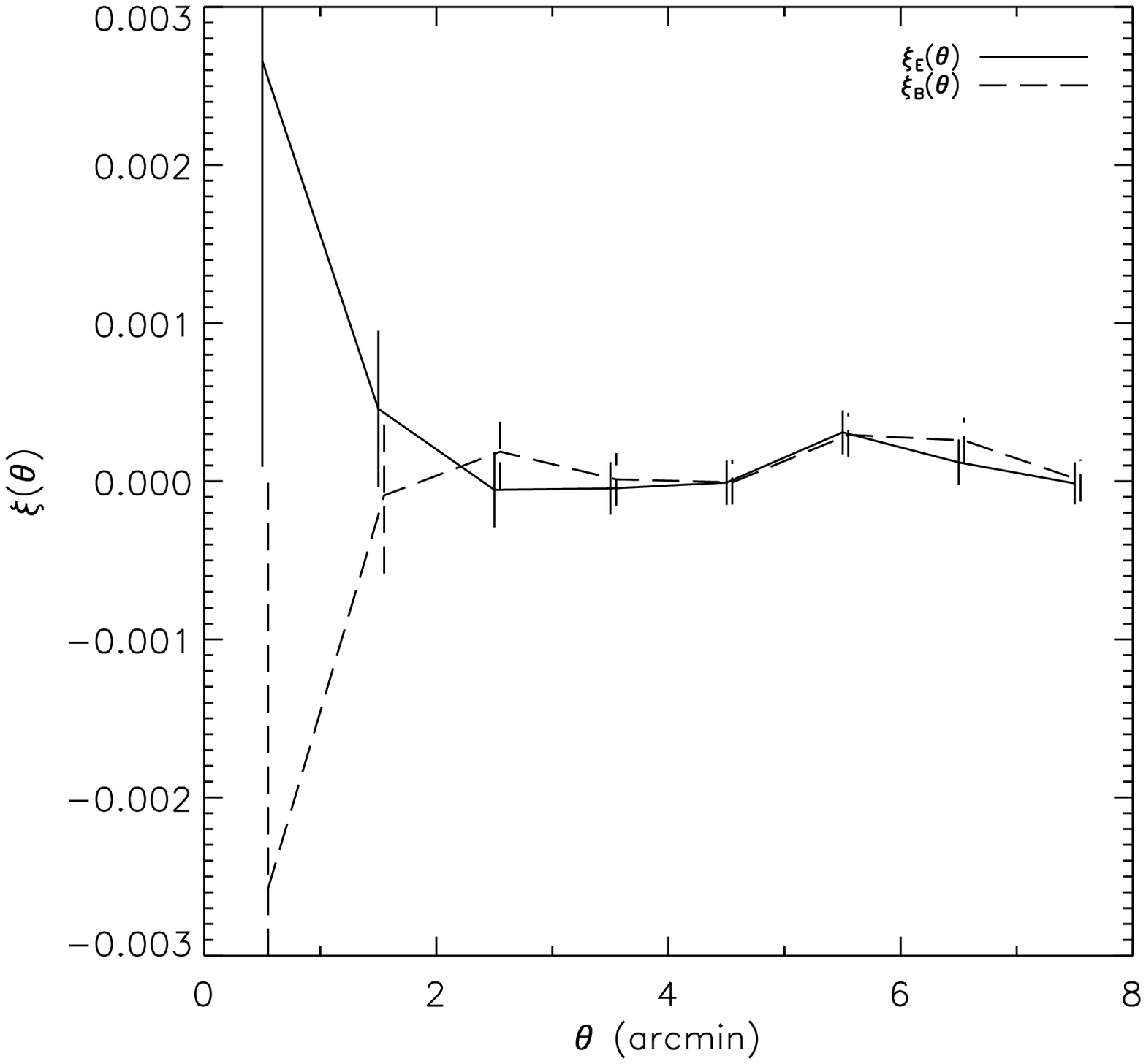,width=80mm}
\caption{\E- and \B- mode shear correlation functions for the gold (top) and silver (midde) radio objects and the optical objects (bottom). In all cases we find systematic contamination on small scales.  \label{fig:alleandb}}
\end{figure}
\begin{figure}
\psfig{figure=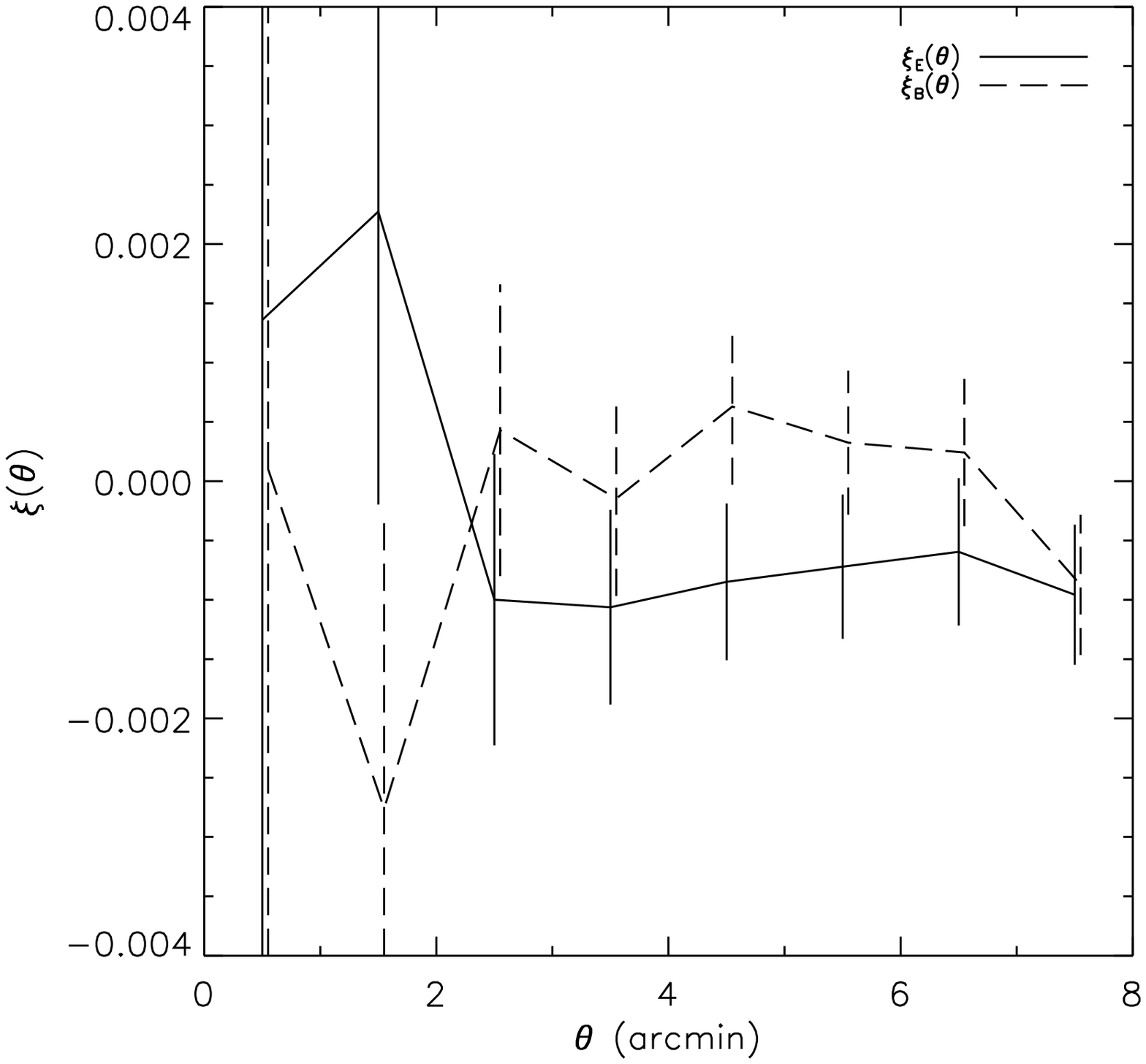,width=80mm}\\\psfig{figure=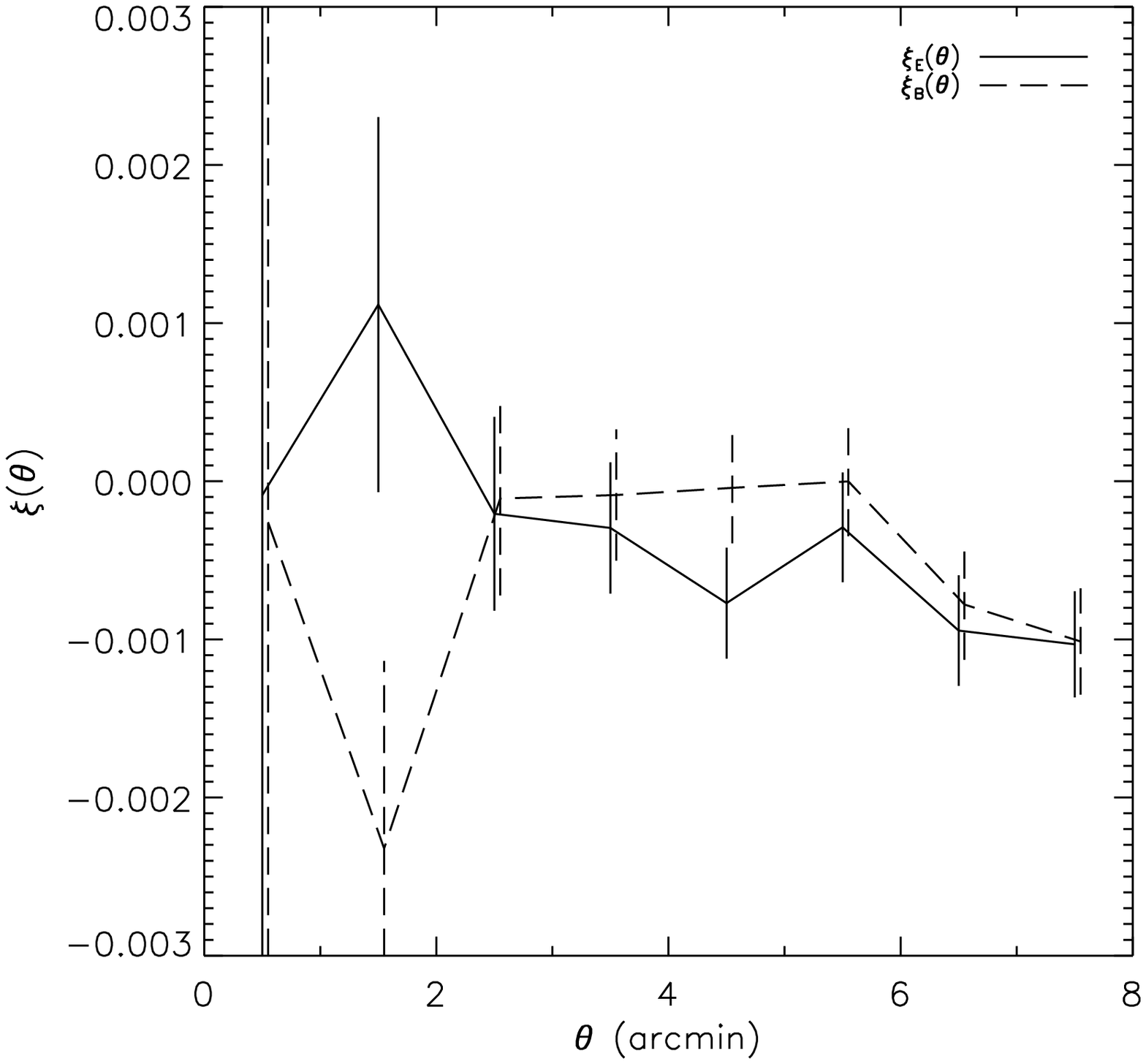,width=80mm}
\caption{Cross correlated \E- and \B- modes, with gold-optical on the top panel and silver-optical on the bottom panel. \label{fig:alleandbcross}}
\end{figure}

We are now in a position to use our shear estimators to constrain cosmological information, using shear correlation functions as described in \S\ref{basics}; recall that the correlation functions are simply related to the shear power spectrum and hence the matter power spectrum; equally, they will respond to any systematics present.

We calculate the $\xi_+$ and $\xi_-$ correlation functions for our gold, silver and optical sets, and their cross-correlations, using equation (\ref{eq:rotcfn}), then convert to \E-mode and \B-mode correlation functions using equations (\ref{eq:cecb}) and (\ref{eq:cprime}). Our results are shown in Figures \ref{fig:alleandb} and \ref{fig:alleandbcross}. 

In each case we see \B-mode contamination on scales $\theta \la 2''$; however, on larger scales there is no evidence of systematics. The gold and silver radio correlation functions have large error bars due to the low number density; in these cases we have not attempted to fit cosmological predictions. The optical-radio cross-correlation functions have significantly smaller error bars and it is these that we use to constrain cosmology; however, even here we are only obtaining an upper bound on the cosmic shear signal, and hence on cosmological parameters.

\begin{figure}
\centering
\psfig{figure=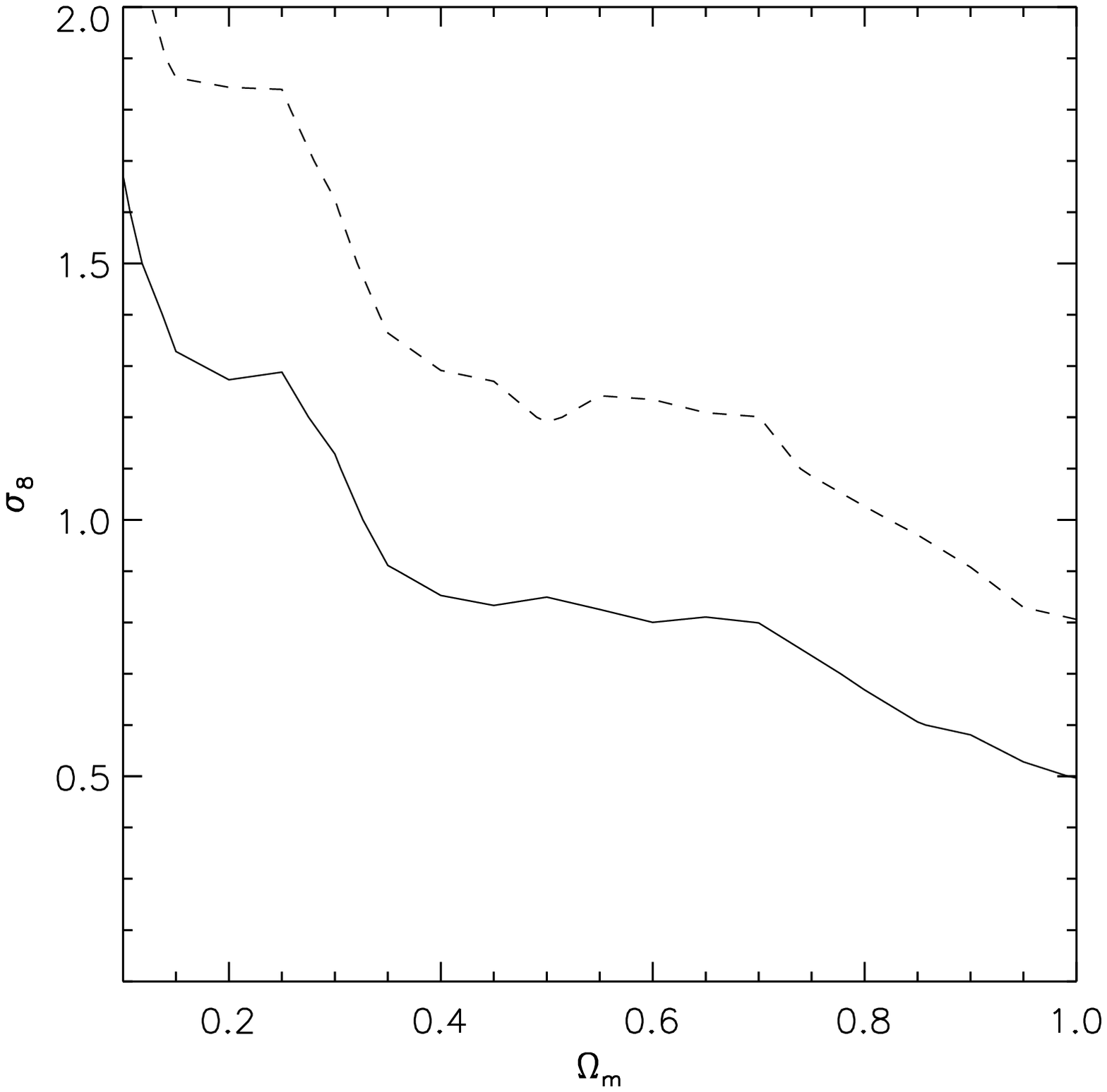,width=80mm}\\\psfig{figure=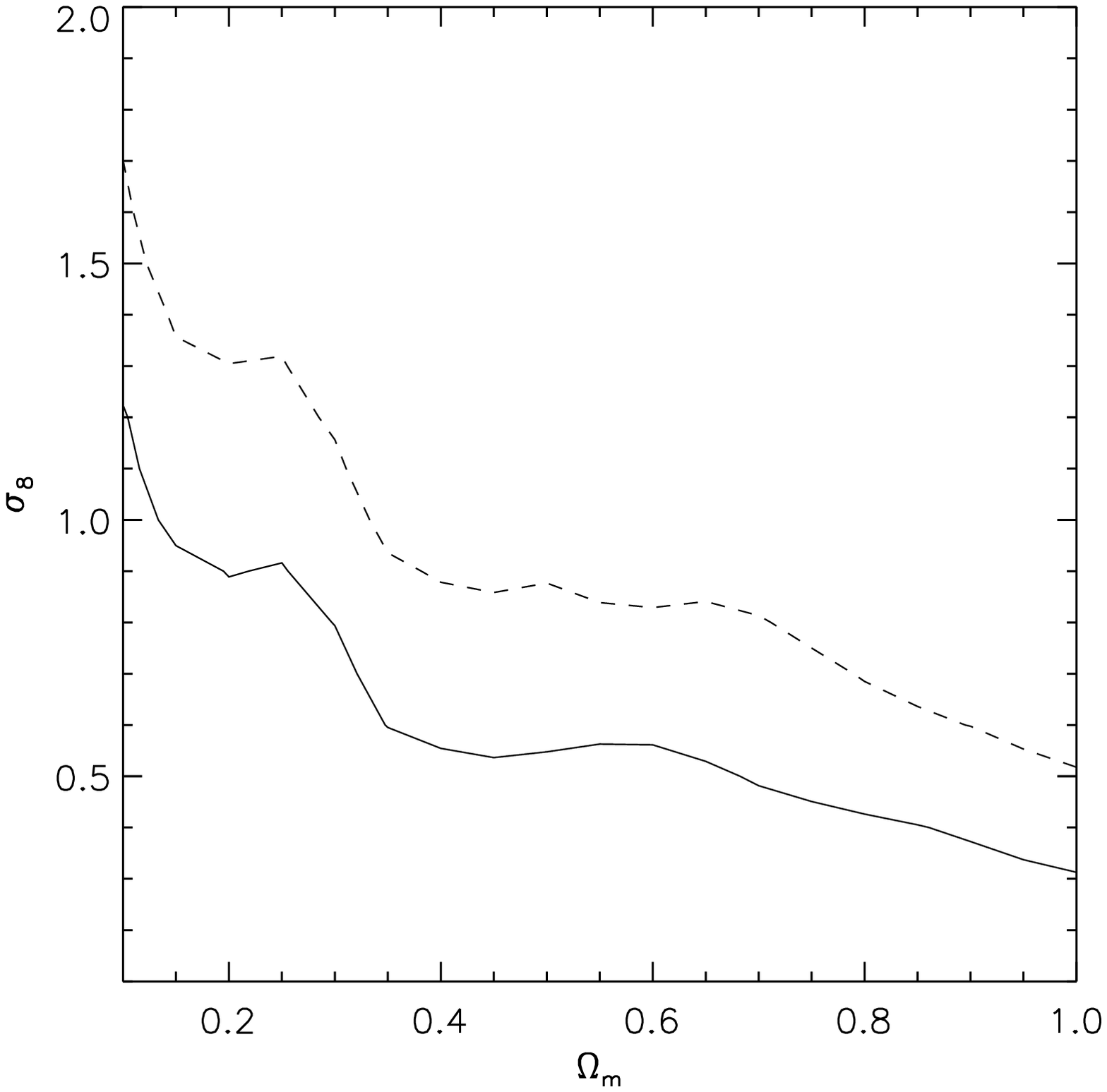,width=80mm}
\caption{Cosmological constraints derived from the gold-optical (top panel) and silver-optical (bottom panel) cross correlation functions at $z_m=1.0.$ The solid and dashed lines are the 1$\sigma$ and 2$\sigma$ contours respectively. \label{fig:cosmo}}
\end{figure}

We calculate correlation function predictions for a flat $\Lambda$CDM model with $H_0=72$ km s$^{-1}$Mpc$^{-1}$ with two varying parameters, $\Omega_{m}$ and $\sigma_{8}$, using equations (\ref{eq:rotcfn}) to (\ref{eq:cecb}). These are $\chi^2$ fit to the measured correlation functions, with resulting constraints shown in Figure \ref{fig:cosmo}. We find upper limits on the cosmological parameters, 

\begin{equation}
\sigma_{8}\left(\frac{\Omega_{m}}{0.25}\right)^{0.5}z_{m}^{1.6}<1.1 \,\,\,\, (1\sigma)
\end{equation}
for the radio-gold optical cross-correlation, while for the radio-silver optical cross-correlation we find

\begin{equation}
\sigma_{8}\left(\frac{\Omega_{m}}{0.25}\right)^{0.5}z_{m}^{1.6}<0.8 \,\,\,\, (1\sigma)
\end{equation}
These results do not include the cosmic variance due to the fact that the HDF-N is a small field, but show that radio-optical weak lensing cross-correlations are already capable of constraining cosmology at an interesting level.

In these constraints we have included the median redshift as one of the parameters, as we do not have a complete redshift sample for our radio catalogue. However, available data suggest that the median redshift is near $z_m\simeq 1$: using 2600 spectroscopic redshifts from \citet{2008ApJ...689..687B}  in the GOODS North with the same median magnitude as our optical sample, we find a median redshift for the optical sample to be $z_{m}=0.84$. This is in close agreement with the median redshift $z_m=0.85$ found from NED\footnote{http://nedwww.ipac.caltech.edu/} for 84 of our matched radio-optical objects; this can be compared with the predicted median redshift from the \citet{2008MNRAS.388.1335W} radio simulations at the relevant flux threshold, which is found to be $z_{m}=1.1$ for $S_{1.4}>50\mu$Jy.

\section{Optical-Radio Cross-Correlations}
\label{cross}

We have seen above that, given the residual systematics in our radio shear data, it is useful to cross-correlate shape information between radio and optical data. The question then naturally arises: are radio and optical shear estimators well correlated for individual galaxies?

The shear estimator of a single object is dominated by its intrinsic ellipticity, so radio and optical shear estimators will be strongly correlated if the radio and optical emission is aligned and has a similar ellipticity. Either a strong or weak correlation would be of interest; a strong correlation would afford a useful check for shear measurement methods, while a weak correlation provides a degree of independence of radio and optical lensing measurements, as we will discuss.

\subsection{Observed Correlation Strength}
\label{crossobs}

\begin{figure}
\centering
\psfig{figure=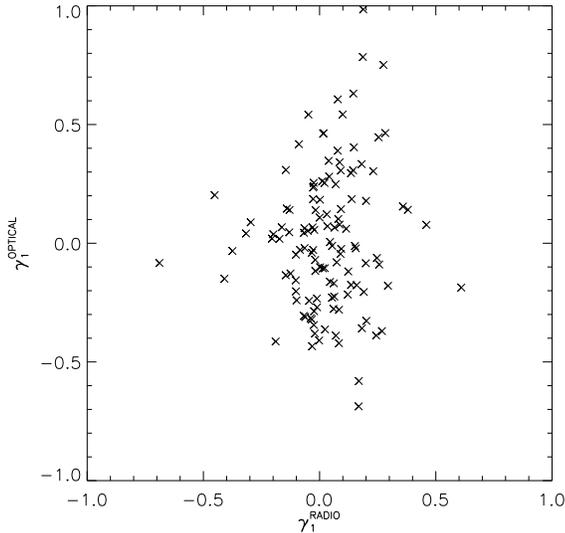,width=80mm}
\caption{Comparison of radio and shear estimators for matched objects.}
\label{fig:grgo}
\end{figure}

We matched our radio and optical catalogues, finding 123 objects with emission in both wavelengths. In our sample there were 4 instances in which multiple radio sources were associated with a single optical one. These have been removed to avoid any ambiguity about which to compare with the optical estimator. We show the shear estimators for these galaxies in both parts of the spectrum in Figure \ref{fig:grgo}. We find that they are in fact not strongly correlated, with a Pearson's correlation coefficient of $0.097\pm0.090$.

\begin{figure}
\centering
\psfig{figure=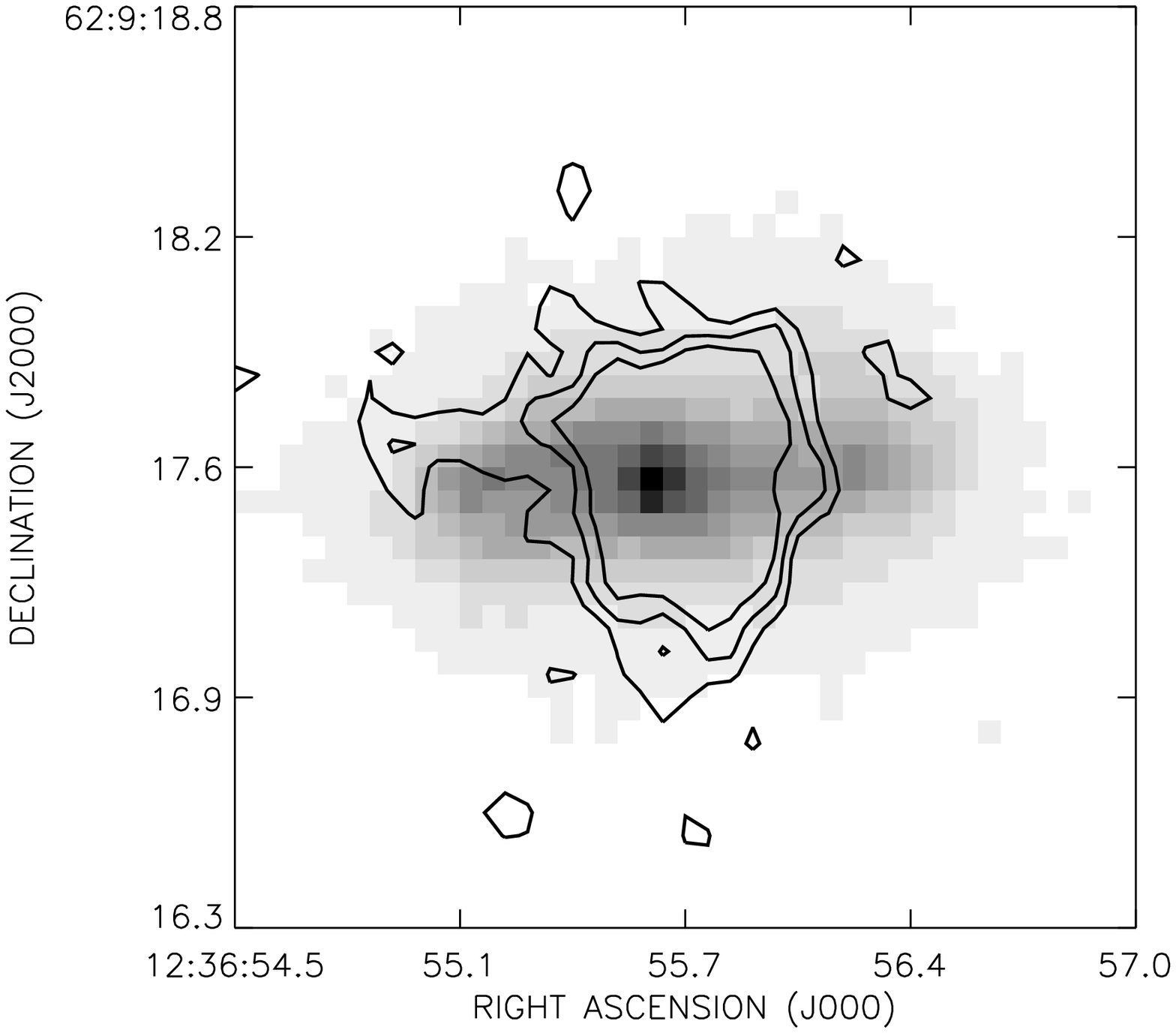,width=40mm}\hspace{3mm}
\psfig{figure=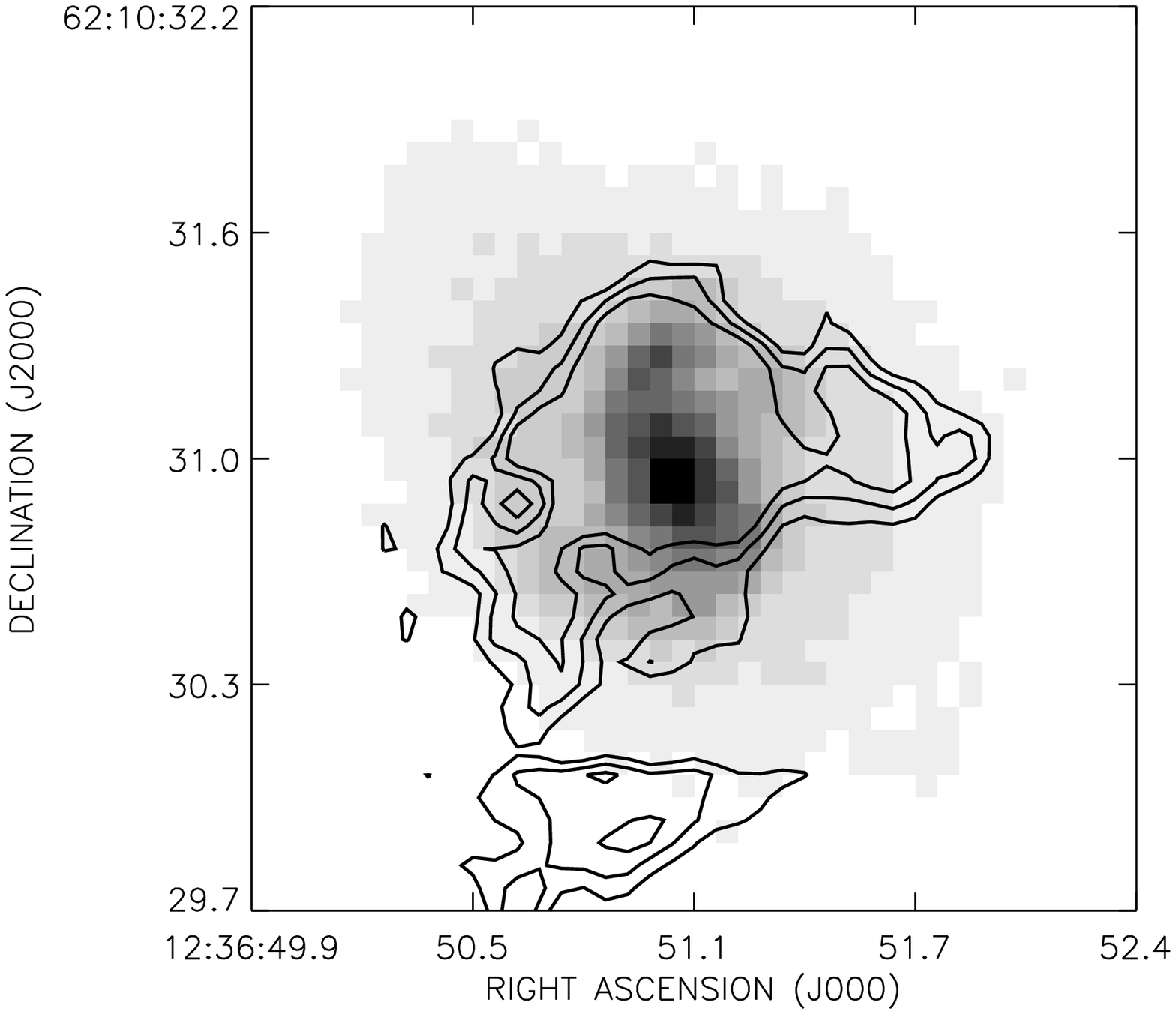,width=40mm}\hspace{3mm}
\psfig{figure=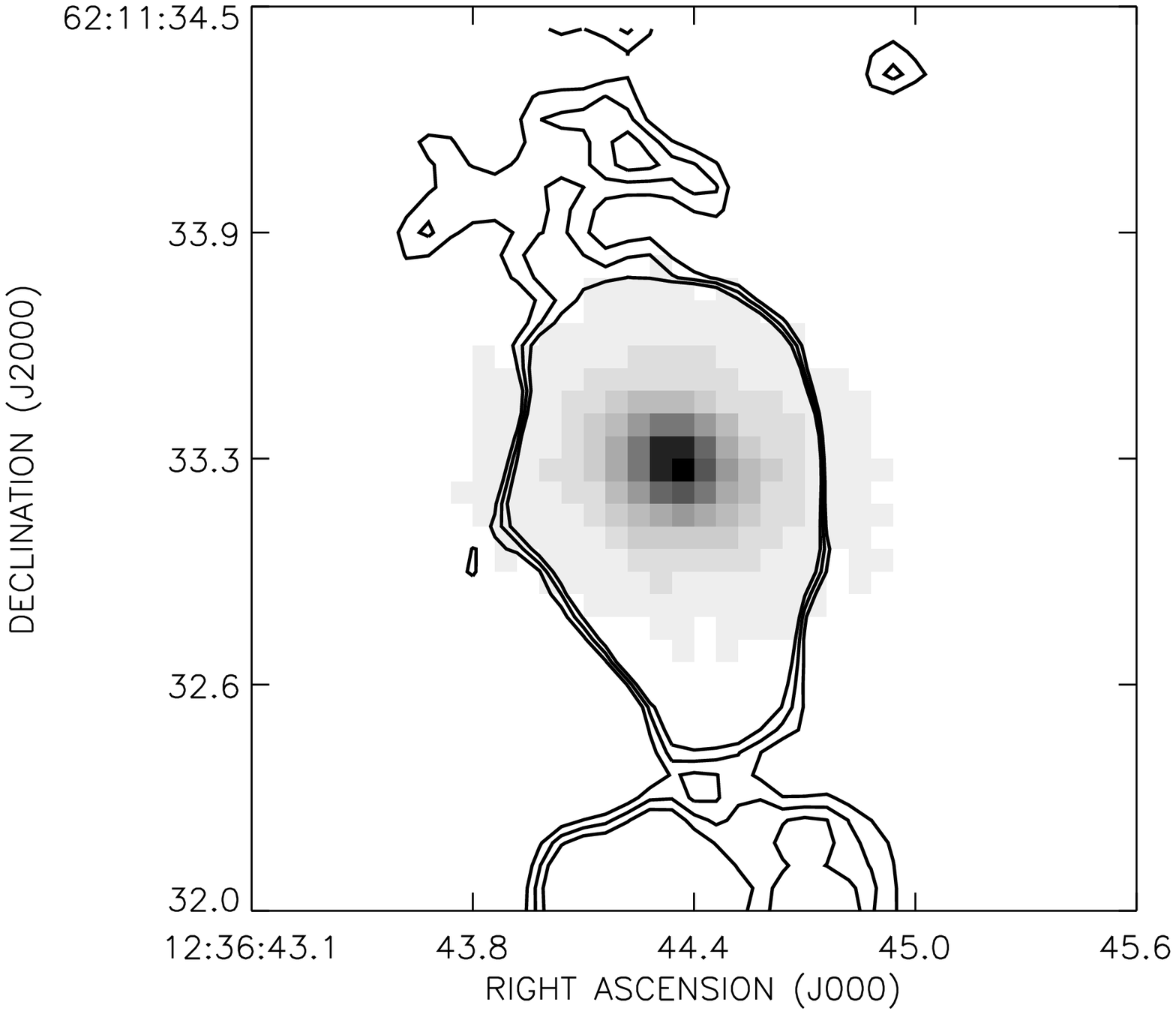,width=40mm}\hspace{3mm}
\psfig{figure=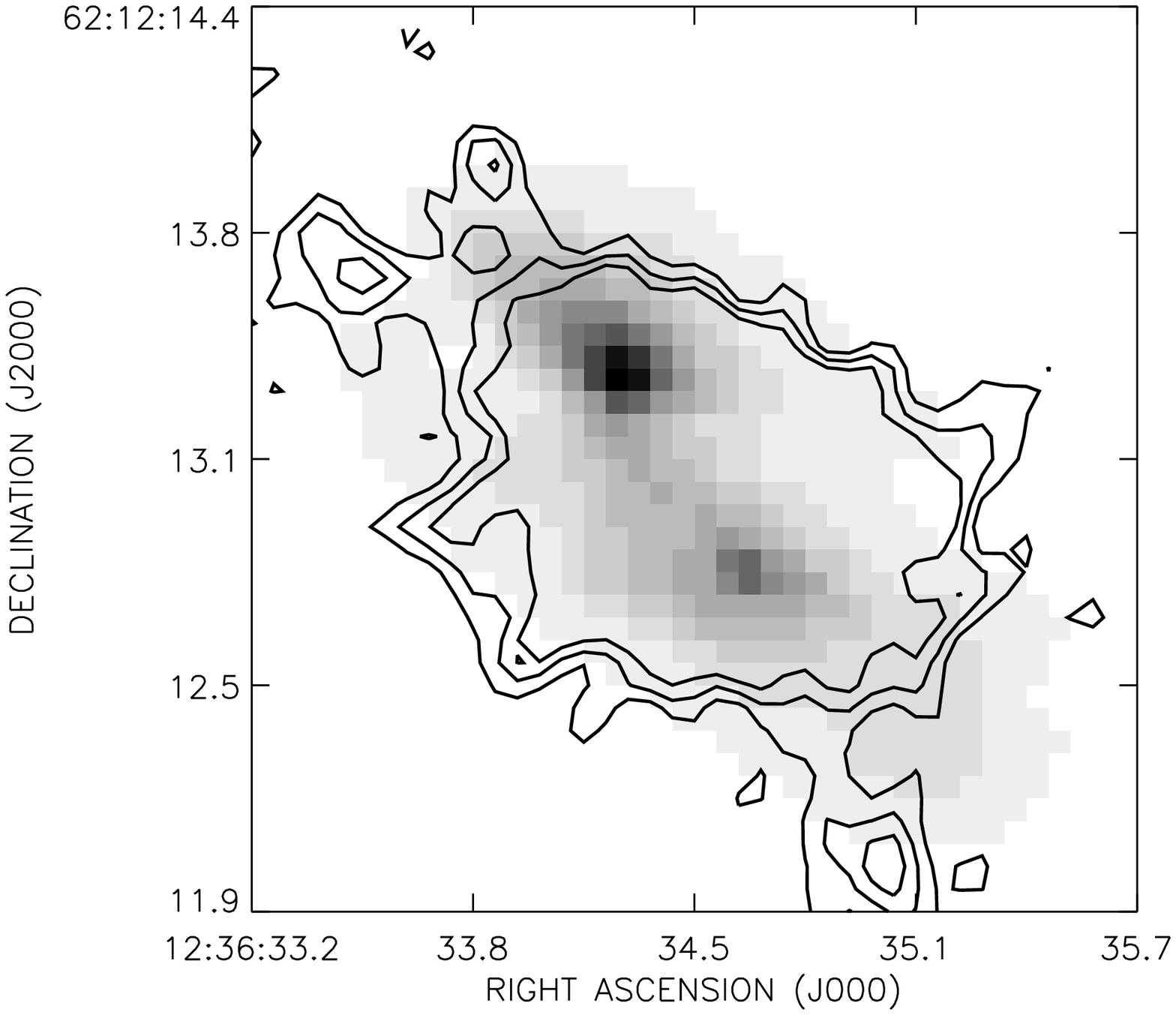,width=40mm}\hspace{3mm}
\psfig{figure=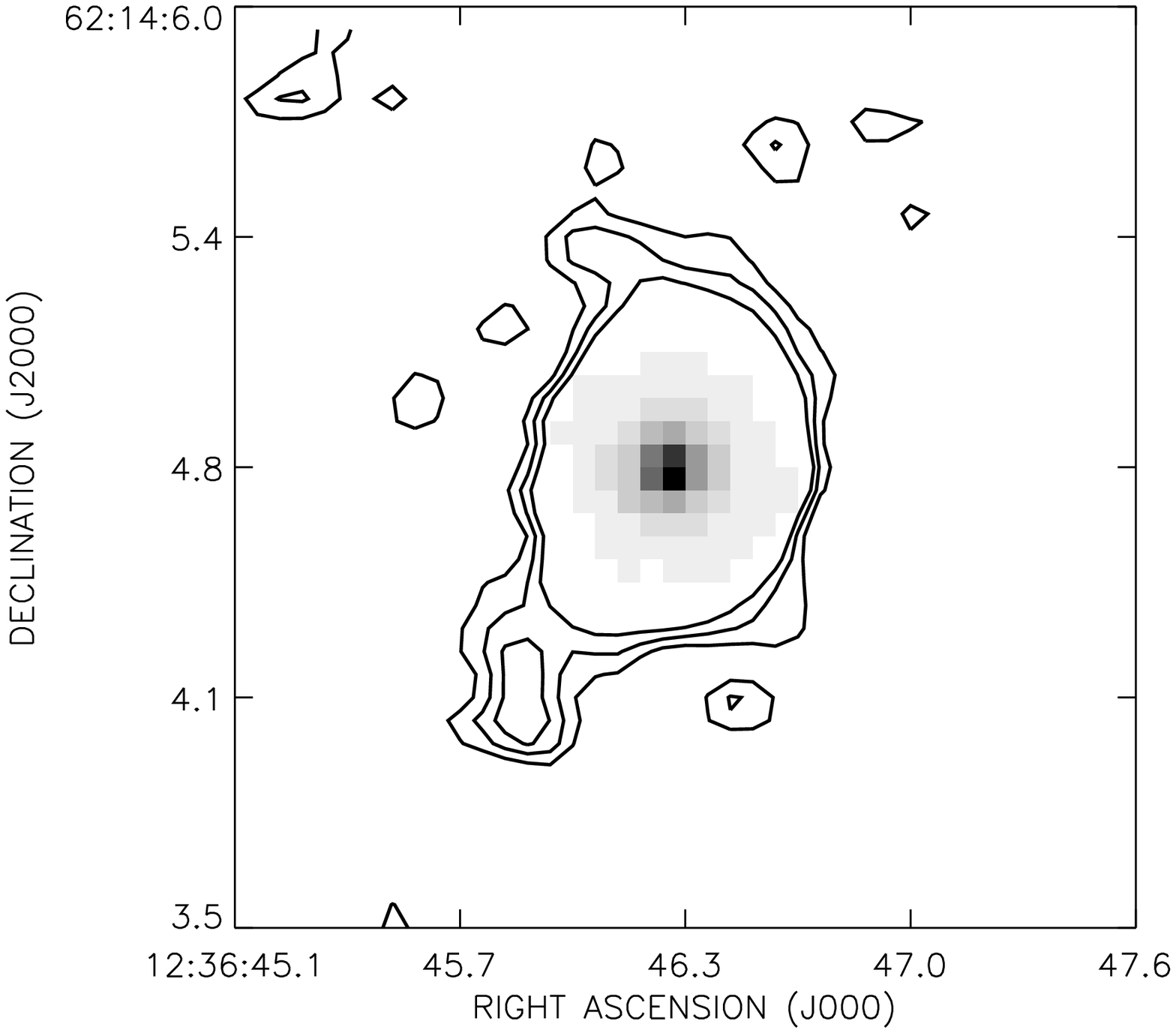,width=40mm}\hspace{3mm}
\psfig{figure=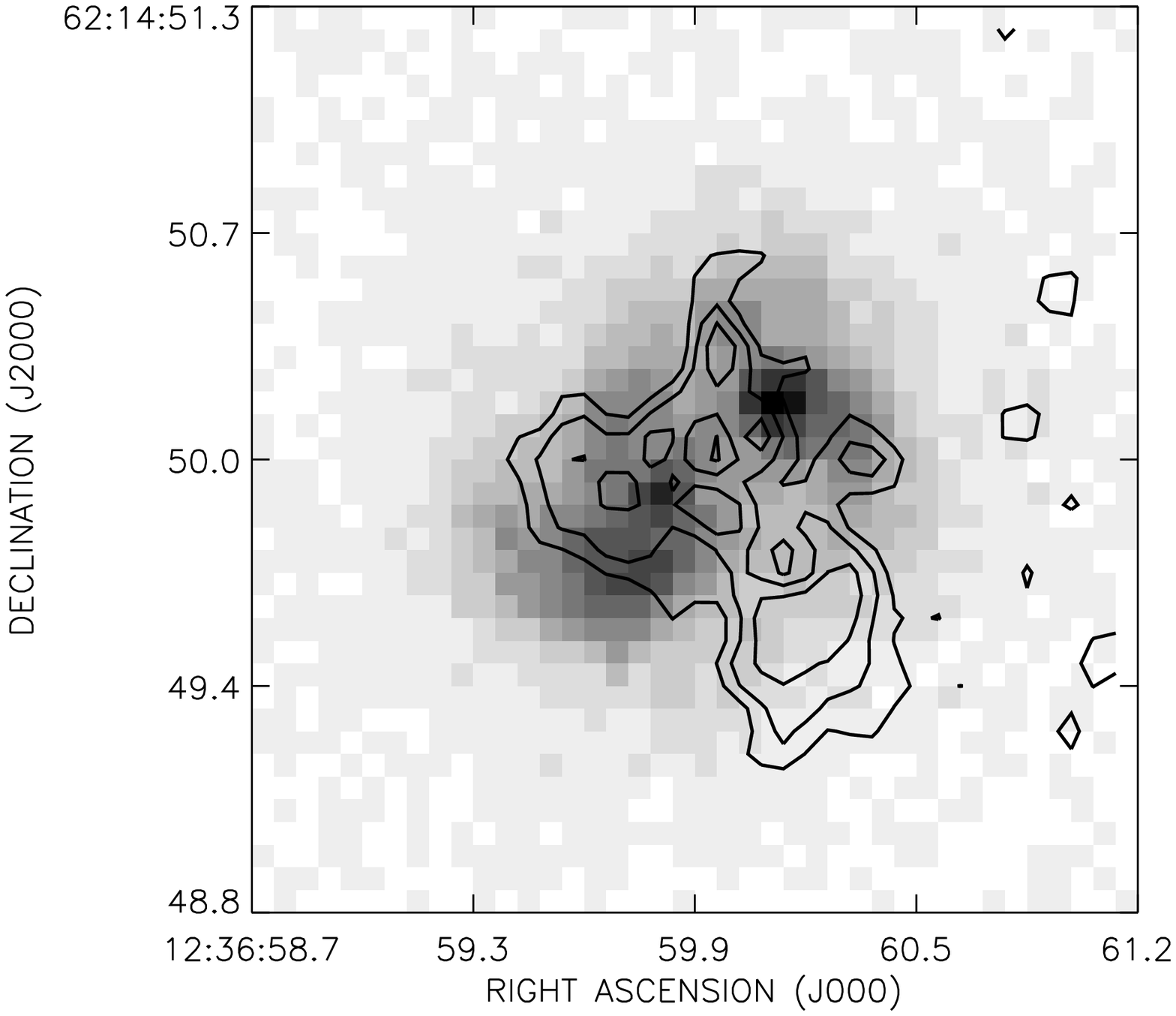,width=40mm}\hspace{3mm}
\psfig{figure=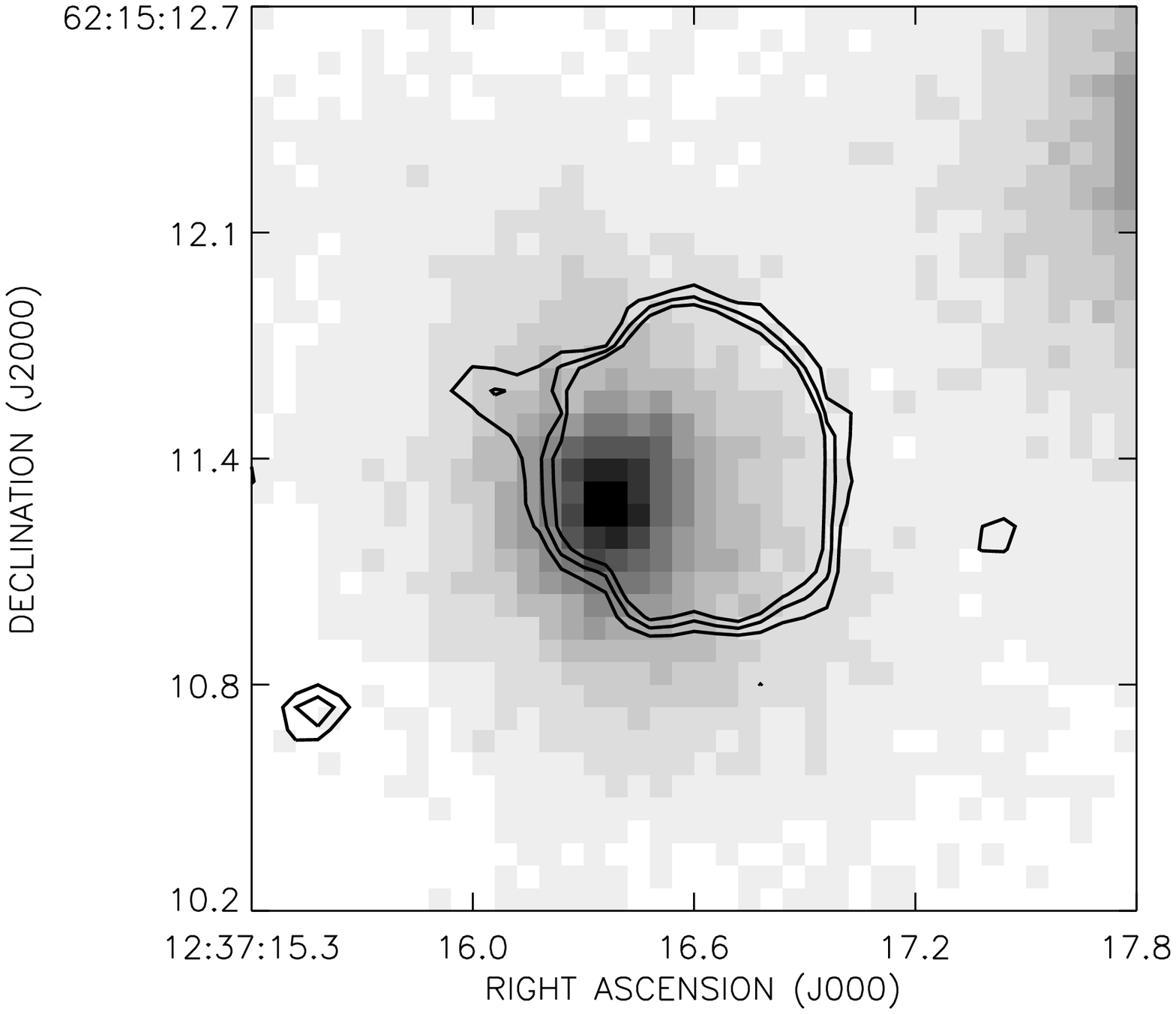,width=40mm}\hspace{3mm}
\psfig{figure=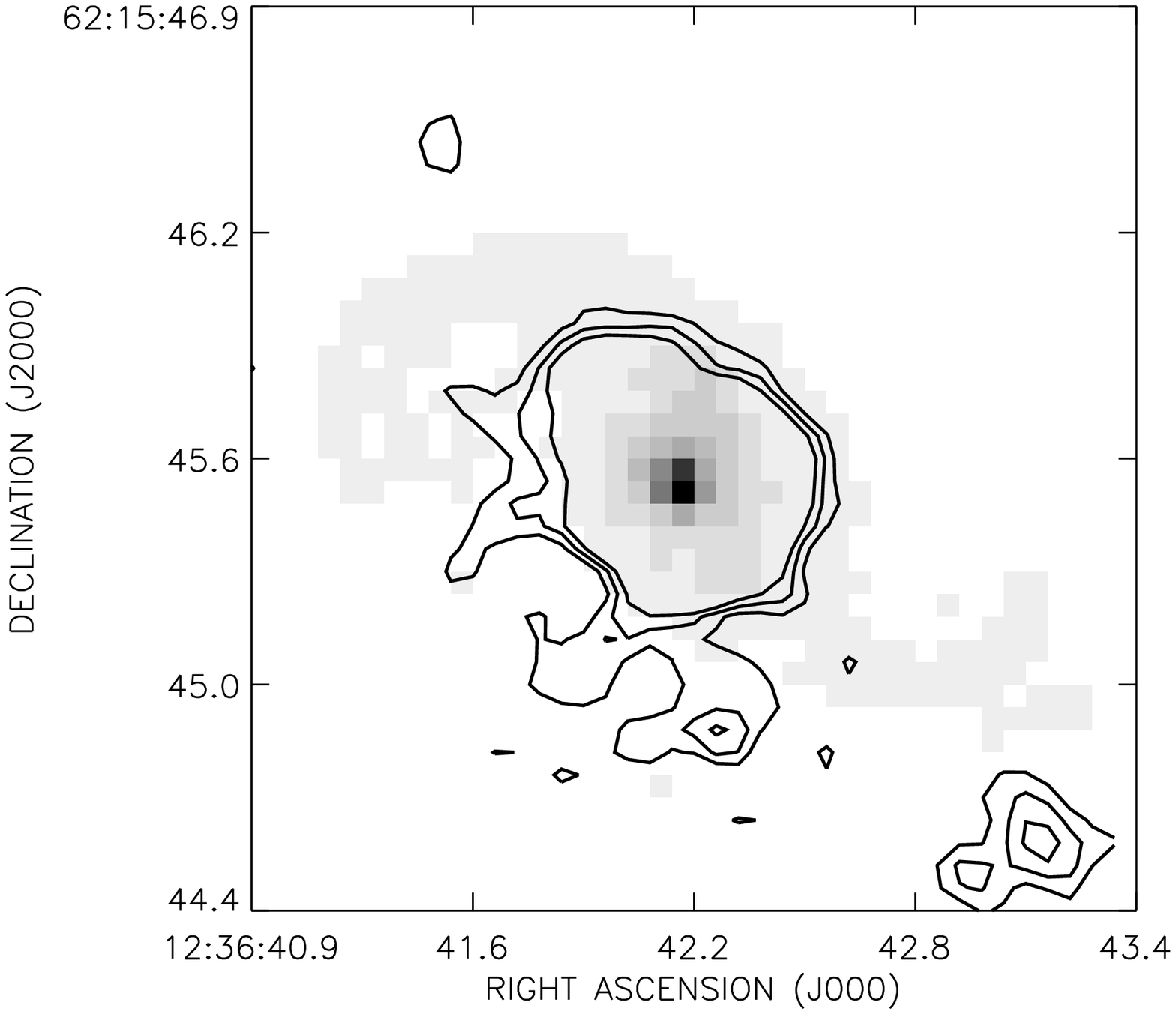,width=40mm}\\
\caption{All matched objects with optical $m_z<21.5$ and radio flux $>140\mu$Jy. The $z$-band optical
  images are overlaid with the 2, 3 and 4$\sigma$ radio
  contours. \label{fig:overlays}}
\end{figure}

This result is initially rather surprising, but is confirmed by
visual inspection of the radio and optical shapes of the objects: as an example, in Figure \ref{fig:overlays} we show the brightest objects in our radio sample that have optical counterparts. We have overlaid the 2, 3 and 4$\sigma$ contours (where $\sigma=3.3\mu$Jy beam$^{-1}$ is the rms noise of the radio image) on the $z$-band optical image. Note that there is little evidence of tracing or alignment between radio and optical emission. This is also confirmed by a statistical analysis in  \S\ref{blind}. We will now discuss the implications of this result, both for weak lensing, and in comparison with other work in the field.


\subsection{Weak Lensing Implications}
\label{cross-wl}
The lack of correlation in radio and optical shapes is of real use for weak lensing studies. The basis of many studies in the field is the measurement of 2-point statistics after measuring shear estimators; in the weak lensing regime, we can write the shear estimator as
\begin{equation}
\tilde{\gamma}=\gamma+\gamma^{i}+\gamma^{s},
\end{equation}
where $\gamma$ is the true gravitational shear signal, $\gamma^{i}$ is proportional to the intrinsic ellipticity of the galaxy, and $\gamma^{s}$ represents any shear contributions induced by systematic effects. If we then write out a correlation function, we obtain the terms
\begin{eqnarray}
\label{eq:correlator}
\langle\tilde{\gamma}\tilde{\gamma}\rangle&=&\langle\gamma\gamma\rangle+\langle\gamma^{i}\gamma^{i}\rangle+2\langle\gamma^{i}\gamma\rangle+\langle\gamma^{s}\gamma^{s}\rangle.
\end{eqnarray}
where we have neglected terms $\langle \gamma \gamma^s \rangle$ and $\langle \gamma^i \gamma^s \rangle$ as evidently zero.

The first term on the right hand side of equation (\ref{eq:correlator}) is the true cosmic shear signal that we desire to measure. The second term arises when physically close galaxies are intrinsically aligned; this is the `II' term \citet{2002MNRAS.332..788M}. The third term can arise when a foreground gravitational potential tidally distorts a neighbouring galaxy, while also lensing a background galaxy; this is the `GI' term \citet{2004PhRvD..70f3526H}. The final term involves only systematic shear. In general, this can be large and requires a good understanding of the experiment and sophisticated techniques to remove any spurious signal.

The disentangling of the pure $\langle\gamma\gamma\rangle$ term from the above equation is therefore very challenging. However, let us now suppose we have a joint radio and optical survey, containing a set of galaxies each of which has a radio or optical shear estimator ($\tilde\gamma_r$ and $\tilde\gamma_o$ respectively), or frequently both. In this case, a cross correlation function can be measured:  

\begin{equation}
\label{eq:crossterm}
\langle\tilde{\gamma}_{o}\tilde{\gamma}_{r}\rangle=\langle\gamma\gamma\rangle+\langle\gamma_{o}^{i}\gamma\rangle+\langle\gamma_{r}^{i}\gamma\rangle+\langle\gamma_{o}^{i}\gamma_{r}^{i}\rangle+\langle\gamma_{o}^{s}\gamma_{r}^{s}\rangle.
\end{equation}
This leads to several advantages. Firstly, it should be noted that if the covariance between shapes in optical and radio is negligible, the error upon the first term is substantially reduced due to an effective increase in galaxy number density, as shown by \citep{2008JCAP...01..003J}; we can effectively consider the optical and radio sources as independent galaxies. This is an advantage over cross-correlating two optical bands, where a strong shape correlation is found and little increase in effective number density results \citep{2008JCAP...01..003J}.

We see that besides the cosmic shear signal, we now have two GI terms, one from the optical and one from the radio data; these will still need to be measured, marginalised over or nulled \citep{2008A&A...488..829J}. However, the other terms are reduced in amplitude by the cross-correlation: in \S\ref{crossobs} we have shown that $\langle\gamma_{o}^{i}\gamma_{r}^{i}\rangle$ is small; there is little correlation between optical and radio shapes.
 
In addition, the last term will be small; the systematics associated with a given optical and radio telescope and ensuing data reduction are so distinct that they could hardly be correlated. This is related to the approach of \citep{2004astro.ph.12234J} who advocate cross-correlating shear measurements in different exposures; however, this might still leave chronic systematics uncorrected, which would be removed by a radio-optical correlation.

We conclude then that cross-correlating future large radio and optical datasets could be a powerful method for lensing studies, substantially reducing the issues associated with intrinsic alignments and systematics. Our results in \S\ref{cosmic} already give some evidence of this but better radio data will be required for the cross-correlation to be fully studied and utilised.

\subsection{Alignment of Optical and Radio Emission}
\label{intali}

\begin{figure} 
\centering
\psfig{figure=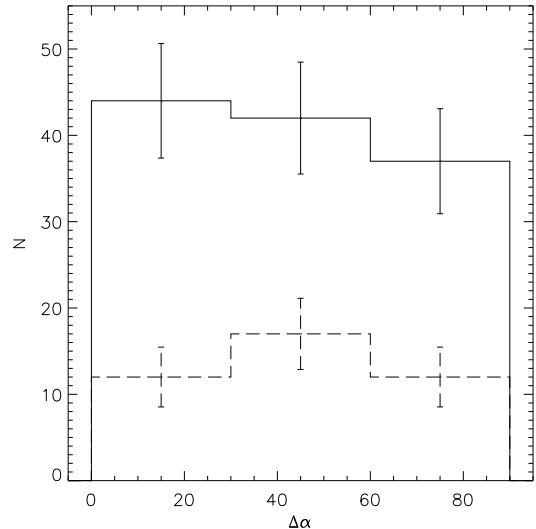,width=80mm}
\caption{Histograms of the matched radio and optical galaxies in our sample binned according to the quantity $\Delta\alpha$, defined as the acute angle between the optical and radio major axes. The solid histogram is the full matched radio-optical sample while the dashed histogram corresponds to the third of the matched objects with the highest signal-to-noise.  \label{fig:orient}}
\end{figure}

Recently \citet{2009arXiv0902.1631B} studied the orientation of radio and optical galaxies in the FIRST and SDSS surveys. They found a clear excess of galaxies in which the major axes of the radio and optical emission were aligned. This may appear to be in tension with what we have shown in Figure \ref{fig:grgo}, but the comparison is not direct; since this figure shows shear estimates, it mixes orientation and flattening information. In order to compare more directly with \citet{2009arXiv0902.1631B}, we calculate for all matched objects (solid histogram) the angle between radio and optical major axes; our results are shown in Figure \ref{fig:orient}. We also plot in Figure \ref{fig:orient} the corresponding histogram for the third of the matched objects with the highest signal-to-noise (dashed histogram).

We see that there is only a $1\sigma$ excess of the total objects with a small
angle between radio and optical major axes. We also see that the histogram for the high signal-to-noise
objects does not show any evidence of such an alignment. Thus there is some level of disparity between our work and
\citet{2009arXiv0902.1631B}. We note that our sample is much smaller
in number than that of \cite{2009arXiv0902.1631B}, and as such, we are
not able to usefully bin our sample more finely than shown in Figure
\ref{fig:orient}. 

Equally it is important to note that this study
involves highly resolved imaging using sub-arcsecond angular
resolution radio and optical data. In comparison, the study by
\citet{2009arXiv0902.1631B} uses much lower resolution data; they use
VLA FIRST radio data with an angular resolution of 5\,arcsec correlated
with SDSS which has typical seeing in the range 1 to 1.5\,arcsec. At
these resolutions both the radio and optical data typically trace the
extended smooth emission from the galaxy and hence its shape on
scales of a few tens of kpc at the typical redshifts of these
sources.  As such
the correlations seen by
\cite{2009arXiv0902.1631B} are not surprising.  The objects studied here in the HDF-N are
much fainter, and at
the 0\farcs4 angular resolution of this study we are typically resolving
structure with linear sizes of just a few kpc. 

Additionally the types
of galaxies involved in each of these two studies differ considerably, with the sources in the
HDF-N sample being predominantly moderate to high redshift systems
with intense young star-formation and large levels of optical
extinction, and the \cite{2009arXiv0902.1631B} sources being lower
redshift, more quiescent galaxies.
 
 Thus there are clear reasons why there may be differences in these
 two studies. However the precise cause of
 this disparity is worthy of further investigation, especially as it has
 strong implications for the design of future combined optical and
 radio weak lensing studies.


\section{Blind Shear Measurements}
\label{blind}

\begin{figure}
\centering
\psfig{figure=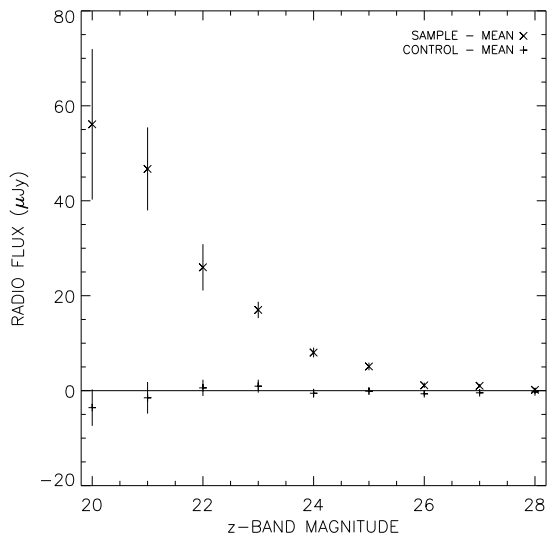,width=80mm}
\caption{The mean 1.4GHz flux density detected within 0.75 arcseconds of the positions of optical galaxies in the field, binned in z-band magnitude. Radio emission is statistically detected from optical galaxies brighter than $m_{z}=25$. Lower data points: results for apertures placed at random on the image.\label{fig:statdet}}
\end{figure}

In this section we describe a supplementary approach to radio shear
measurements. We are motivated by \citet{2007ASPC..380..199M} who find
92 radio sources at a detection threshold of 40$\mu$Jy within a
$10^{\prime}\times 10^{\prime}$ region centered on the HDF-N. This
should be compared with the size of the HST ACS $z$-band catalogue,
which contains $\simeq13000$ galaxies brighter than $m_{z}=28.3$ in the
same region. Although a vast majority of these sources are not detected individually at radio wavelengths, it is possible to statistically detect these very weak radio sources. Figure \ref{fig:statdet} shows the mean radio flux measured within a 0.75 arcsecond radius of the positions of $\simeq8000$ of these sources, as a function of their optical magnitude. Radio emission at the level of a few $\mu$Jy is statistically detected from optical systems as faint as $m_{z}=25$; this is in good agreement with the analysis conducted by \citet{2007ASPC..380..199M}. The lower data points in Figure \ref{fig:statdet} show mean flux when aperture positions are chosen at random.

\begin{figure}
\centering
\psfig{figure=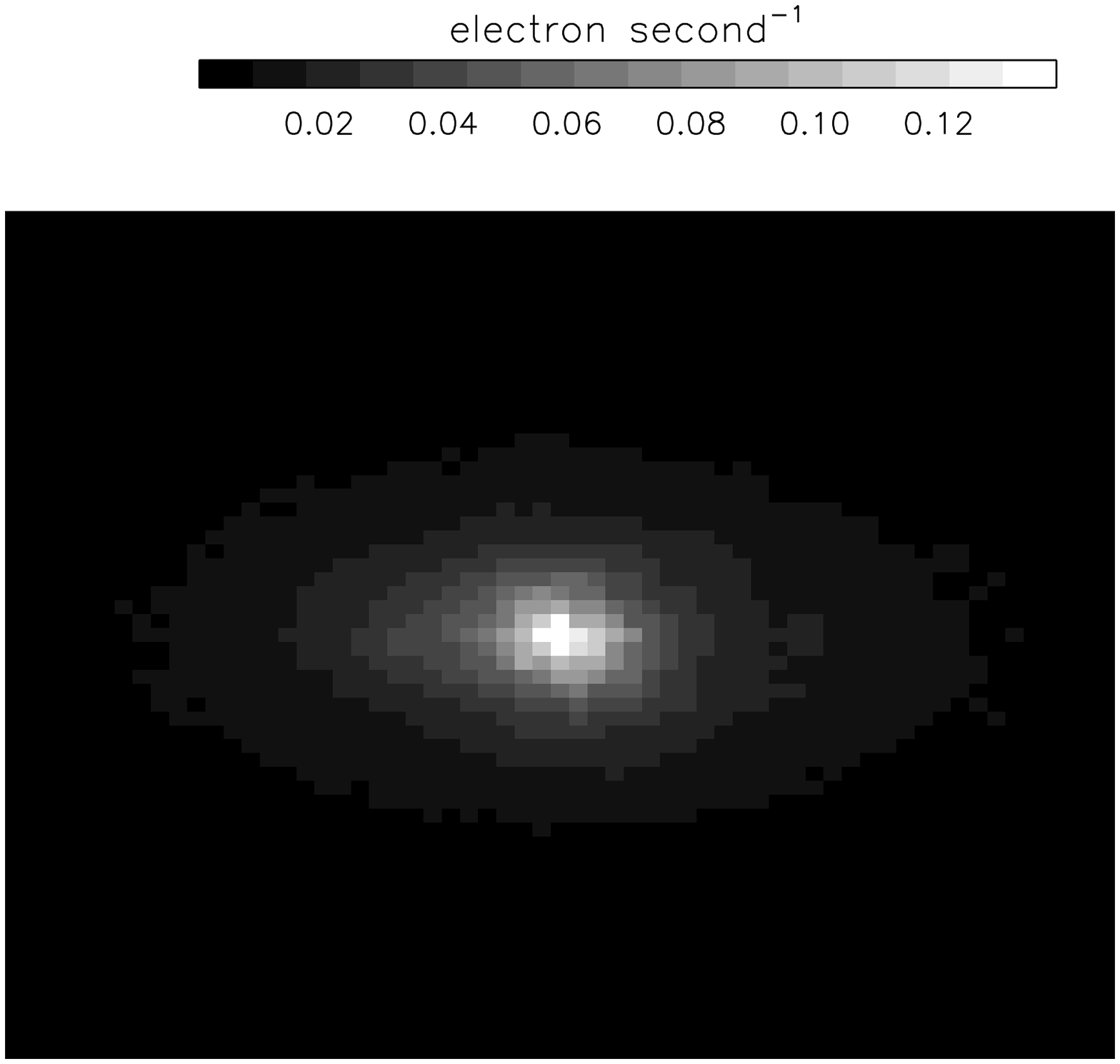,width=35mm,height=45mm}\psfig{figure=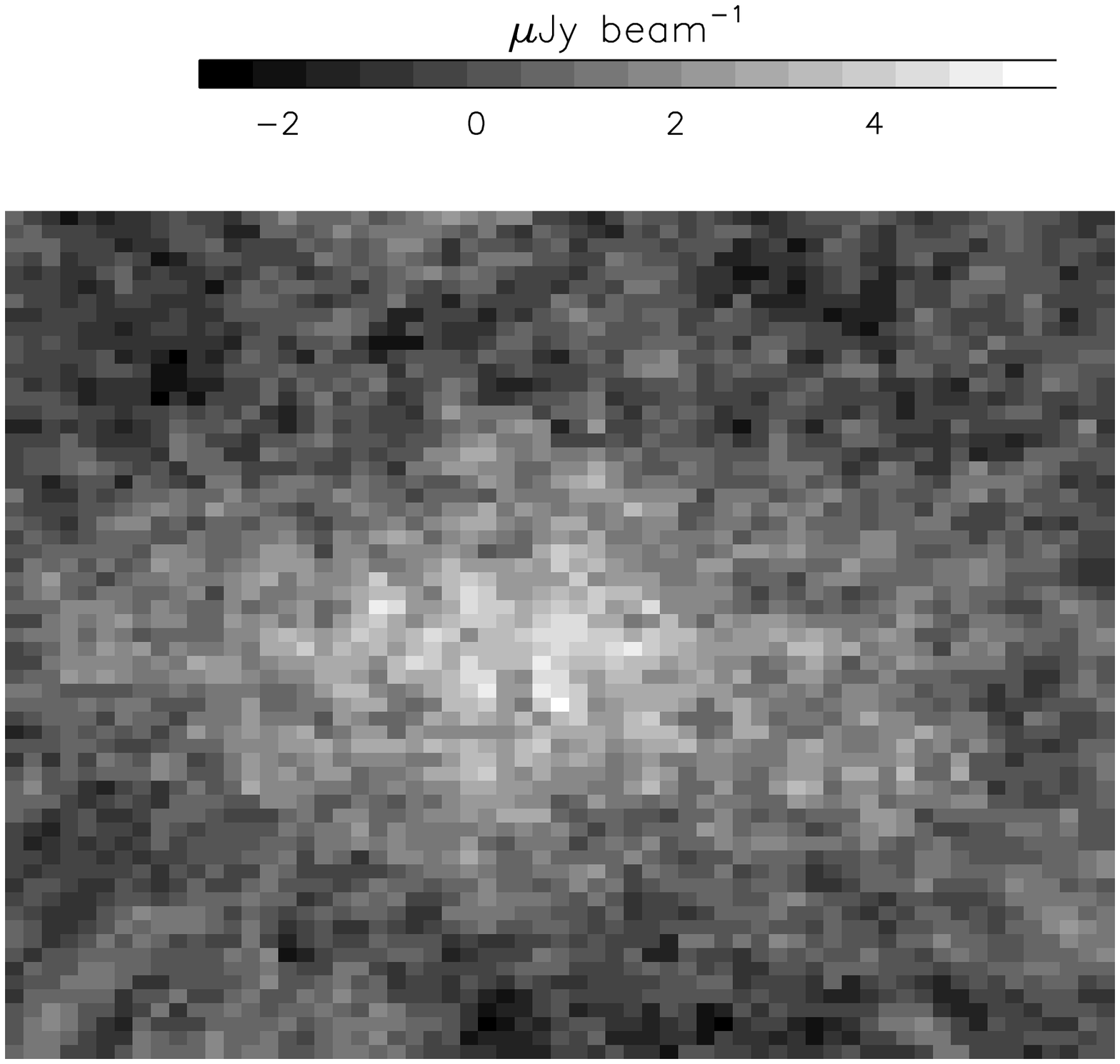,width=35mm,height=45mm}\\\vspace{10mm}
\psfig{figure=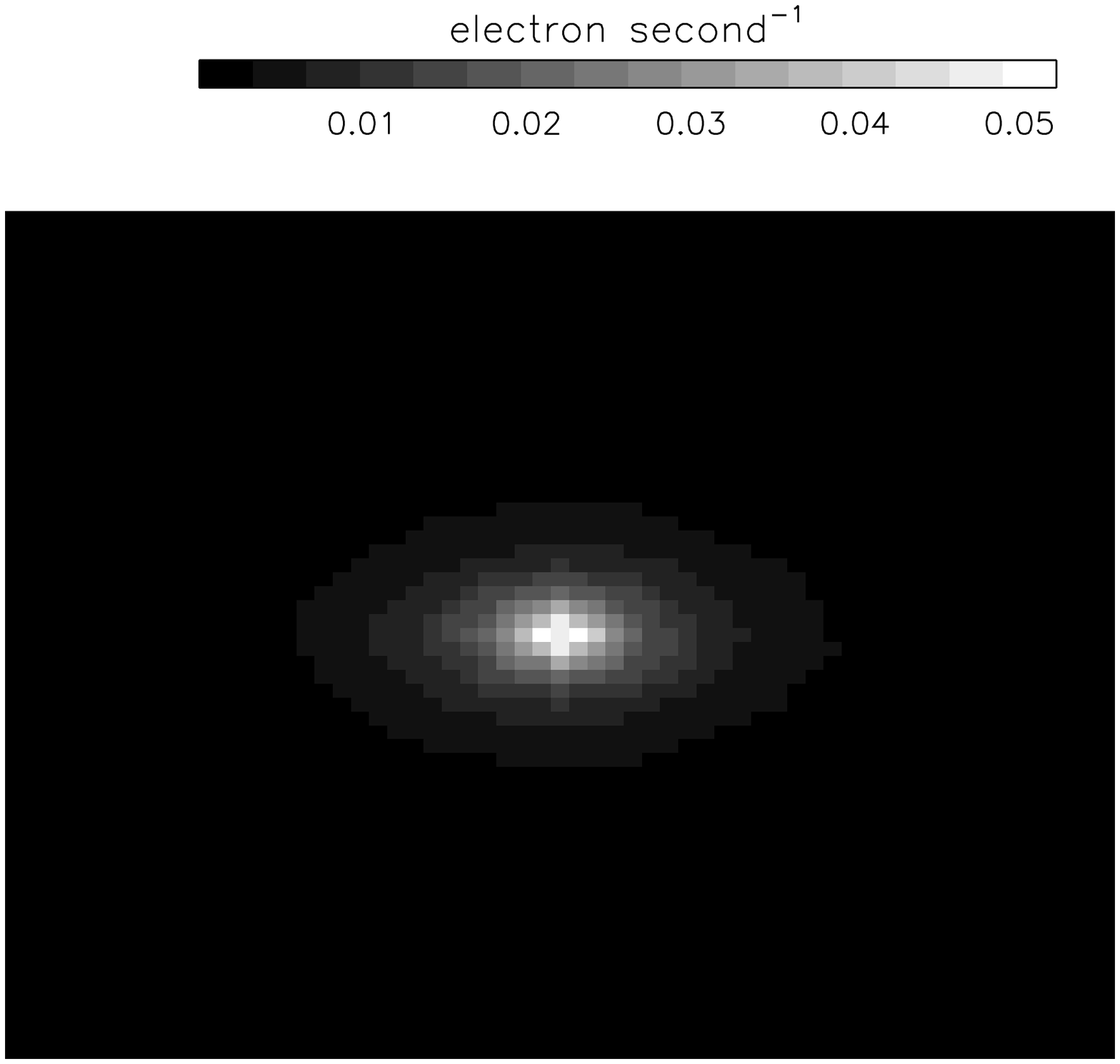,width=35mm,height=45mm}\psfig{figure=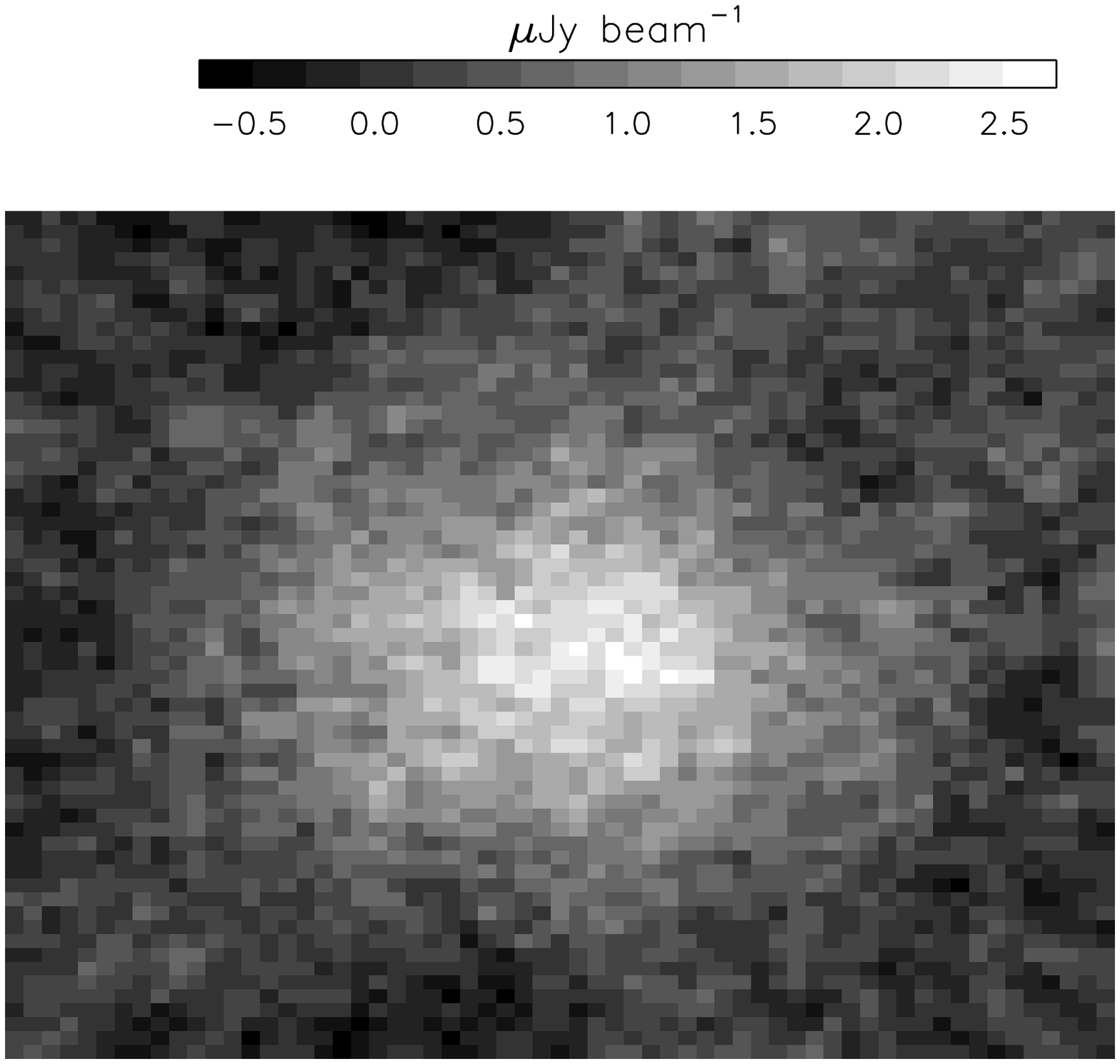,width=35mm,height=45mm}\\\vspace{10mm}
\psfig{figure=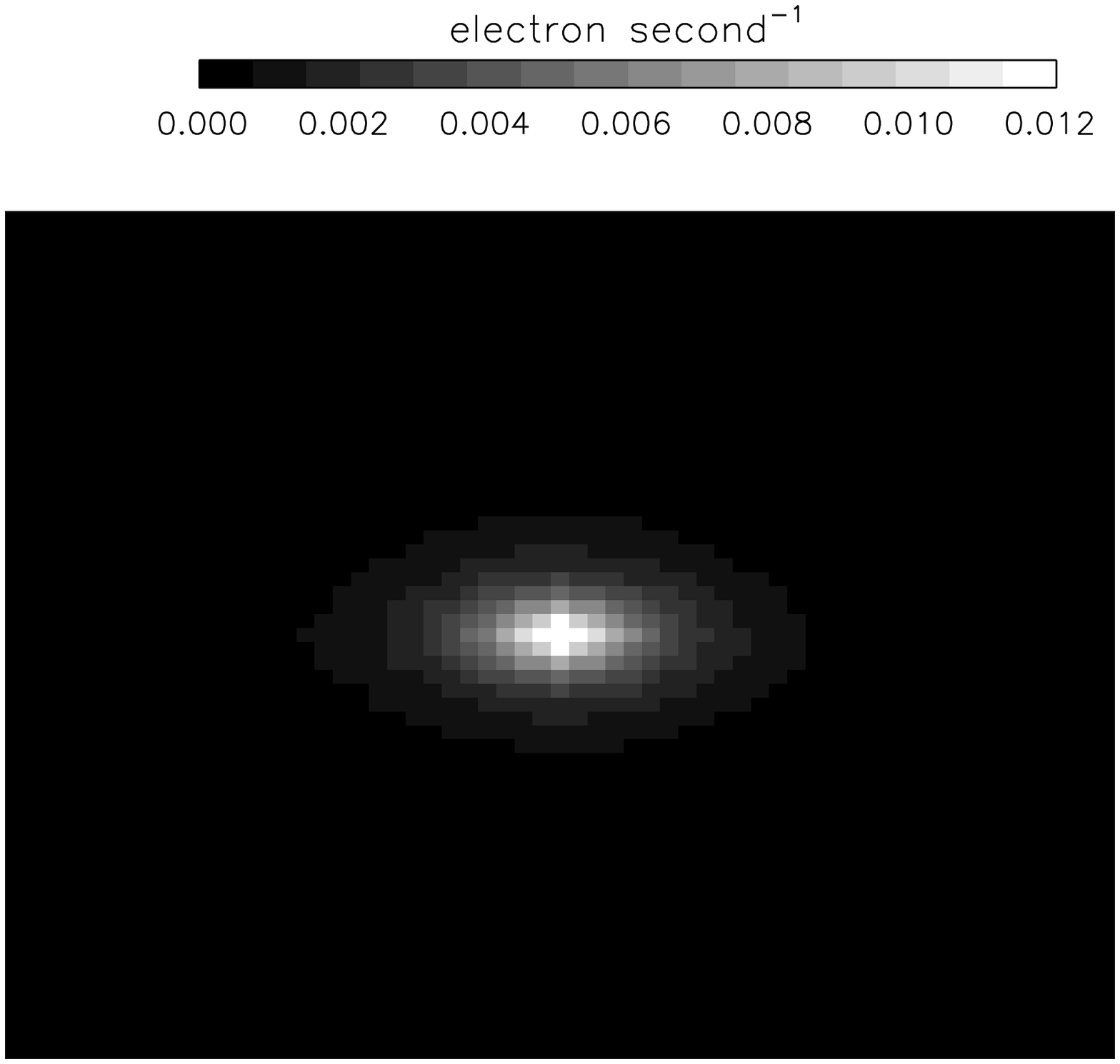,width=35mm,height=45mm}\psfig{figure=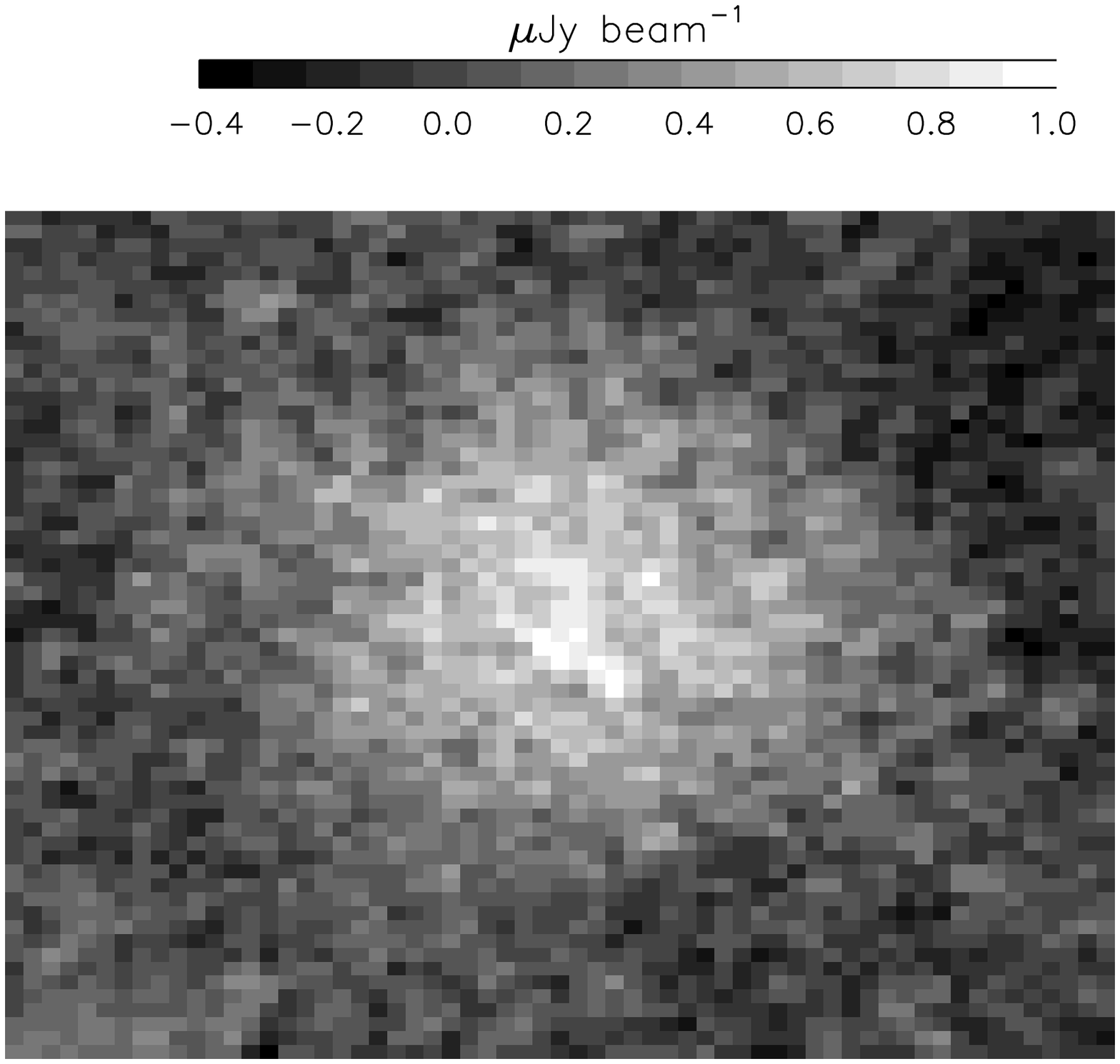,width=35mm,height=45mm}\\\vspace{10mm}
\psfig{figure=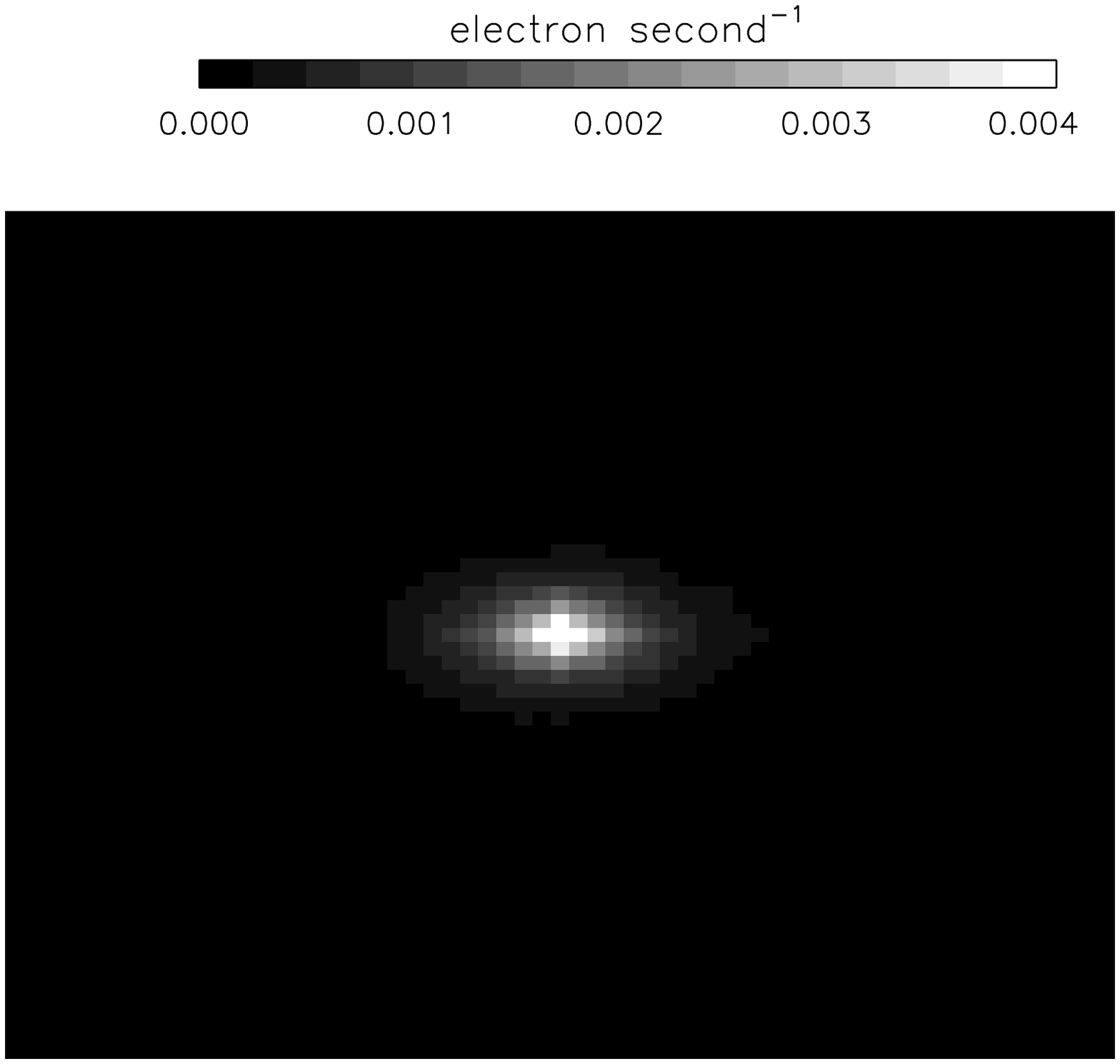,width=35mm,height=45mm}\psfig{figure=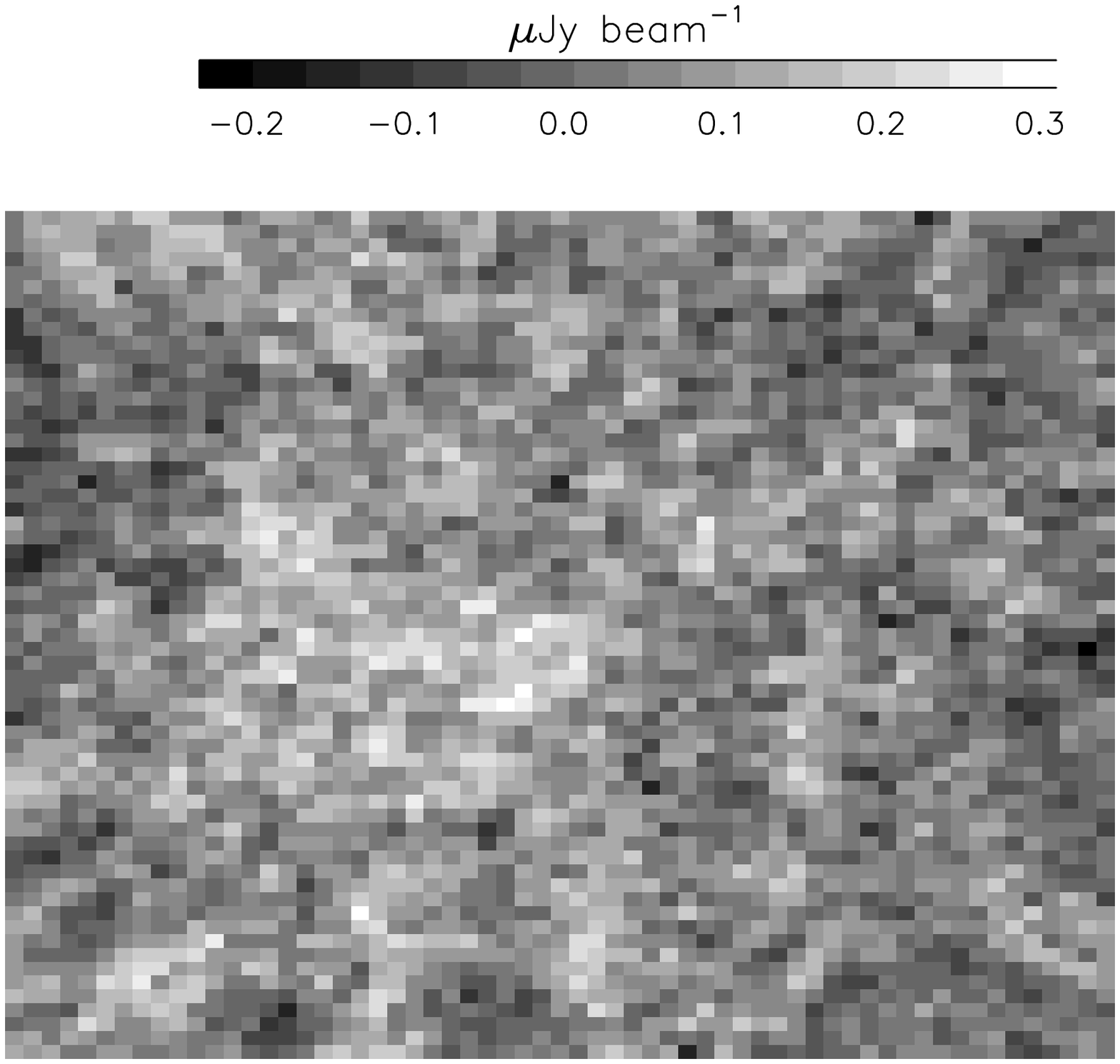,width=35mm,height=45mm}\\\vspace{10mm}
\caption{Left: median stacked, rotated optical images. Right: corresponding median stacked, rotated radio postage stamps. From top to bottom: $m_{z} \leq 20$, $20 < m_{z} \leq 22$, $22 < m_{z} \leq 24$, $24 < m_z \leq 26$. \label{fig:blinde}}
\end{figure}

Since the surface brightness of these objects can be statistically detected, we can also attempt to statistically quantify their ellipticities. In order to do so, we choose optical galaxies in a particular magnitude bin, rotate them so that their major axes are parallel, and create a composite optical galaxy by finding the median value of all galaxy surface brightnesses on a pixel by pixel basis. We then use the positions of the optical galaxies to extract postage-stamps from the radio image of diameter 3.75$''$. We rotate these by the same amount as the relevant optical image, and median average these also. The resulting stacked optical and radio images are shown in Figure \ref{fig:blinde}.

\begin{figure}
\centering
\psfig{figure=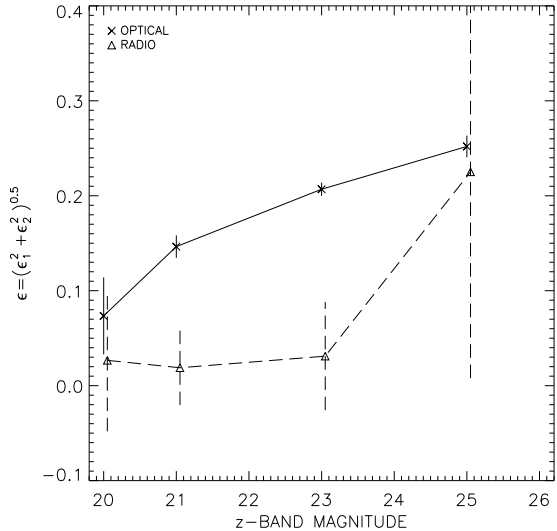,width=80mm}
\caption{Ellipticities of the rotated optical (solid line) and radio (dashed line) composite images in Figure \ref{fig:blinde}. \label{fig:blindeplot}}
\end{figure}

As expected the stacked rotated optical objects are elliptical in shape. If radio object ellipticities were well-aligned with optical ellipticities, we would see the stacked rotated radio objects having an ellipticity on our plots; instead, they are rather circular in shape, suggesting a low correlation in alignment, in confirmation of our result in \S\ref{cross}. Quantitatively, we measure the ellipticity of the images in Figure \ref{fig:blinde}. We calculate the quadrupole moments of the images,
\begin{equation}
Q_{ij}=\frac{\int d^{2}\vec{x}\,I(\vec{x})W(\vec{x})x_{i}x_{j}}{\int d^{2}\vec{x}\,I(\vec{x})},
\end{equation}
where $I(\vec{x})$ is the intensity profile of the object and $W(\vec{x})$ is a weighting function, which here is a top-hat function with radius equal to a third of the radius of the postage stamp. The 2-component ellipticity can then be calculated as
\begin{equation}
\epsilon_{1,2}=\frac{\{Q_{11}-Q_{22},2Q_{12}\}}{Q_{11}+Q_{22}}.
\end{equation}
The calculated ellipticities are shown as a function of the $z-$band magnitude in Figure \ref{fig:blindeplot}, with errors calculated via jack-knifing of the object ellipticities in each magnitude bin. We see that the ellipticities of the rotated stacked radio images are much lower than the optical stacked images; again, this acts as a confirmation of our results in \S\ref{cross}, as it means that radio emission is not coherently oriented with respect to the optical emission. We see that at optical magnitudes greater than $m_{z}=23$ the error on the radio ellipticity is substantial, as the radio emission is not well characterized beyond this magnitude limit (see bottom panel of Figure \ref{fig:blinde}). We have therefore applied a magnitude cut of $m_{z}\leq23$ in what follows.

\begin{figure}
\centering
\psfig{figure=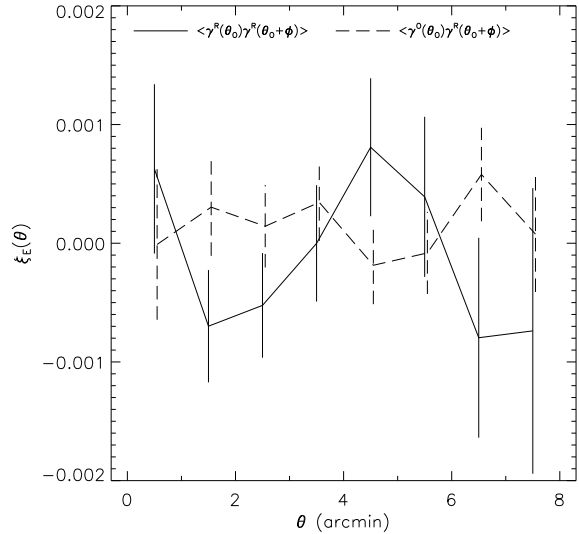,width=80mm}\\\psfig{figure=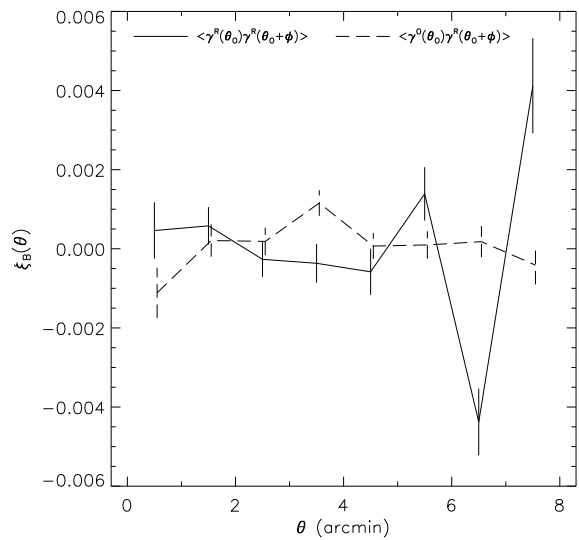,width=80mm}
\caption{Blind radio shear correlation functions, showing the \E-mode (top) and \B-mode (bottom). Solid line: using optical positions only, with radio ellipticities; dashed line: cross-correlating optical shear estimators with blind radio ellipticities. \label{fig:blindcfn}}
\end{figure}

\begin{figure}
\centering
\psfig{figure=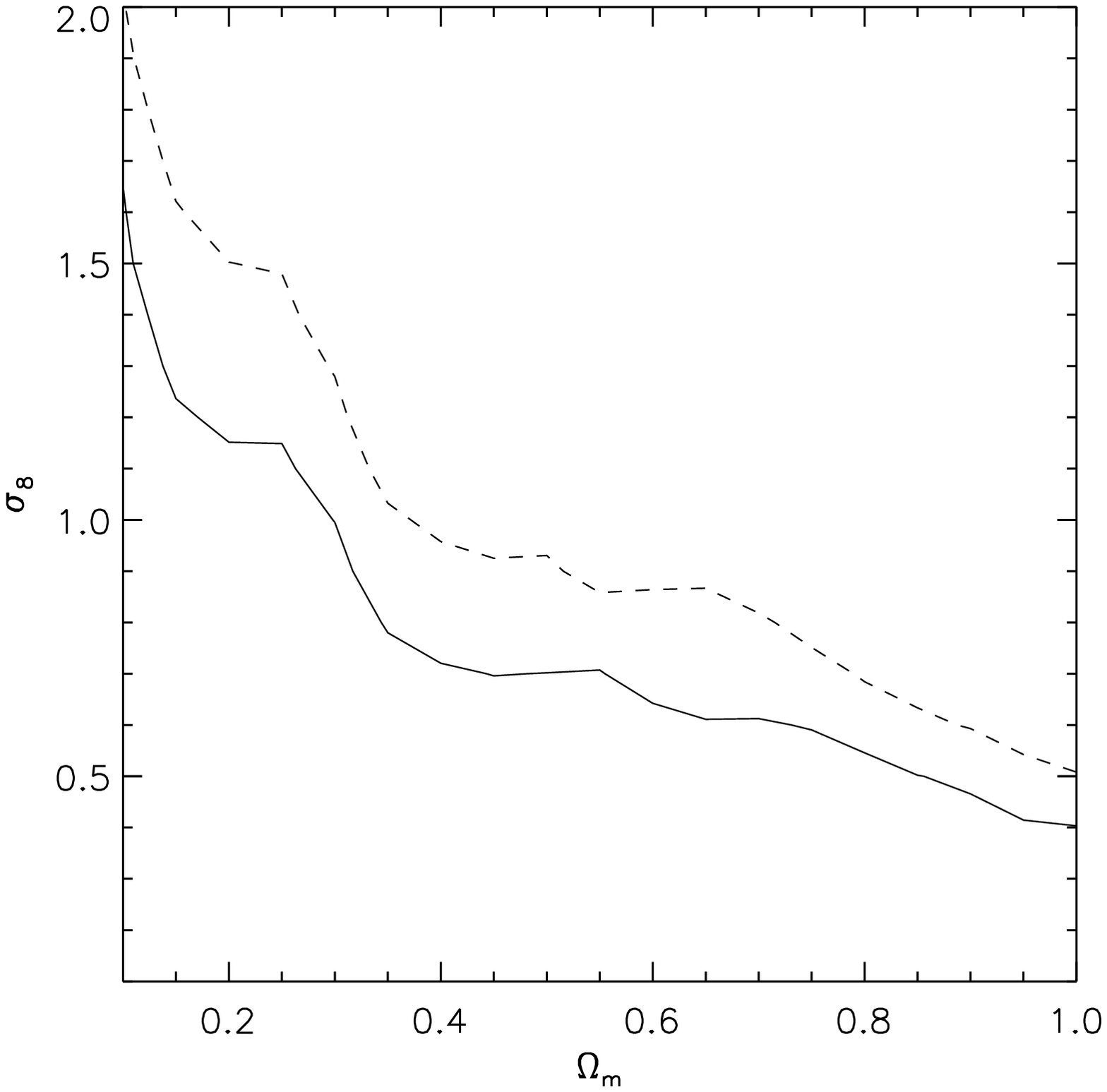,width=80mm}\\\psfig{figure=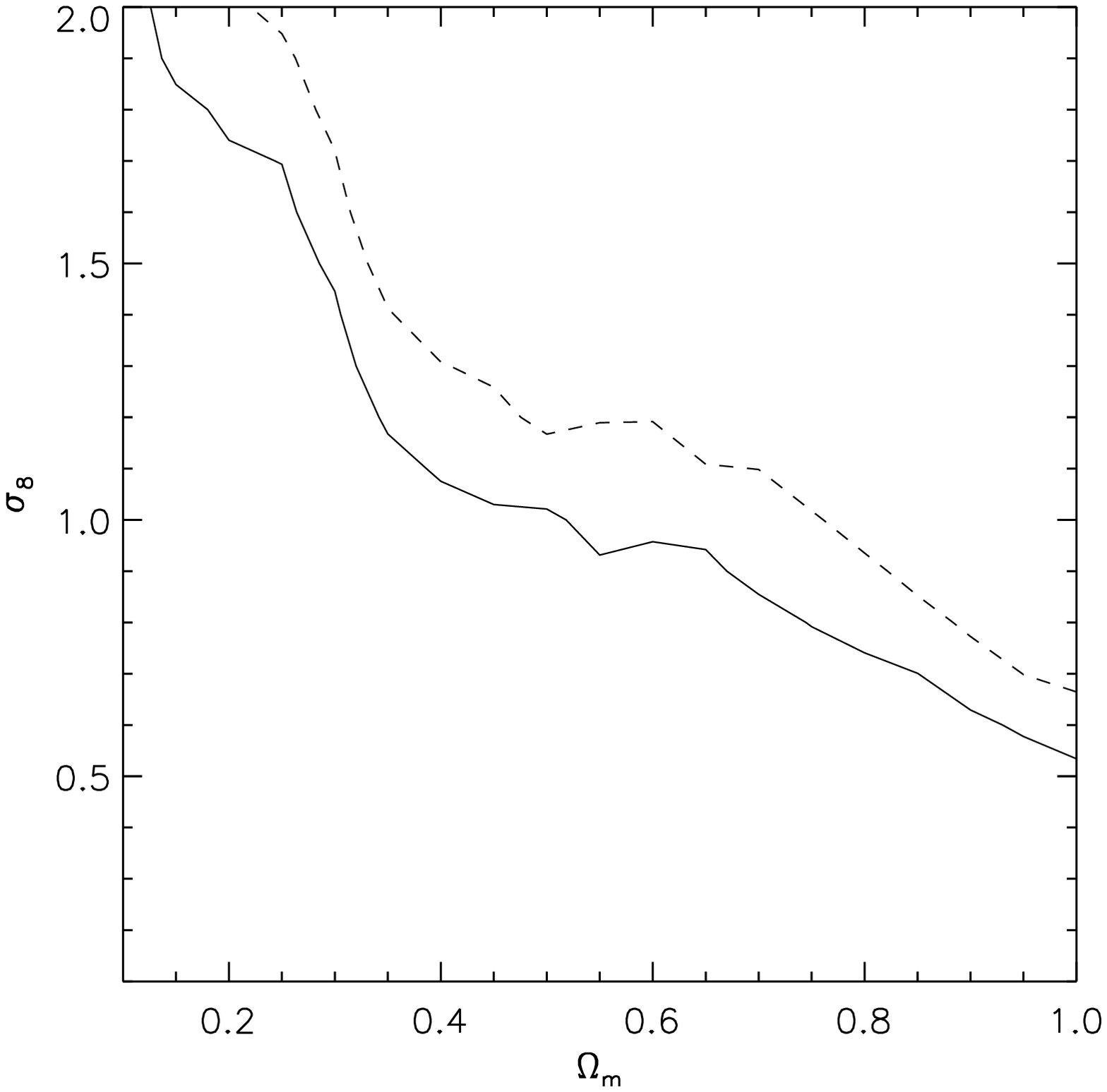,width=80mm}
\caption{Cosmological constraints from the blind-blind (top panel) and blind-optical cross correlation functions with $z_m=1.0$. \label{fig:blindcosmo}}
 \end{figure}

The natural extension to this analysis is to form blind radio shear correlation functions. By using only the positions of the optical objects we can extract postage stamps from the radio image and again calculate the ellipticites using the quadrupole approach above; we can then calculate the correlation functions for the simple shear estimator $\gamma\simeq\epsilon/(2-\langle\epsilon^2\rangle)$. The resulting correlation functions are shown as solid lines in Figure \ref{fig:blindcfn}. The errors from this technique are comparable with those found for optical-radio correlations in section \ref{cosmic}; resulting cosmological constraints are shown in Figure \ref{fig:blindcosmo}, amounting to

\begin{equation} 
\sigma_{8}\left(\frac{\Omega_{m}}{0.25}\right)^{0.5}z_{m}^{1.6}<1.0 \,\,\,\, (1\sigma)
\end{equation}
We can also make use of the optical shear estimators we have already calculated. We cross-correlate these optical shears with the blind radio ellipticities, showing the resulting correlation functions as the dashed lines in Figure \ref{fig:blindcfn} and cosmological constraints in Figure \ref{fig:blindcosmo}, giving
\begin{equation}
\sigma_{8}\left(\frac{\Omega_{m}}{0.25}\right)^{0.5}z_{m}^{1.6}<1.4 \,\,\,\, (1\sigma)
\end{equation}
The fact that these constraints are comparable with those from the more conventional radio-optical correlation function in \ref{cosmic} will lead us to pursue both techniques in future work.

\section{Conclusions}
\label{conclusions}

In this paper we have endeavoured to measure some key properties of
weak gravitational lensing at radio wavelengths. The motivation for
this is the fact that to date, almost all
\citep[apart from][]{2004ApJ...617..794C} weak lensing studies have been carried
out at optical wavelengths. Forthcoming radio telescopes such as {\it e}-MERLIN, LOFAR and ultimately the SKA with their massively increased sensitivity, will be able to detect a vast number of radio sources that will make weak lensing accessible at radio wavelengths, so this work makes some early steps towards the full exploitation of those future surveys.

In this work we have taken some of the most sensitive radio data currently available and measured the shear estimator distribution of the detected radio sources. Making use of the publicly available optical HST ACS data we were able to directly compare the properties of the radio sources and the corresponding optical galaxies. We have used the shapelets method of \citet{2003MNRAS.338...35R} to measure the shapes of our sources while deconvolving the beam/PSF. For the radio data the beam was exactly known, whereas for the optical data the PSF was estimated using the stars in the image. We found a useable radio number density for shear analysis of $n=3.8 \mbox{ arcmin}^{-2}$ compared to that of the optical $n=40.7 \mbox{ arcmin}^{-2}$. The radio number density is therefore still rather low compared to optical, but is much larger than that available to \citet{2004ApJ...617..794C}.

We used a simple shear estimator for our radio and optical objects and found comparable shear distributions; the dispersion $\sigma_{\gamma}$ governs the error on the derived cosmic shear correlation functions, and we find its value in the optical and radio is comparable ($\sigma_\gamma=0.3$ to 0.4 depending on the particular sample).

We present various tests of systematics, including measuring mean shear as a function of image position, flux, and FWHM, finding little evidence of systematics in our gold radio set, but substantial remaining systematics in the silver radio and optical sets, particularly at a low flux limit. In the silver radio case, this may be due to the effect of anisotropic noise affecting the measured ellipticity of objects. 

We measure shear correlation functions for the gold radio, silver radio and optical datasets, finding B-mode contamination on scales $<2'$; the uncertainty on the radio correlation functions is too large to provide significant cosmological constraints, whereas the cross-correlation of radio and optical shear estimators provides an upper bound $\sigma_8 (\Omega_m/0.25)^{0.5} z_m^{1.6} < 0.8$ ($1\sigma$, not including cosmic variance). This cross-correlation also removes the observed systematic error in the $\xi_{tr}$ correlation function.

When we compared the shear estimators in optical and radio for the matched objects, we found very little correlation between the shear estimators of the radio source and the optical counterparts, with a Pearson's correlation coefficient of $\sigma_{P}=0.097\pm0.090$. This is in apparent contrast to \citet{2009arXiv0902.1631B} who recently showed that there is an excess of galaxies in which the major axes of the optical and radio emission are aligned when comparing optical galaxies from SDSS and radio sources from FIRST. However, we have argued that our very different regime of depth and resolution accounts for this disparity.

We have discussed how the low correlation can be beneficial in extracting a cleaner cosmic shear signal given the availability of both radio and optical data sets, by measuring the radio-optical shear cross-correlation function; this is borne out by the improvement in systematics this provides in our measured cosmic shear constraints. Additionally, the fact that the galaxies have uncorrelated shapes in the radio and optical data increases the effective source density.

Finally, motivated by \citet{2007ASPC..380..199M}, we made blind measurements of radio shear at the known positions of optical galaxies. We confirmed the result of \citet{2007ASPC..380..199M} that radio emission down to a few $\mu $Jy is statistically detected from optical galaxies as faint as $m_{z}=25$. Building on this, we proceeded to measure the average shape of these statistically detected objects. By stacking optical images that had been rotated by their orientation angle alongside postage stamps of the radio image with the same optical position and rotation, we found very little association between the ellipticities of the resulting optical stacks and the radio stacks, confirming our earlier optical-shear correlation result.

We constrained the shear correlation functions in this blind approach; using either the radio ellipticities found at the optical positions, or using the cross-correlation between optical shear estimators and radio blind ellipticities, we obtain errors comparable with the more conventional approach above.

The principal goal of this paper was to determine whether weak lensing at radio wavelengths is possible and promising. The improved number densities and suitable $\sigma_\gamma$ reported here, together with the reasonable cosmological constraints measured, argue towards this being the case. Indeed, the precise knowledge of the beam for the deconvolution step provides radio weak lensing with a potential advantage over optical studies. However, a careful analysis of systematics will be required, as we have shown that they are currently dominant unless one cross-correlates with optical data. Perhaps this is a satisfying conclusion: that in the future, we may not favour optical or radio shear surveys, but both together.


\section*{Acknowledgments} 
We would like to thank Richard Battye, Karen Masters and Charles Shapiro for useful comments and suggestions regarding the progress of this work. PP would like to thank Mathew Smith, Cristiano Sabiu \& Janine Pforr for insightful discussions. DB is supported by an STFC Advanced Fellowship and an RCUK Research Fellowship. PP is funded by a STFC PhD studentship.

This research has made use of the NASA/IPAC Extragalactic Database (NED) which is operated by the Jet Propulsion Laboratory, California Institute of Technology, under contract with the National Aeronautics and Space Administration.


\label{lastpage}

\end{document}